\documentclass[11pt,a4paper]{article}

\pdfoutput=1
\usepackage[left=2.5cm,right=2.5cm,top=2.5cm,bottom=2.5cm]{geometry}
\usepackage[onehalfspacing]{setspace}
    \usepackage[pdftex,     
            plainpages=false,   
            breaklinks=true,    
            colorlinks=true
           ]{hyperref} 
\usepackage{thumbpdf}
\usepackage{amsmath}
\usepackage{amssymb}
\usepackage{amsfonts}
\usepackage{array}
\usepackage{bbold}
\usepackage{cancel}
\usepackage[hang,small]{caption2}
\usepackage{subfigure}
\usepackage{verbatim}
\usepackage{slashed}
\usepackage{graphicx}
\usepackage{url}

\numberwithin{equation}{section}

\newcommand{\be}{\begin{equation}}
\newcommand{\ee}{\end{equation}}

\makeatletter
\def\maketag@@@#1{\hbox{\m@th\normalfont\normalsize#1}}
\makeatother

\begin{document}

\begin{titlepage}

\vspace*{1cm}

\begin{center}{
\bfseries\LARGE
STR: a Mathematica package for the method of uniqueness}\\[8mm]
Michelangelo~Preti $^{a,b}$ \\[1mm]
\end{center}

\vspace*{0.50cm}

\centerline{\itshape
$\,^{a}$Laboratoire de Physique Th\'eorique, \'Ecole Normale Sup\'erieure,}

\centerline{\itshape 24 rue Lhomond 75005 Paris, France \footnote{E-mail: \texttt{michelangelo.preti@lpt.ens.fr}}}

\centerline{\itshape $\,^{b}$PSL Universit\'es, Sorbonne Universit\'es, UPMC Universit\'e Paris 6, CNRS}

\vspace*{1cm}
\begin{abstract}
We present \texttt{STR} (\textbf{S}tar-\textbf{T}riangle \textbf{R}elations), a \textit{Mathematica}\textsuperscript{\textregistered} package designed to solve Feynman diagrams by means of the method of uniqueness in any Euclidean spacetime dimension. The method of uniqueness is a powerful technique to solve multi-loop Feynman integrals in theories with conformal symmetry imposing some relations between the powers of propagators and the spacetime dimension. In our algorithm we include both identities for scalar and Yukawa type integrals. The package provides a graphical environment in which it is possible to draw the desired diagram with the mouse input and a set of tools to modify and compute it. Throughout the use of a graphic interface, the package should be easily accessible to users with little or no previous experience on diagrams computation. This manual includes some pedagogical examples of computation of Feynman graphs as the scalar two-loop kite master integral and a fermionic diagram appearing in the computation of the spectrum of the $\gamma$-deformed $\mathcal{N}=4$ SYM in the double-scaling limit.
\end{abstract}
\end{titlepage}

\section*{Program Summary}\noindent
\textbf{Author:} Michelangelo Preti\\\noindent
\textbf{Title:} STR\\\noindent
\textbf{Licence:} GNU LGPL\\\noindent
\textbf{Programming language / external routines:} Wolfram \textit{Mathematica}\textsuperscript{\textregistered} version 8.0 or higher\\\noindent
\textbf{Operating system:} cross-platform\\\noindent
\textbf{RAM:} $>512$ MB RAM recommended\\\noindent
\textbf{Current version:} 1.0\\\noindent
\textbf{Nature of the problem:} 
The main goal of a perturbative quantum field theory (QFT) is the exact computation of multi-loop Feynman diagrams that are crucial in many applications in particle, statistical and condensed matter physics. Since the number of diagrams grows rapidly with the perturbation theory order and the precision of numerical calculations is not satisfactory, the development of analytic tools to compute multi-loop integrals plays a central role in QFT. One of those powerful techniques is the \textit{method of uniqueness} that provides a reduction method for involved Feynman diagrams as a sequence of simple transformations without performing any explicit integration. Considering that such a sequence could be extremely long and not unique, the development of a simple and automatic method is needed.
\\\noindent
\textbf{Solution method:} The aim of the package \texttt{STR} is to solve Feynman diagrams by means of the method of uniqueness for any Euclidean spacetime dimension $D$.
Its main feature is to provide a simple graphical environment to draw Feynman integrals and interact with them only by the use of the mouse. All the operations of the method of uniqueness are implemented in the same window as well as for the drawing and output tools.
\\\noindent
\textbf{Web page:}  \url{https://github.com/miciosca/STR}\\\noindent
\textbf{Contact:} \url{michelangelo.preti@gmail.com} --- for bugs, possible improvements and questions.

\section{Introduction}
\label{sec:intro}

In recent years, remarkable progresses have been achieved in evaluating higher-order corrections to various quantities in terms of multi-loop Feynman diagrams in perturbative quantum field theories (QFT). This investigation is stimulated by the efforts to make more precise predictions as well as a deeper understanding of the structure of perturbation series in QFT by attempting to resum it. 

With this purpose, different methods of multi-loop calculations have been developed. 
One of the most efficient technique is the \textit{integration by parts} method (IBP) \cite{Vasiliev:1981yc,Tkachov:1981wb,Chetyrkin:1981qh} 
that allows to reduce an involved Feynman integral in terms of a base of a finite number of \textit{master integrals}. This method is implemented in \textit{Mathematica}\textsuperscript{\textregistered} and \texttt{C++} with the package \texttt{FIRE5} \cite{Smirnov:2014hma} by means of the Laporta algorithm \cite{Laporta:2001dd} . Sometimes the IBP method is not sufficient to solve the diagrams and then it has to be combined with other techniques as the one of \textit{Gegenbauer polynomial} \cite{Chetyrkin:1980pr}, or \textit{Mellin transform} \cite{Bergere:1973fq,Usyukina:1975yg}, or \textit{HQET} (see \cite{Grozin:2014hna,Grozin:2015kna,Bianchi:2017svd} for recent applications in 4 and 3 dimensions), or \textit{differential equations} \cite{Remiddi:1997ny}. For a review of the various methods see also \cite{Smirnov:2012gma}.

Another very powerful technique in multi-loop computations is the \textit{method of uniqueness}, also known as \textit{star-triangle relation}. In the following we will focus on this method. 
This relation was used the first time to solve three-dimensional integrals in the computation of the critical exponents of Bose liquids \cite{DEramo:1971hnd}.
In the context of conformal field theories, the star-triangle relation was firstly considered in \cite{Fradkin:1978pp} and successively in the framework of multi-loop calculations in \cite{Vasiliev:1981yc,Usyukina:1983gj,Kazakov:1984km}.
This method provides a reduction tool for complicated Feynman diagrams encoded in a sequence of simple transformations without performing any explicit integration.
Once the appropriate sequence is found, the diagram can be easily integrated. For example, the analytic expression of the $\beta$-function at 5-loop in the $\varphi^4$-theory was computed with the uniqueness method in \cite{Kazakov:1983ns,Kazakov:1983pk}.
Moreover, in \cite{Zamolodchikov:1980mb} was pointed out that the star-triangle relation is a kind of Yang-Baxter equation. Indeed, this intuition was crucial for the computation of the sufficiently large \textit{fishnet} diagrams in the $\varphi^6$, $\varphi^4$ and $\varphi^3$ theories in $D=3$, $D=4$ and $D=6$ respectively and also for the R-matrix formulation \cite{Derkachov:2001yn} of the integrable non-compact Heisenberg spin-chain proposed in \cite{Lipatov:1993qn,Lipatov:1993yb}.
The method of uniqueness hold also in the case we are considering a Yukawa theory with $\psi\varphi\bar{\psi}$ vertices. Its generalization, including $\gamma$-matrices and propagators of spin particles, were considered in \cite{Symanzik:1972wj,Fradkin:1978pp,Isaev:2003tk,Chicherin:2012yn}.

In this paper we present \texttt{STR} (\textbf{S}tar-\textbf{T}riangle \textbf{R}elations), a \textit{Mathematica}\textsuperscript{\textregistered} package designed to solve Feynman diagrams by means of the method of uniqueness. We are interested to solve integrals in position space in any Euclidean spacetime dimension $D$ both in the cases of scalar and fermionic star-triangle relations. The main feature of the package is to provide an user-friendly graphical way to solve Feynman integrals interacting with the algorithm only by the use of the mouse. Indeed, the main function of the package will furnish a graphical environment in which it is possible to draw any kind of Feynman diagram left- or right-clicking and dragging the mouse. Once the graph is set, the user can interact with it with the help of many tools designed to modify it, identify unique star or triangles and compute it (or part of it) by means of the uniqueness relations. At any step of the process, the user can also print or export the main data (list of the uniqueness equations for the weights, the result of the computations and the integral representation of the diagram) and the graph itself. 

The paper is organized as follows. In section 2 we review the method of uniqueness for the scalar and fermionic cases. In section 3 we present in detail the manual of the \texttt{STR} package, while in section 4 we show some simple applications. An appendix follows, which contain a list of notation and conventions.

\section{The star-triangle relations (method of uniqueness)}
\label{sec:startriangle}

In this paper we will consider the usual Feynman diagrams with vertices connected by lines labeled by an index (weight). Each vertex represents a point in the $D$-dimensional Euclidean space $\mathbb{R}^D$ while the lines of the graph (with weights $\alpha_i$) are associated with the following massless propagators ($x_{ij}\equiv x_i-x_j$)
\begin{equation}\begin{split}\label{prop}
\vcenter{\hbox{\includegraphics[width=3.7cm]{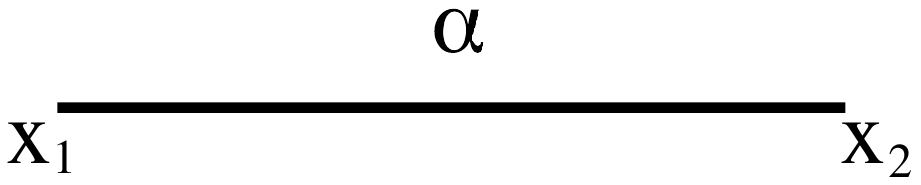}}}=\frac{1}{(x_{12}^2)^\alpha}\qquad\text{and}\qquad\vcenter{\hbox{\includegraphics[width=3.7cm]{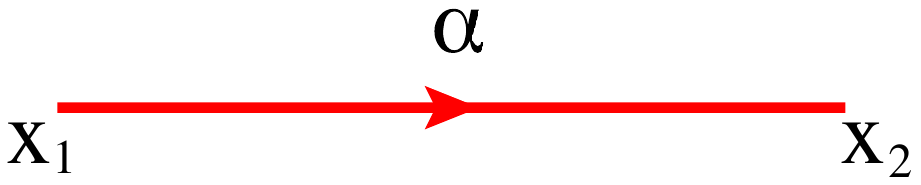}}}=\frac{\slashed{x}_{12}}{(x_{12}^2)^{\alpha+1/2}}\,,
\end{split}\end{equation}
respectively the scalar and spin-1/2 fermionic massless propagators. The symbol $\slashed{x}$ stands for the contraction between the position $x^\mu$ and the element of the Clifford algebra in $D$ dimensions, namely $\sigma_\mu x^\mu$ or $\bar{\sigma}_\mu x^\mu$ depending if the fermion is chiral or anti-chiral in even dimensions and $\gamma_\mu x^\mu$ in odd dimensions (see Appendix \ref{sec:appendixA}). 
The black dot vertices denote that the corresponding points are integrated over $\mathbb{R}^D$.
All the computation will be performed in coordinate space. However one can also choose to compute the Feynman integrals in momentum space using the following Fourier transforms
\begin{equation}\label{Fourier}
\frac{1}{(x_{12}^2)^\alpha}=\frac{\mathbb{a}_0(\alpha)}{4^\alpha\pi^{D/2}}\int d^Dk\frac{e^{ik\cdot x_{12}}}{(k^2)^{D/2-\alpha}}\qquad\text{and}\qquad
\frac{\slashed{x}_{12}}{(x_{12}^2)^{\alpha+1/2}}=\frac{-i\,\mathbb{a}_{1/2}(\alpha)}{4^\alpha\pi^{D/2}}\int d^Dk\frac{e^{ik\cdot x_{12}}\slashed{k}}{(k^2)^{D/2-\alpha+1/2}}\,,
\end{equation}
where
\begin{equation}\label{defa}
\mathbb{a}_{\ell}(\alpha)=\frac{\Gamma\left(\frac{D}{2}-\alpha+\ell\right)}{\Gamma(\alpha+\ell)}\qquad\text{with}\qquad\mathbb{a}_{\ell}(\alpha)\mathbb{a}_{\ell}(D/2-\alpha)=1\,.
\end{equation}

We call the \textit{weight of the diagram} (or of a portion of it) the sum of all the weights of the constituent lines. 
For instance, the weight of a three-leg vertex (also called \textit{star}) is defined as the sum of the weights of the three lines that converge at it, and the weight of a \textit{triangle} is defined as the sum of the weights of the three lines that form it. We shall say that a line, star and triangle are \textit{unique} if their weights are 0, $D$ and $D/2$ respectively. If a Feynman diagram contains unique stars or triangles, its computation is drastically simplified. Indeed the so-called \textit{method of uniqueness}, based on a set of relations between unique stars and triangles, provides a simple evaluation of massless Feynman diagrams according to the rules listed below. In this section we are reviewing \cite{Kazakov:1984km,Kazakov:1983pk,Usyukina:1983gj,Belokurov:1983km,Chicherin:2012yn}. 

\paragraph{1. Merging rules:} In the following, we list a set of identities to represent simple loop of propagators as a single line with different weight. It is straightforward to see that in the case in which more than two lines are passing through the same two points, the loop can be written as a combination of the following simple rules.
\begin{itemize}
\item The contribution of a simple loop of bosonic propagators is an ordinary product
\begin{equation}\label{merge1}
\vcenter{\hbox{\includegraphics[width=3.7cm]{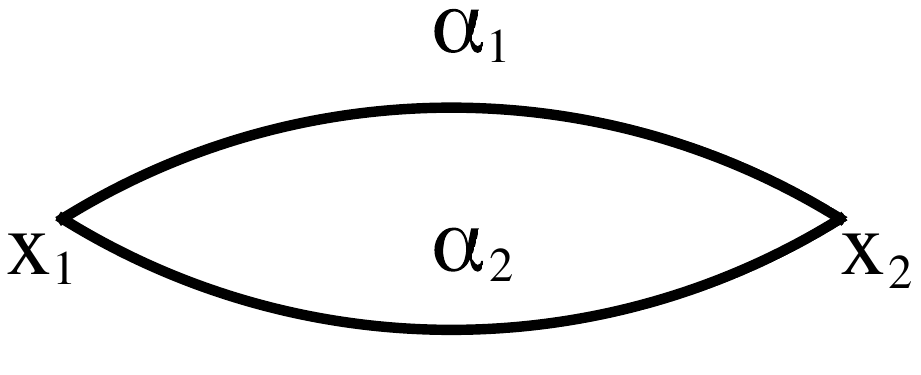}}}=\vcenter{\hbox{\includegraphics[width=3.7cm]{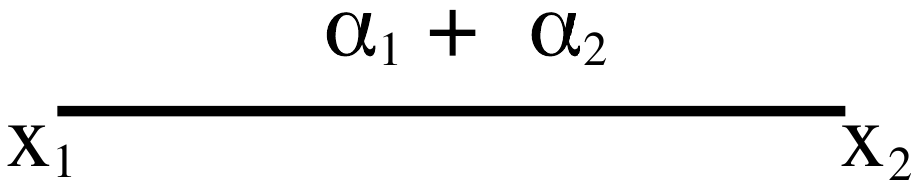}}}\,.
\end{equation}
\item The contribution of a simple loop of a bosonic and a fermionic propagator is a fermionic line with weight given  by the sum of the original weights
\begin{equation}\label{merge2}
\vcenter{\hbox{\includegraphics[width=3.7cm]{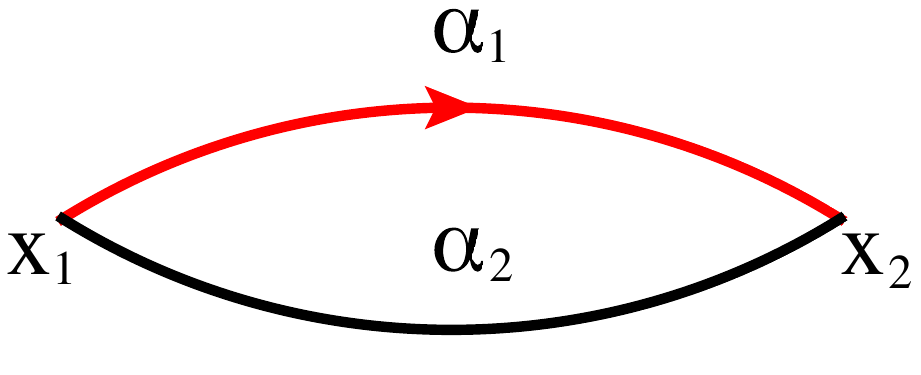}}}=\vcenter{\hbox{\includegraphics[width=3.7cm]{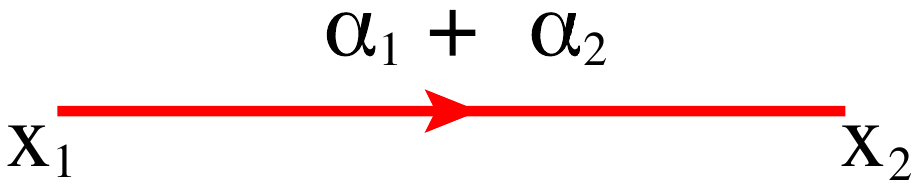}}}\,.
\end{equation}
\item The contribution of a simple loop of fermionic propagators, since we are considering the spin structure to be contracted, is equal to a bosonic line with weight given by the sum of the original weights 
\begin{equation}\label{merge3}
\vcenter{\hbox{\includegraphics[width=3.7cm]{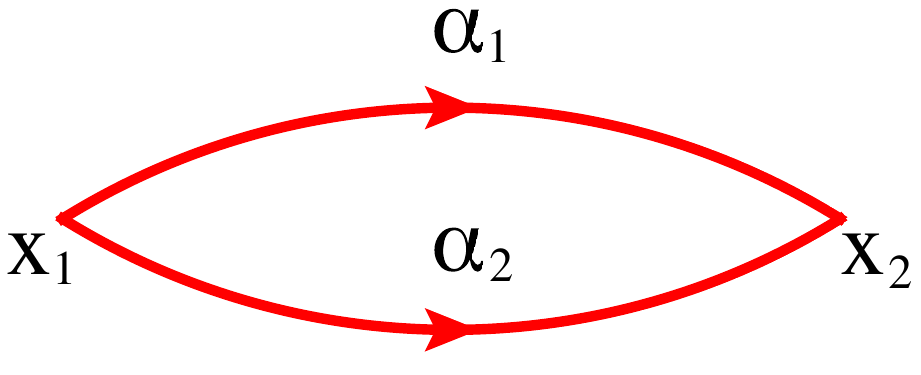}}}=\mathbb{1}\;\vcenter{\hbox{\includegraphics[width=3.7cm]{merge_bosbos2}}}\,.
\end{equation}
To compute it we used the $D$-dimensional Clifford algebras \eqref{clifford} for odd $D$ and \eqref{symferm} for even $D$. We are considering that adjacent fermionic lines are contracted in the spin structure alternating $\sigma$'s  and $\bar{\sigma}$'s (even $D$) or simply contracting $\gamma$'s (odd $D$)\footnote{\label{foot:2}For instance in this case the numerator is: for even $D$ $\slashed{x}_{12}\slashed{x}_{12}=(\sigma_\mu)_{\alpha\dot{\alpha}}(\bar{\sigma}_{\nu})^{\dot{\alpha}\beta}x_{12}^\mu x_{12}^\nu=x_{12}^2(\mathbb{1}_{2^{D/2-1}})_{\alpha}^{\;\,\beta}$ and for odd $D$ $\slashed{x}_{12}\slashed{x}_{12}={(\gamma_\mu)_{\alpha}}^\beta{(\gamma_{\nu})_{\beta}}^\gamma x_{12}^\mu x_{12}^\nu=x_{12}^2(\mathbb{1}_{2^{(D-1)/2}})_{\alpha}^{\;\,\gamma}$.}, then the bosonic propagator in the right-hand side of the formula will carry the spin indices in the identity matrix $\mathbb{1}$ that corresponds to $\mathbb{1}_{2^{D/2-1}}$ for even $D$ or $\mathbb{1}_{2^{(D-1)/2}}$ for odd $D$ (see Appendix \ref{sec:appendixA}). 
\end{itemize}

\paragraph{2. Star-triangle relations:} Those are the main identities of the uniqueness method that allows to integrate a unique star into a unique triangle. In this paper we will consider only unique stars and triangles, even if there are also some \textit{semi-unique} \footnote{\label{foot:semiunique}By \textit{semi-uniqueness} we mean star and triangles one-step-deviating from uniqueness, in other words they have weight $D-1$ and $D/2+1$ respectively \cite{Usyukina:1983gj}.}  identities allowing to integrate a star in a linear combination of other stars and triangles with different weights. In the following we will present the relations in the case of a full bosonic and a Yukawa type stars/triangles.   
\begin{itemize}
\item The bosonic star-triangle relation can be written in the following graphical representation
\begin{equation}\label{STRbos}
\vcenter{\hbox{\includegraphics[width=3.7cm]{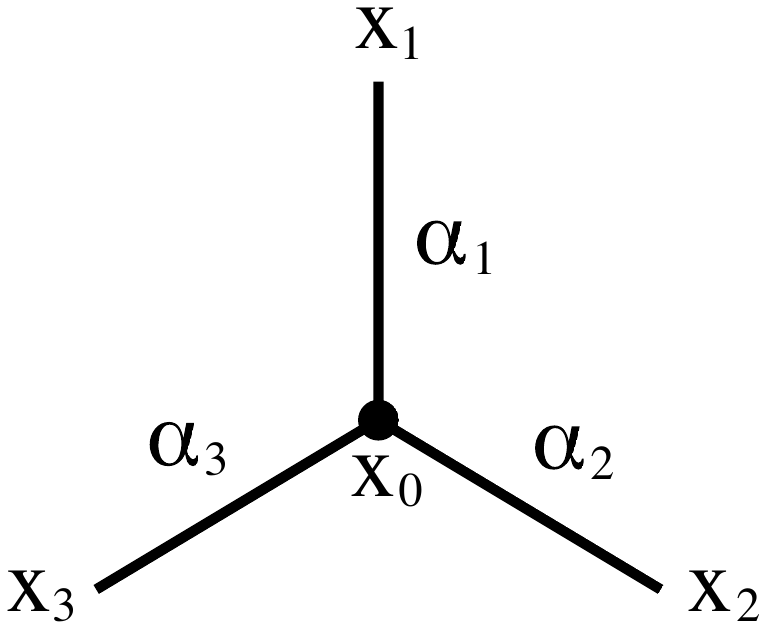}}}\overset{\sum_k\alpha_k=D}{=}
\quad\pi^{D/2}\;\mathbb{a}_{0}(\alpha_1,\alpha_2,\alpha_3)\;\vcenter{\hbox{\includegraphics[width=3.7cm]{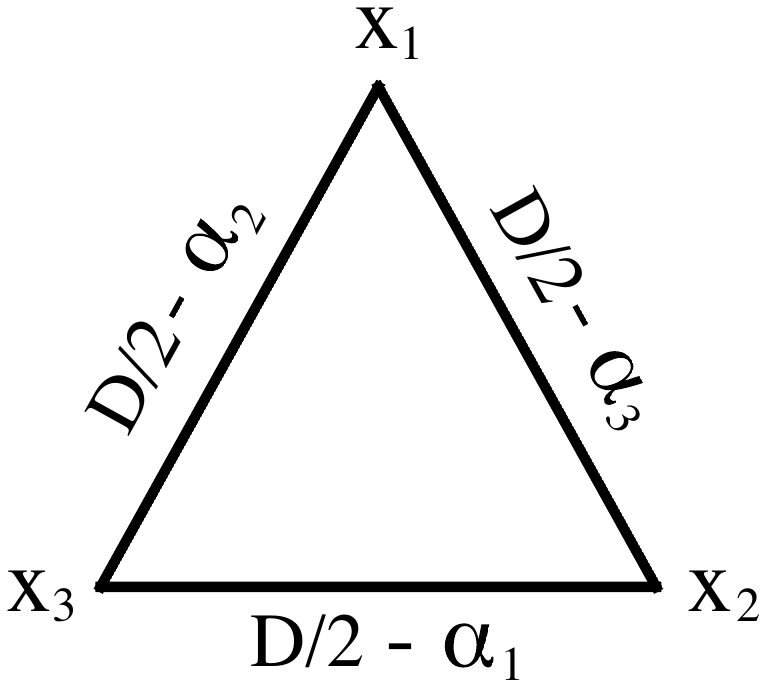}}}\,,
\end{equation}
where the function $\mathbb{a}_{\ell}$ of many arguments has the following property
\begin{equation}\label{amultiple}
\mathbb{a}_{\ell}(\alpha_1,\alpha_2,...,\alpha_n)=\prod_{k=1}^n \mathbb{a}_{\ell}(\alpha_k)\,.
\end{equation}
The same relation can be written in its integral form
\begin{equation}
\int \frac{d^Dx_0}{(x_{10}^2)^{\alpha_1}(x_{20}^2)^{\alpha_2}(x_{30}^2)^{\alpha_3}}\overset{\sum_k\alpha_k=D}{=}
\frac{\pi^{D/2}\;\mathbb{a}_{0}(\alpha_1,\alpha_2,\alpha_3)}{(x_{12}^2)^{D/2-\alpha_3}(x_{23}^2)^{D/2-\alpha_1}(x_{31}^2)^{D/2-\alpha_2}}\,.
\end{equation}
\item  The Yukawa star-triangle relation can be written in the following graphical representation 
\begin{equation}\label{STRferm}
\vcenter{\hbox{\includegraphics[width=3.7cm]{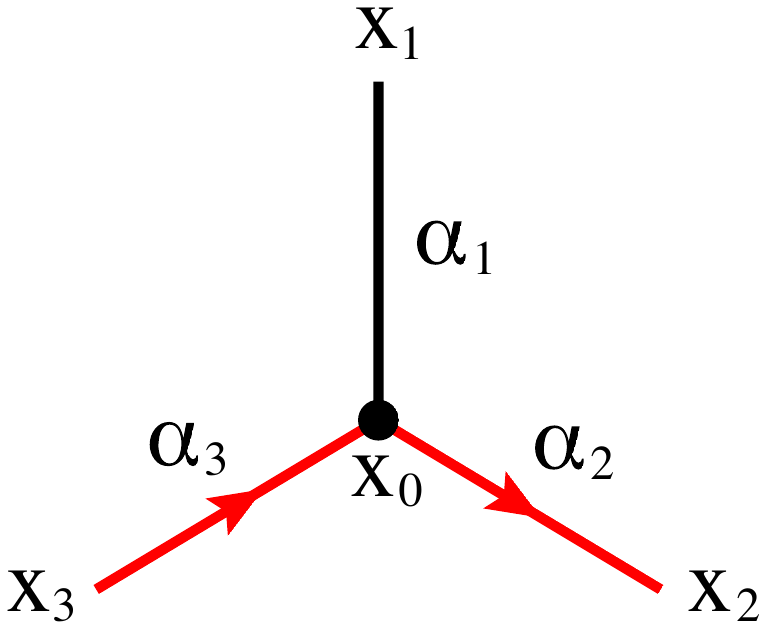}}}\overset{\sum_k\alpha_k=D}{=}
\quad\!\!\pi^{D/2}\;\mathbb{a}_{0}(\alpha_1)\;\mathbb{a}_{1/2}(\alpha_2,\alpha_3)\;\vcenter{\hbox{\includegraphics[width=3.7cm]{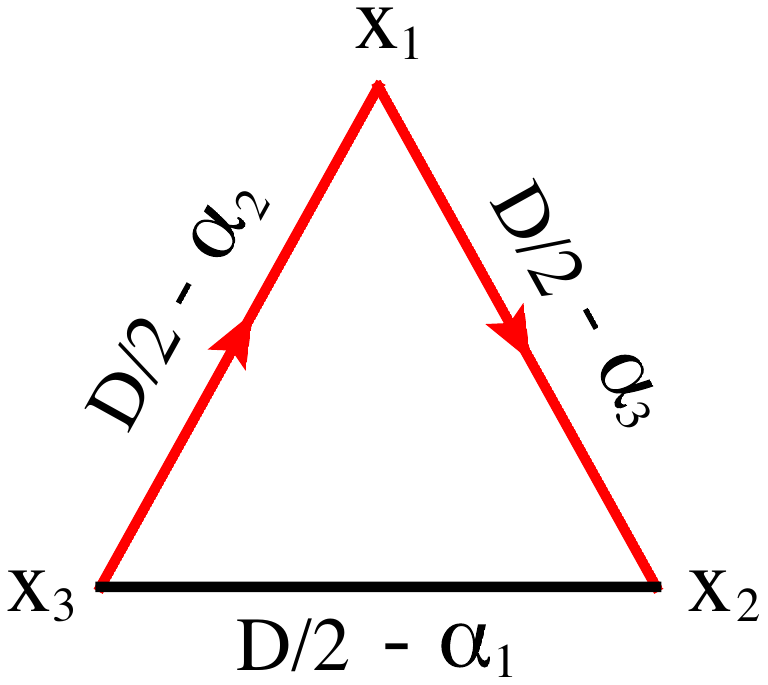}}}\,.
\end{equation}
The same relation can be written in its integral form
\begin{equation}
\int \frac{d^Dx_0\;\slashed{x}_{02}\,\slashed{x}_{30}}{(x_{10}^2)^{\alpha_1}(x_{02}^2)^{\alpha_2+1/2}(x_{30}^2)^{\alpha_3+1/2}}\overset{\sum_k\alpha_k=D}{=}
\!\!\frac{\pi^{D/2}\;\mathbb{a}_{0}(\alpha_1)\;\mathbb{a}_{1/2}(\alpha_2,\alpha_3)\slashed{x}_{12}\,\slashed{x}_{31}}{(x_{12}^2)^{D/2-\alpha_3+1/2}(x_{23}^2)^{D/2-\alpha_1}(x_{31}^2)^{D/2-\alpha_2+1/2}}\,.
\end{equation}
Notice that the order of contraction of the $\sigma$ matrices is not changing after the integration, indeed
\begin{equation}\begin{split}
\text{even}\;D\qquad&\slashed{x}_{02}\,\slashed{x}_{30}=(\sigma_\mu\bar{\sigma}_\nu)_\alpha^{\;\,\beta} x_{02}^\mu x_{30}^\nu\Longrightarrow(\sigma_\mu\bar{\sigma}_\nu)_\alpha^{\;\,\beta} x_{12}^\mu x_{31}^\nu=\slashed{x}_{12}\,\slashed{x}_{31}\,,\\
\text{odd}\;D\qquad&\slashed{x}_{02}\,\slashed{x}_{30}=(\gamma_\mu\gamma_\nu)_\alpha^{\;\,\beta} x_{02}^\mu x_{30}^\nu\Longrightarrow(\gamma_\mu\gamma_\nu)_\alpha^{\;\,\beta} x_{12}^\mu x_{31}^\nu=\slashed{x}_{12}\,\slashed{x}_{31}\,,
\end{split}\end{equation}
and the same if we are choosing the case in which $\sigma\leftrightarrow\bar{\sigma}$. 
\end{itemize}

\paragraph{3. Chain rules:} This set of identities is needed to integrate two propagators meeting in one internal point (a simple loop in momentum space) in terms of a single propagator. Multiple chains are integrated successively by means of a combination of the following rules. 

\begin{itemize}
\item A chain of bosonic propagators is integrated by means of the following identity
\begin{equation}\label{chain1}
\vcenter{\hbox{\includegraphics[width=3.7cm]{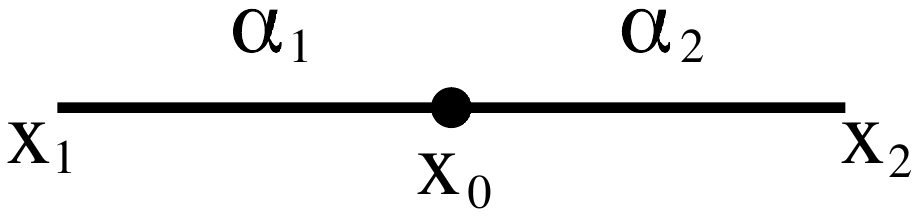}}}=\pi^{D/2}\;\mathbb{a}_{0}(\alpha_1,\alpha_2,D-\alpha_1-\alpha_2)\;\vcenter{\hbox{\includegraphics[width=3.7cm]{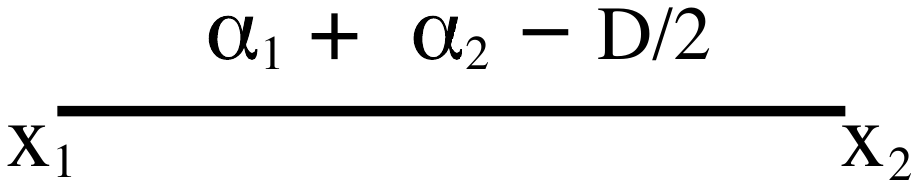}}}\,,
\end{equation}
that in the integral form is
\begin{equation}
\int \frac{d^Dx_0}{(x_{10}^2)^{\alpha_1}(x_{02}^2)^{\alpha_2}}=
\frac{\pi^{D/2}\;\mathbb{a}_{0}(\alpha_1,\alpha_2,D-\alpha_1-\alpha_2)}{(x_{12}^2)^{\alpha_1+\alpha_2-D/2}}\,.
\end{equation}
Notice that this identity can be interpreted as the star-triangle relation \eqref{STRbos} in which the missing weight $\alpha_3=D-\alpha_1-\alpha_2$ is determined by the uniqueness requirement.
\item A chain of a fermionic and a bosonic propagator is integrated by means of the following identity
\begin{equation}\label{chain2}
\vcenter{\hbox{\includegraphics[width=3.7cm]{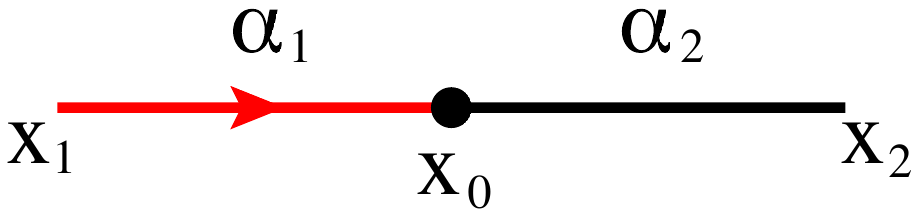}}}=\pi^{D/2}\;\mathbb{a}_{0}(\alpha_2)\,\mathbb{a}_{1/2}(\alpha_1,D-\alpha_1-\alpha_2)\;\vcenter{\hbox{\includegraphics[width=3.7cm]{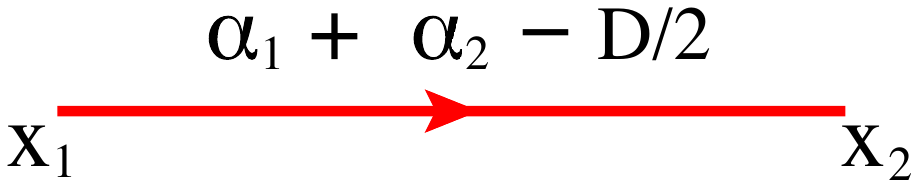}}}\,,
\end{equation}
that in the integral form is
\begin{equation}\begin{split}
\int \frac{d^Dx_0\; \slashed{x}_{10}}{(x_{10}^2)^{\alpha_1+1/2}(x_{02}^2)^{\alpha_2}}&=-\frac{i\,\mathbb{a}_{1/2}(\alpha_1)\,\mathbb{a}_{0}(\alpha_2)}{4^{\alpha_1+\alpha_2-D/2}}\int \frac{d^Dk\; e^{ik\cdot x_{12}}\slashed{k}}{(k^2)^{D-\alpha_1-\alpha_2+1/2}}\\
&=\frac{\pi^{D/2}\;\mathbb{a}_{0}(\alpha_2)\,\mathbb{a}_{1/2}(\alpha_1,D-\alpha_1-\alpha_2)\slashed{x}_{12}}{(x_{12}^2)^{\alpha_1+\alpha_2-D/2+1/2}}\,,
\end{split}\end{equation}
where in the intermediate step we have used the Fourier representation of the propagators \eqref{Fourier} and then, transforming back to coordinate space, the identity \eqref{defa}. Notice that, as the bosonic case, this identity can be interpreted as the star-triangle relation \eqref{STRferm} in which, after relabeling the weights as $\alpha_3\rightarrow\alpha_1$, $\alpha_1\rightarrow\alpha_2$ and $\alpha_2\rightarrow\alpha_3$, the missing weight $\alpha_3=D-\alpha_1-\alpha_2$ is determined by the uniqueness requirement.
\item A chain of fermionic propagators is integrated by means of the following identity
\begin{equation}\label{chain3}
\!\!\vcenter{\hbox{\includegraphics[width=3.7cm]{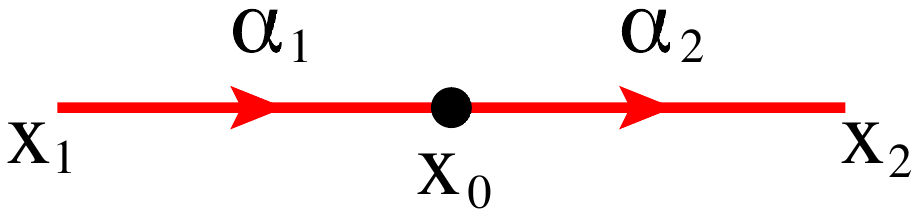}}}\!=\!-\pi^{D/2}\,\mathbb{a}_{0}(D-\alpha_1\!-\alpha_2)\,\mathbb{a}_{1/2}(\alpha_1,\alpha_2)\,\mathbb{1}\vcenter{\hbox{\includegraphics[width=3.7cm]{chain_bosbos2}}}\,.
\end{equation}
Since we are considering that adjacent fermionic lines are contracted in the spin structure (see footnote \ref{foot:2}), in the right-hand side of the formula, it appears the identity matrix $\mathbb{1}$ that corresponds to $\mathbb{1}_{2^{D/2-1}}$ for even $D$ or $\mathbb{1}_{2^{(D-1)/2}}$ for odd $D$ (see Appendix \ref{sec:appendixA}).
In the integral form the previous identity is
\begin{equation}\begin{split}
\int \frac{d^Dx_0\; \slashed{x}_{10}\slashed{x}_{02}}{(x_{10}^2)^{\alpha_1+1/2}(x_{02}^2)^{\alpha_2+1/2}}&=-\frac{\mathbb{a}_{1/2}(\alpha_1,\alpha_2)}{4^{\alpha_1+\alpha_2-D/2}}\;\mathbb{1}\;\int \frac{d^Dk\; e^{ik\cdot x_{12}}}{(k^2)^{D-\alpha_1-\alpha_2}}\\
&=-\mathbb{1}\frac{\pi^{D/2}\;\mathbb{a}_{0}(D-\alpha_1-\alpha_2)\,\mathbb{a}_{1/2}(\alpha_1,\alpha_2)}{(x_{12}^2)^{\alpha_1+\alpha_2-D/2}}\,,
\end{split}\end{equation}
where in the intermediate step, we have used the Fourier representation of the propagators \eqref{Fourier} and then, transforming back to coordinate space, the identity \eqref{defa}. Notice that this identity can be interpreted as the star-triangle relation \eqref{STRferm} in which, after relabeling the weights with $\alpha_3\leftrightarrow\alpha_1$, the missing weight $\alpha_3=D-\alpha_1-\alpha_2$ is determined by the uniqueness requirement.
\end{itemize}

\section{The \texttt{STR} \textit{Mathematica}\textsuperscript{\textregistered} package}\label{sec:sec2}

In the following we introduce the \textit{Mathematica}\textsuperscript{\textregistered} package \texttt{STR} and its usage.
The purpose of the package is to simplify the computation of Feynman integrals by means of the \textit{uniqueness method} that we reviewed in the previous section, using a graphical interactive approach. In this section we present how to setup the package in a \textit{Mathematica}\textsuperscript{\textregistered} notebook and a detailed manual.

\subsection{Download and installation}

The set up of the package is very simple\footnote{For more detailed set up instructions see also \cite{Preti:2017fjb}}.
The package files can be downloaded from the arXiv servers in the source directory of this paper but, considering the possibility of updates of the code to correct bugs or to add new features, we suggest to download it from the GitHub repository at the following address 

 \url{https://github.com/miciosca/STR} .

In order to load the package in a \textit{Mathematica}\textsuperscript{\textregistered} notebook, 
one has to save the file \texttt{STR.m} in the same directory of the notebook and run the command 

\vspace{.1cm}
\noindent\hspace{.5cm}\includegraphics[scale=.8]{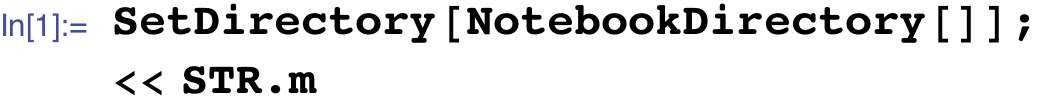}
\vspace{.1cm}

\subsection{Manual}\label{sec:manual}

Running the following line 

\vspace{.1cm}
\noindent\hspace{.5cm}\includegraphics[scale=.8]{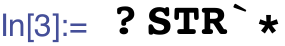}
\vspace{.1cm}

it is possible to list the new functions available with the package. Those functions are the following
\begin{itemize}
  \item \texttt{STR[\textit{dimension}]}: specifying the dimension of the Euclidean spacetime \texttt{\textit{dimension}}, the function opens a graphical panel in which is possible to draw and modify the desired Feynman diagrams;
  \item \texttt{STRrelation}: it is a list of relations that identify the unique stars and triangles in the \texttt{STR} graphical environment;
  \item \texttt{STRintegral}: it shows the integral representation of the integral drawn in the \texttt{STR} graphical environment;
  \item \texttt{STRprefactor}: it shows the prefactor of the integral \texttt{STRintegral} that contains all the functions generated by acting with the uniqueness method in the \texttt{STR} graphical environment;
  \item \texttt{STRgraph}: it generates a modifiable version of the diagram drawn in the \texttt{STR} graphical environment;
    \item \texttt{STRSimplify[\textit{expr},\textit{dimension}]}: specifying the dimension of the Euclidean spacetime \texttt{\textit{dimension}}, it rewrites \textit{expr} (the output of \texttt{STRprefactor}) in terms of Euler gammas by means of \eqref{defa} and \eqref{amultiple}.
\end{itemize}
In the following paragraphs we will present in detail all those functions.

\paragraph{STR:} After the package is loaded, run the function \texttt{STR} specifying the dimension of the spacetime without semicolon at the end. For example

\vspace{.1cm}
\noindent\hspace{.5cm}\includegraphics[scale=.8]{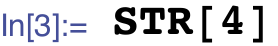}
\vspace{.1cm}

where we selected the dimension of the Euclidean spacetime to be $D=4$. In general the user can choose any dimension in the argument of \texttt{STR}, even with some non-numerical character as needed, for instance, in dimensional regularization. The output will be the following interactive panel
\begin{center}\label{STRpanel}
\includegraphics[width=\textwidth]{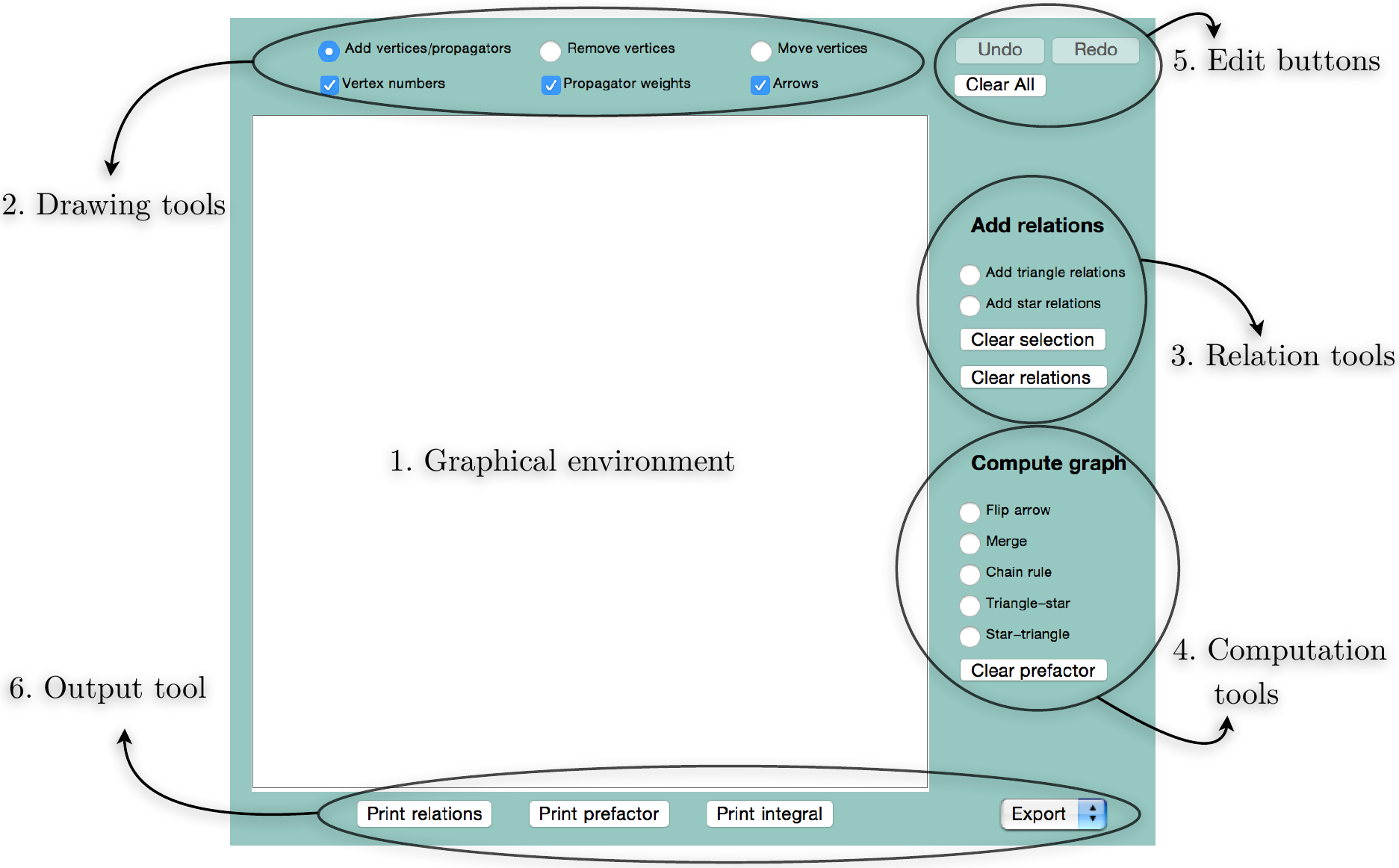}
\end{center}
The purpose of this function is to allow the user to draw a diagram only using the mouse and interact with it to compute it (or part of it) by means of the uniqueness method (see section \ref{sec:startriangle}). Any operation in this panel will affect the diagrams and the output in real time.
This panel can be divided in six groups of functions (as highlighted above in the figure) presented in details in the following list. 

\begin{enumerate}
\item \textbf{Graphical environment}: This white window is the portion of the interactive panel in which the user is allowed to draw Feynman diagrams interacting with the mouse. The algorithm tracks the position of the mouse cursor and draws vertices and propagators clicking or dragging (see "\textit{Drawing tools}" for a detailed guide to drawing). This space is obviously not continuos but it is a lattice with spacing optimized considering the dimensions of the graphical elements that represents vertices and propagators. All the operations that the user can do on the graphs can be done in this space selecting the appropriate option in the tools around the window. 
\item \textbf{Drawing tools}: These are the main tools to draw and modify diagrams in the graphical environment. Any modification made when those option are selected will affect the content of the functions \texttt{STRintegral} and \texttt{STRgraph}.
\begin{itemize}
\item \textbf{Add vertices/propagators}: When this option is selected the user can draw Feynman diagrams in the graphical environment only using the mouse. A single left- or right-click in a empty space of the window will place an isolated external (not integrated) point represented by a white dot and labeled by a number $k$ identifying its position $x_k$. Clicking on an already placed external point will turn the white dot into a black one that represent and internal (integrated on $\mathbb{R}^D$) point. Repeating this action again, it is possible to turn the color of the vertices any number of times. 

To add a propagator, the user has to left- or right-click, drag and release the mouse on the desired points of the interactive window. Doing this procedure with the left button of the mouse one will place a scalar propagator represented by a black line as in \eqref{prop}, with the right button one will place a fermionic propagator represented by a red line with an arrow as in \eqref{prop}.
The starting/ending point can be any empty spot of the window or even an already placed white or black vertex. When the mouse is dragged it will appear a dashed temporary line (black or red depending if the propagator is bosonic of fermionic) in order to track the creation of the line. When the mouse is released. if the spot is empty, a vertex will be drawn in the releasing position. Moreover, the dashed line will disappear and it will be substituted with the final propagator. In the case of a fermionic propagator, the arrow points always to the releasing point. 
In the case in which between two vertices there are more then one line, those will be represented as curved lines giving the possibility to the user to distinguish all of them and see their related weights. 

When a propagator is placed, it will appear a pop-up window with an input field in it. The user has to specify there the name of the weight of the propagator (namely $\alpha$ in \eqref{prop}) that can contains any number, letter and arithmetic operation. Pressing \texttt{Enter} or clicking on the "\textit{OK}" button, the name of the edge weight will be updated on the graphical environment. If the weight of the propagator is not specified, it will be named as $w(k)$ with $k$ an increasing integer.

With the standard appearance options, all the vertices will appear with an unique number $k=1,...,n$ with $n$ the total number of vertices, all the fermionic propagators with an arrow pointing in the chosen direction and all the propagator with a label representing their weights. 

As an example see the following diagram
\begin{center}
\includegraphics[trim={1.5cm 0 0 0},clip,width=6cm]{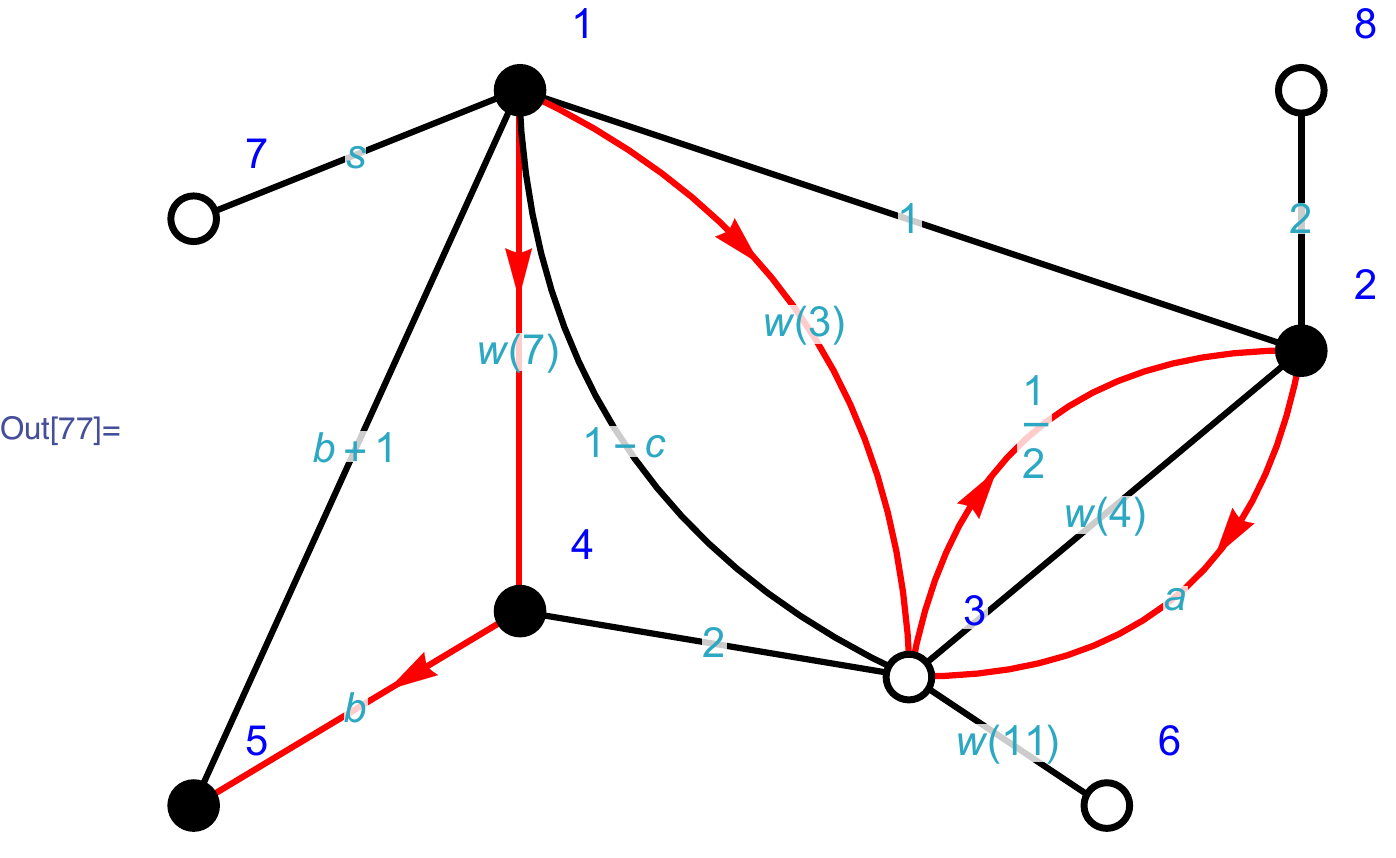}
\end{center}
which contains 4 internal integrated points (black), 4 external non-integrated points (white), 8 scalar propagators (black lines) and 5 fermionic propagators (red lines with arrows). Notice that adjacent fermionic lines are considered contracted in their spin indices carried by $\sigma$ or $\gamma$ matrices. If more than two fermionic lines are merging in the same vertex (see the external point labeled by 3), the algorithm doesn't know which propagators are contracted between themselves or if, for example, the ones in the loop are traced.
In this case, since all the possible operations with multiple fermions are proportional to $\mathbb{1}$ (see section \ref{sec:startriangle}), the identity will be explicitly written in the final result and then the user can choose if contract it with the other fermionic structures or trace it.
\item \textbf{Remove vertices}: Selecting this tool, the user can erase from the diagram an internal or external vertex and all the propagators attached to it by left-clicking on it. The remaining vertices will be renamed such that their labels will run from 1 to the total number of vertices.
For example
\begin{equation*}
\vcenter{\hbox{\includegraphics[trim={1.2cm 0 0 0},clip,width=4.5cm]{EXgraph}}}\qquad\overset{\text{Remove }x_4}{\Longrightarrow}\qquad
\vcenter{\hbox{\includegraphics[trim={1.2cm 0 0 0},clip,width=4.5cm]{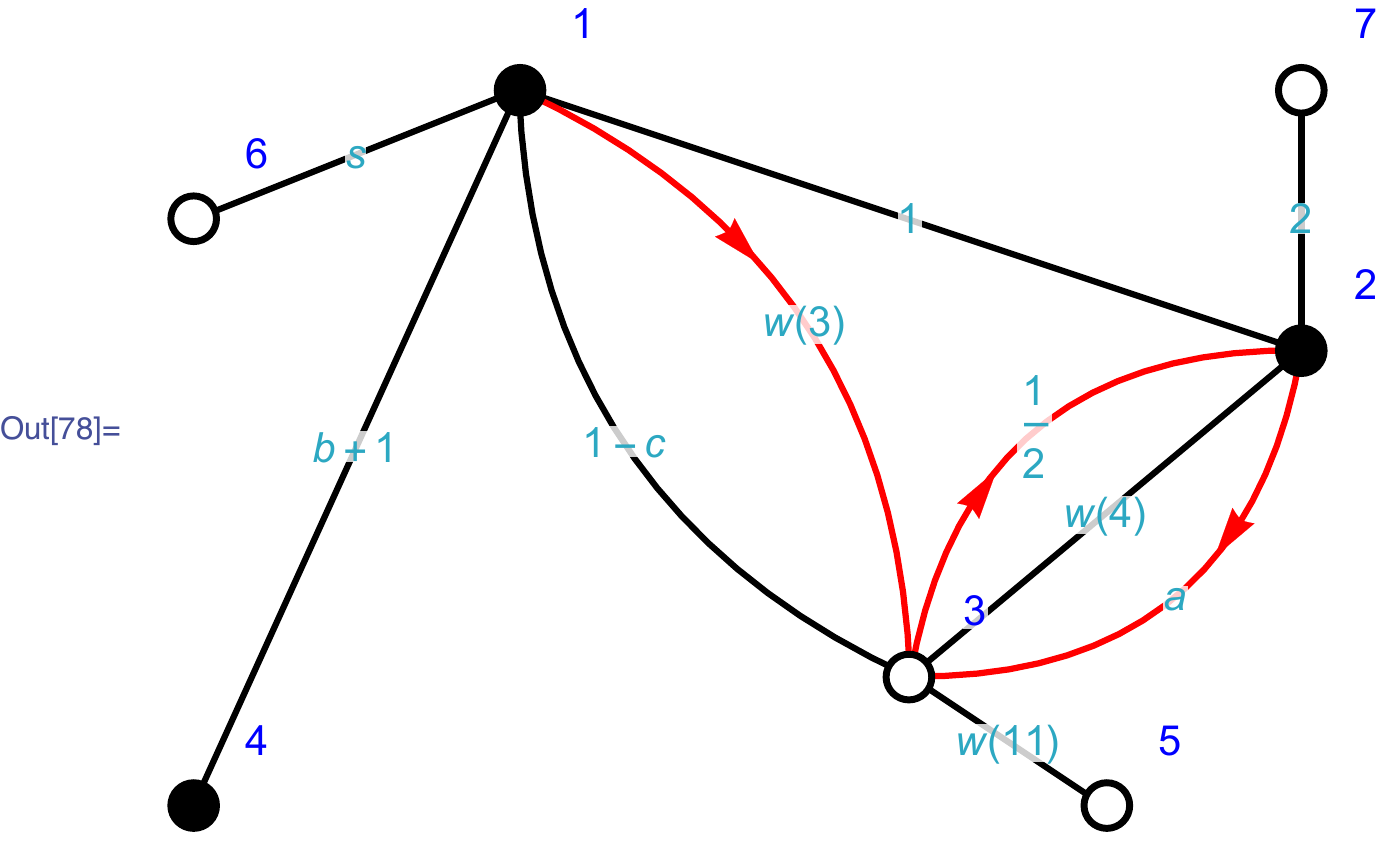}}}
\end{equation*}
where we removed the vertex in the position $x_4$.
\item \textbf{Move vertices}: Selecting this tool, the user can modify the shape of an existing diagram moving the vertices in the graphical environment. Left-clicking on a vertex and dragging it, it will placed in a new position together with the propagators connected to it. When the node is in the final desired position, the user has to release the mouse. The positions of propagator weights, vertex numbers and arrows will be updated in real-time tracking the mouse cursor.
For example
\begin{equation*}
\vcenter{\hbox{\includegraphics[trim={1.2cm 0 0 0},clip,width=4.5cm]{EXgraph}}}\qquad\overset{\text{Move }x_{2},\,x_5,\,x_6}{\Longrightarrow}\qquad
\vcenter{\hbox{\includegraphics[trim={1.2cm 0 0 0},clip,width=4.5cm]{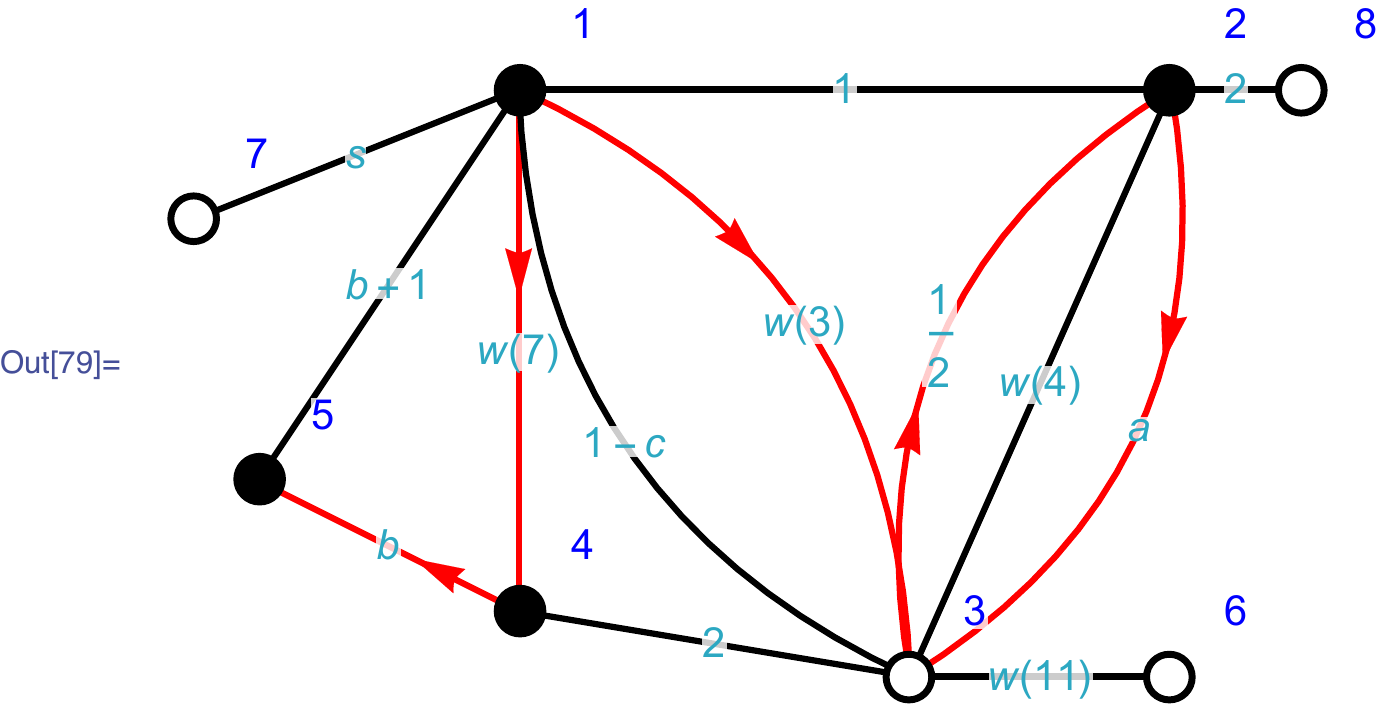}}}
\end{equation*}
where we moved the vertices in the position $x_2$, $x_5$ and $x_6$.
\item \textbf{Vertex numbers}: Checking/unchecking this option, the numerical labels written near the vertices will appear/disappear from the graphical environment. 
\item \textbf{Propagator weights}: Checking/unchecking this option, the labels representing the propagator weights will appear/disappear from the graphical environment. 
\item \textbf{Arrows}: Checking/unchecking this option, the arrows of the fermionic propagators will appear/disappear from the graphical environment. 
\end{itemize}
\item \textbf{Relation tools}: These tools allow the user to identify unique stars or triangles in the graph. Those options are helpful only when the weights of stars or triangles are not automatically true or false: for instance when the weights of the propagators are non-numeric symbols and we want to impose a relation between those labels and the spacetime dimension.
When a star or a triangle is identified  as unique by those options, the function \texttt{STRrelations} is updated. If the user will add a new uniqueness relation that is automatically false or incompatible with the other in \texttt{STRrelations}, he will be notified by a pop-up window that the selected star or triangle cannot be unique.
\begin{itemize}
\item \textbf{Add triangle relations}: 
Selecting this tool the user can identify a triangle in the diagram imposing the uniqueness relation on its propagator weights $\alpha_k$ such that $\sum_k\alpha_k=D/2$ with $k=1,2,3$ as presented in section \ref{sec:startriangle}.

In order to select the desired triangle directly on the graph, after selecting \texttt{Add triangle relations}, the user has to left- or right-click on the three black or white dots at the vertices of it. Notice that in this package we are providing solutions for scalar  and Yukawa triangles as reviewed in section \ref{sec:startriangle}. When a vertex is selected, it will be highlighted by changing the color of its border from black to red. In the case in which the vertices are not connected by propagators or if they don't identify a bosonic or Yukawa triangle, a pop-up window will notify the user about the issue, and the selection will be reset. If the selected vertices identify a proper triangle, the uniqueness relation for its weights will be registered updating the function \texttt{STRrelations}. Finally the user will be notified by a pop-up window that the relation is added successfully. Clicking on the "\textit{OK}" button or pressing \texttt{Enter}, the window will be closed and all the vertices will be deselected turning back their borders from red to black.

If between two selected points there are more than one propagator, it could happen that the same set of vertices identify more than one triangle. In this case, it will appear a pop-up window containing a list of buttons that represent all the possible scalar and Yukawa \textit{sub-triangles} passing for the selected set of vertices. Any button appears with a schematic representation of the related sub-triangle: the position of the vertices $x_k$ in the order selected by the user connected by black double arrows or red standard arrows representing respectively scalar and fermionic propagators with their directions. Any arrows is labelled by the related propagator weight. 

For example
\begin{equation*}
\vcenter{\hbox{\includegraphics[trim={1.2cm 0 0 0},clip,width=4.5cm]{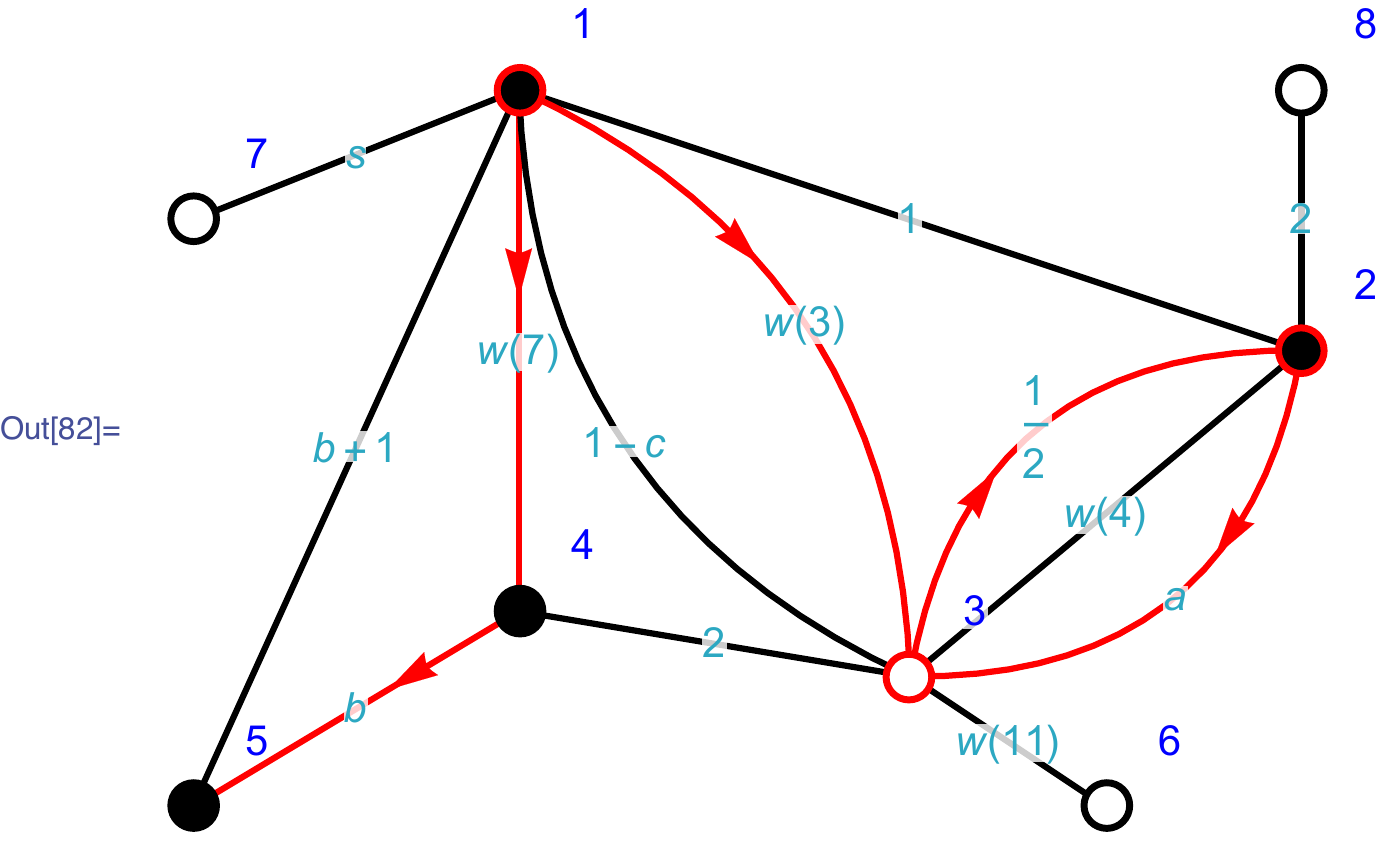}}}\qquad
\vcenter{\hbox{\includegraphics[width=8.5cm]{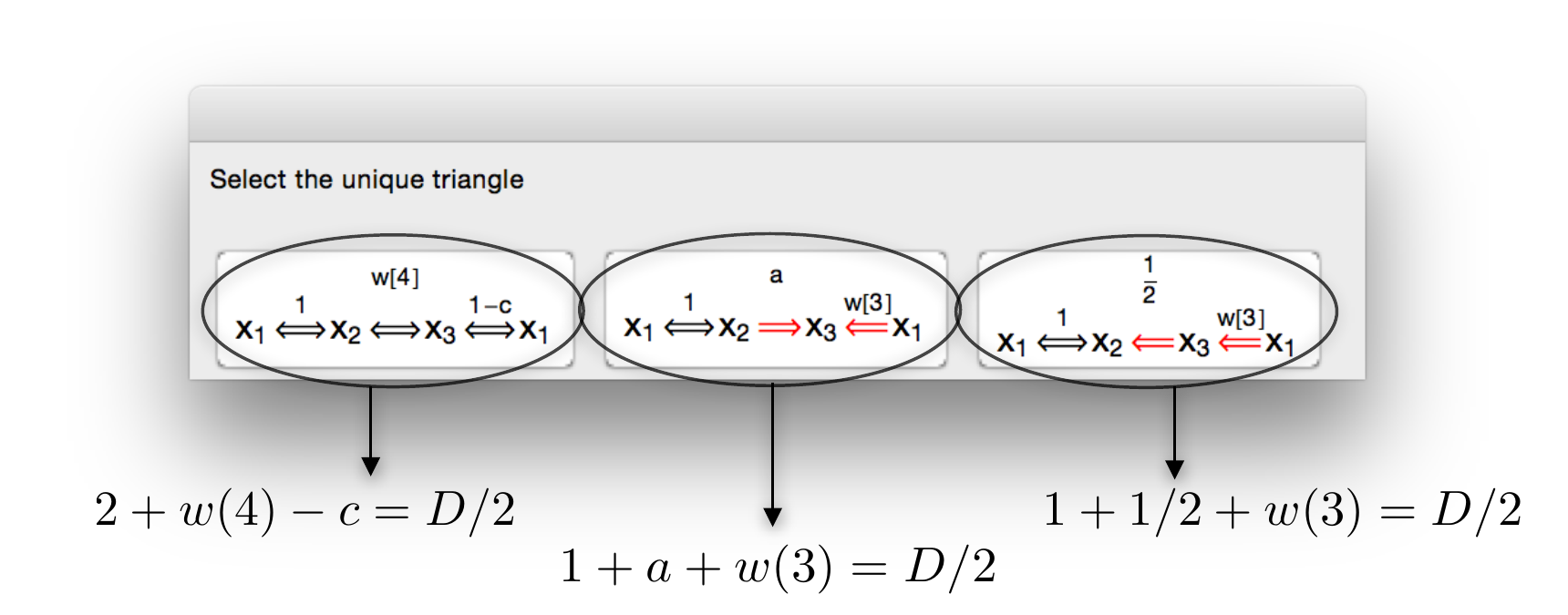}}}
\end{equation*}
we want to identify as an unique triangle the Yukawa one with weights $1$, $a$ and $w(3)$ in the diagram considered above. Selecting the three vertices at the positions $x_1$, $x_2$ and $x_3$, the color of the vertices border turns to red as in figure and the pop-up window in the right will appear. The possible triangles for those vertices are three and our choice corresponds to the one in the middle. Then the \texttt{STRrelations} function will be updated with the uniqueness relation $1+a+w(3)=D/2$ with a notification of the successful addition.
\item \textbf{Add star relations}: 
Selecting this tool the user can identify a star in the diagram imposing the uniqueness relation on its propagator weights $\alpha_k$ such that $\sum_k\alpha_k=D$ with $k=1,2,3$ as presented in section \ref{sec:startriangle}.

In order to select the desired star directly on the graph, after selecting \texttt{Add star relations}, the user has to left- or right-click on a black (integrated) vertex in which three propagators merge. Notice that in this package we are providing solutions for scalar  and Yukawa stars as reviewed in section \ref{sec:startriangle}. Once the vertex is selected, it will be highlighted by changing the color of its border from black to red. In the case in which the vertex is not integrated (white dot) or more than three propagators are meeting in it or it doesn't identify a scalar or Yukawa star, a pop-up window will notify the user about the issue, and the selection will be reset. If the selected vertex identifies a proper star, the uniqueness relation for its weights will be registered updating the function \texttt{STRrelations}. Finally the user will be notified by a pop-up window that the relation is added successfully. Clicking on the "\textit{OK}" button or pressing \texttt{Enter}, the window will be closed and all the vertices will be deselected turning back their borders from red to black.

For example
\begin{equation*}
\vcenter{\hbox{\includegraphics[trim={1.2cm 0 0 0},clip,width=4.5cm]{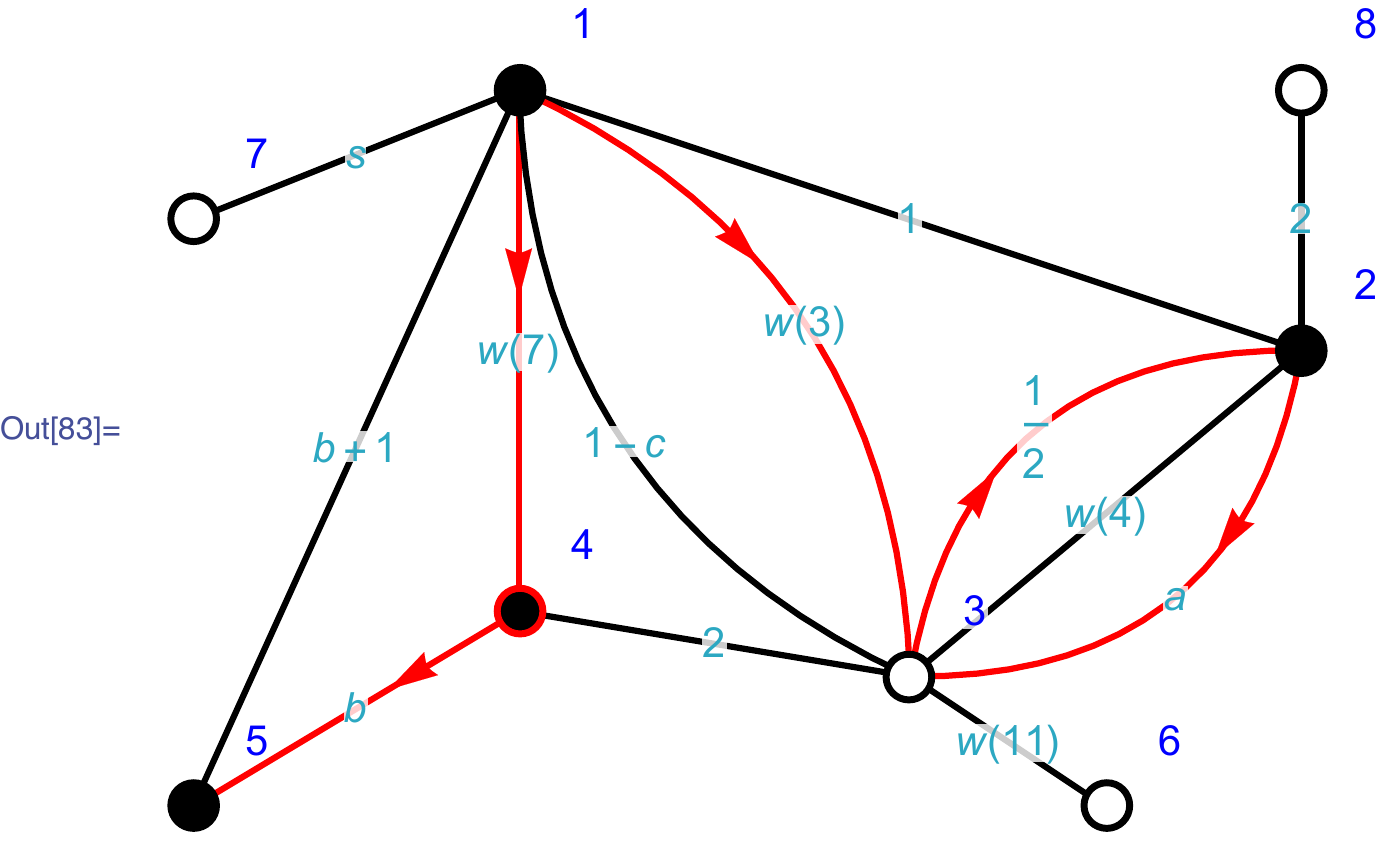}}}\qquad\Longrightarrow\qquad
2+b+w(7)=D
\end{equation*}
we want to identify as an unique star, the one with weights $2$, $b$ and $w(7)$ in the diagram considered above. Selecting the vertex at the positions $x_4$, the color of its border turns to red as in figure and the \texttt{STRrelations} function is updated with the uniqueness relation $2+b+w(7)=D$ with a notification of the successful addition.

\item \textbf{Clear selection}: Clicking on this button, the user can deselect the vertices previously selected. In other words if some vertex has red border, the border will be turned back to black.
\item \textbf{Clear relations}: Clicking on this button, the user can reset the \texttt{STRrelations} function erasing all the uniqueness relations saved in it.
\end{itemize} 
\item \textbf{Computation tools}: 
All the main operations that the user can do on the graph are included in this set of tools. Indeed, it allows to use the uniqueness method reviewed in section \ref{sec:startriangle} on the diagram drawn with the \texttt{Drawing tools}. All those operations update in real-time the functions \texttt{STRprefactor}, \texttt{STRintegral} and \texttt{STRgraph}.
\begin{itemize}
\item \textbf{Flip arrow}:
Selecting this tool, the user can flip the direction of the arrow of a chosen fermionic propagator. In order to select the propagator, one has to left- or right-click on the two vertices at the endpoints of it. The border of the selected vertices will turn from black to red to enlightening it and the arrow will be flipped updating the \texttt{STRprefactor} function with a minus sign. 

If between the selected vertices there are more than one fermionic propagators, the user has to choose which arrow has to be flipped between them. This choice can be done through a pop-up window containing a number of buttons equal to the number of fermionic propagators between the two points. Any button contains a schematic representation of the related propagator, in particular the direction of the arrow and its weight. Once the user selects the desired button, the arrow is flipped and the function \texttt{STRprefactor} is updated in the same way of the previous case.

For example
\begin{equation*}
\vcenter{\hbox{\includegraphics[trim={1.2cm 0 0 0},clip,width=4.5cm]{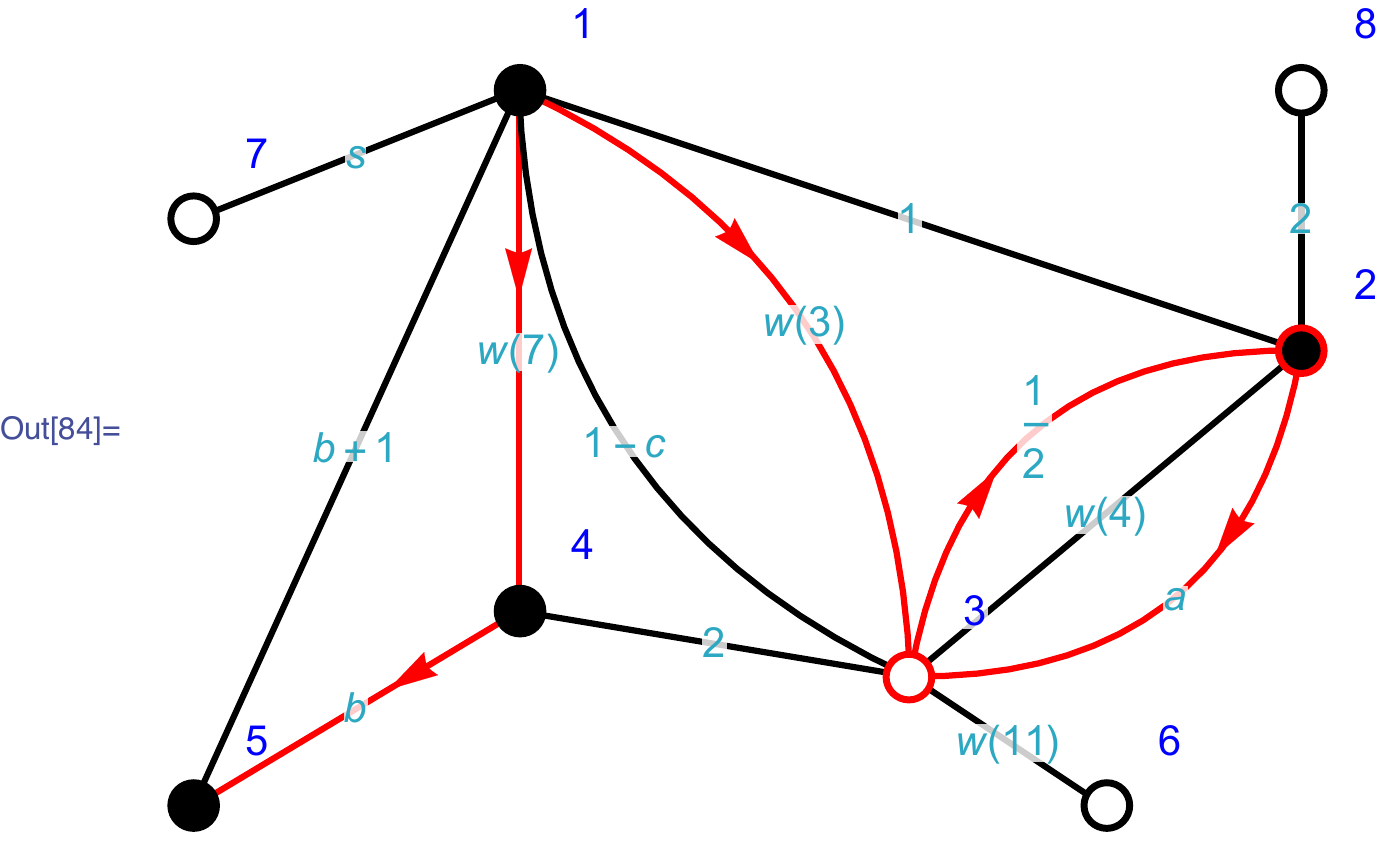}}}\quad\quad
\vcenter{\hbox{\includegraphics[trim={1.2cm 0 0 0},clip,width=3.5cm]{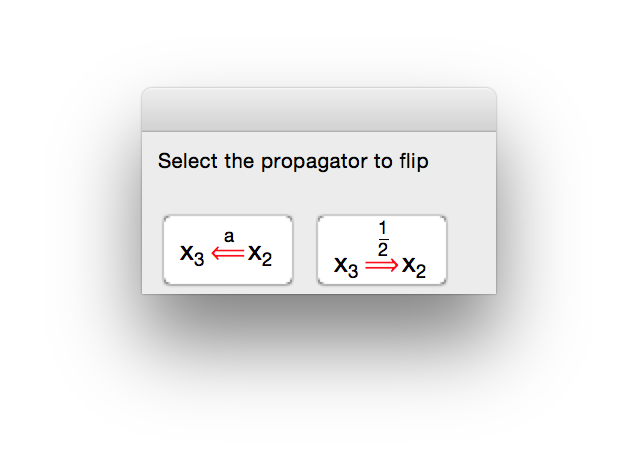}}}\;\quad
\vcenter{\hbox{\includegraphics[trim={1.2cm 0 0 0},clip,width=4.5cm]{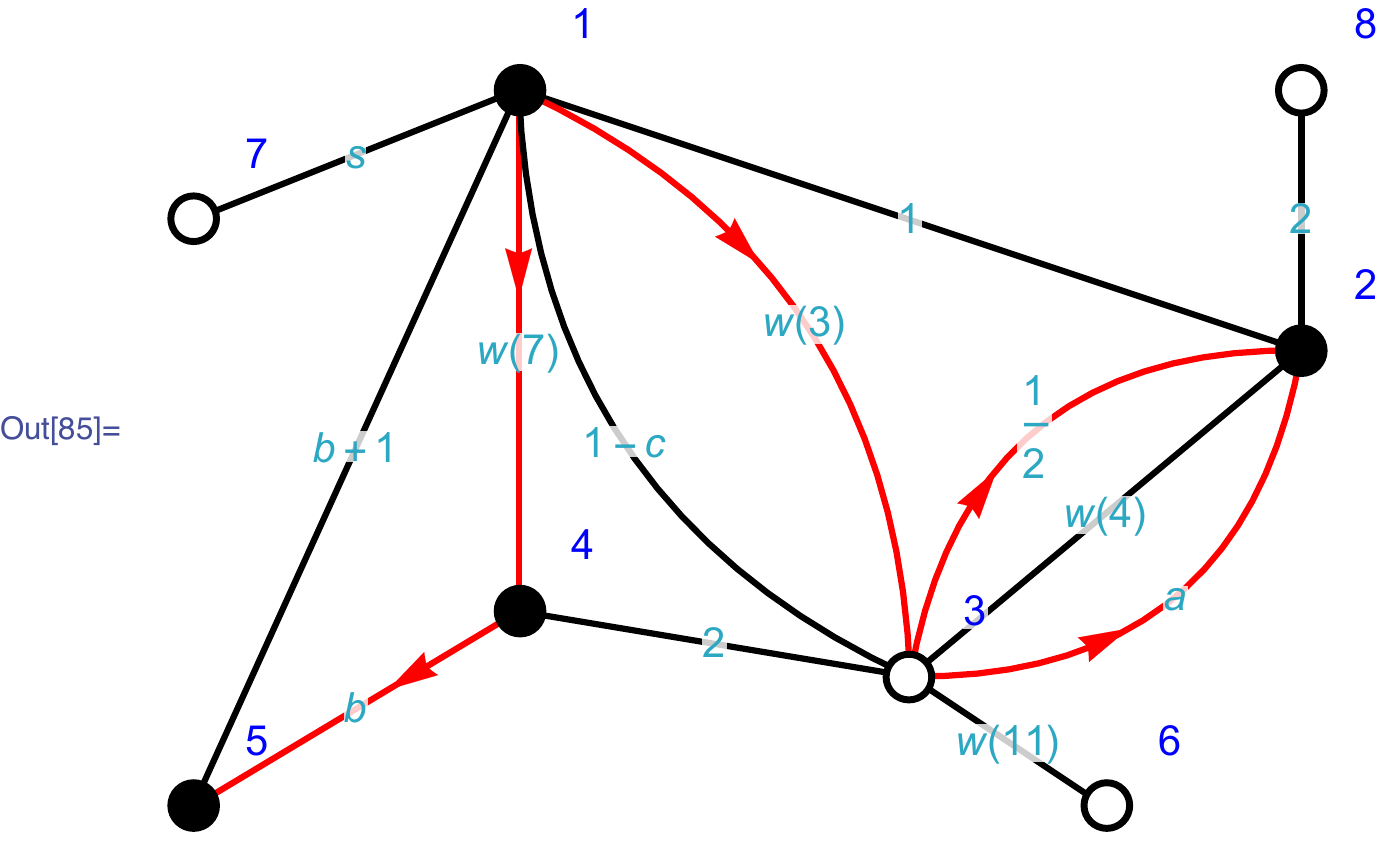}}}
\end{equation*}
where we flipped the fermionic line passing through the vertices in the positions $x_2$ and $x_3$ with weight $a$.
\item \textbf{Merge}:
With this tool, the user can combine together multiple propagators lying between two vertices. Selecting two vertices connected by more than one propagator, the color of their border will turn from black to red to highlight them and all the propagators between them will collapse in a single one following the rules give in \eqref{merge1}, \eqref{merge2} and \eqref{merge3}. In the case in which the fermionic propagators have different arrow orientations, the algorithm will take into account the phase generated by the flipping. All the functions \texttt{STRprefactor}, \texttt{STRintegral} and \texttt{STRgraph} will be updated.
For example
\begin{equation*}
\vcenter{\hbox{\includegraphics[trim={1.2cm 0 0 0},clip,width=4.5cm]{graph_flip}}}\qquad\Longrightarrow\qquad
\vcenter{\hbox{\includegraphics[trim={1.2cm 0 0 0},clip,width=4.5cm]{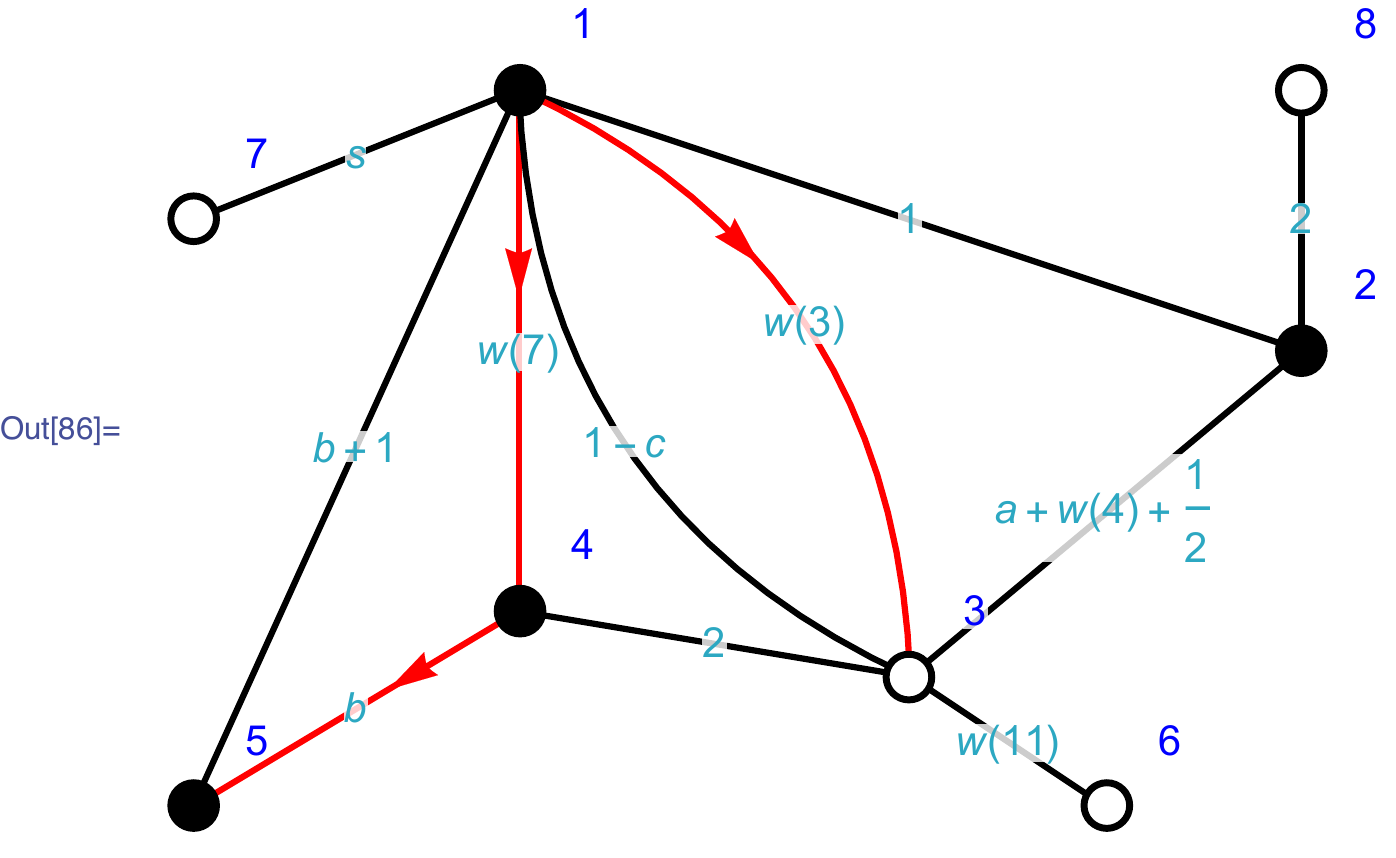}}}
\end{equation*}
where we merged the propagators between the vertices in the positions $x_2$ and $x_3$ according with the rules \eqref{merge2} and \eqref{merge3}. Notice that this operation generates a scalar propagator from a fermionic loop, then in the updated version of \texttt{STRprefactor} it will appear the identity $\mathbb{1}$.
\item \textbf{Chain rule}:
Selecting this tool, the user can solve the integral related to one internal (black) vertex connected with two propagators using the chain rules \eqref{chain1}, \eqref{chain2} and \eqref{chain3}. In order to solve the integral, one has only to left- or right-click on the desired internal vertex on the diagram. Then the graph will be modified and the functions \texttt{STRprefactor}, \texttt{STRintegral} and \texttt{STRgraph} will be updated accordingly. For example
\begin{equation*}
\vcenter{\hbox{\includegraphics[trim={1.2cm 0 0 0},clip,width=4.5cm]{EXgraph}}}\qquad\overset{\text{Chain rule in }x_5}{\Longrightarrow}\qquad
\vcenter{\hbox{\includegraphics[trim={1.2cm 0 0 0},clip,width=4.5cm]{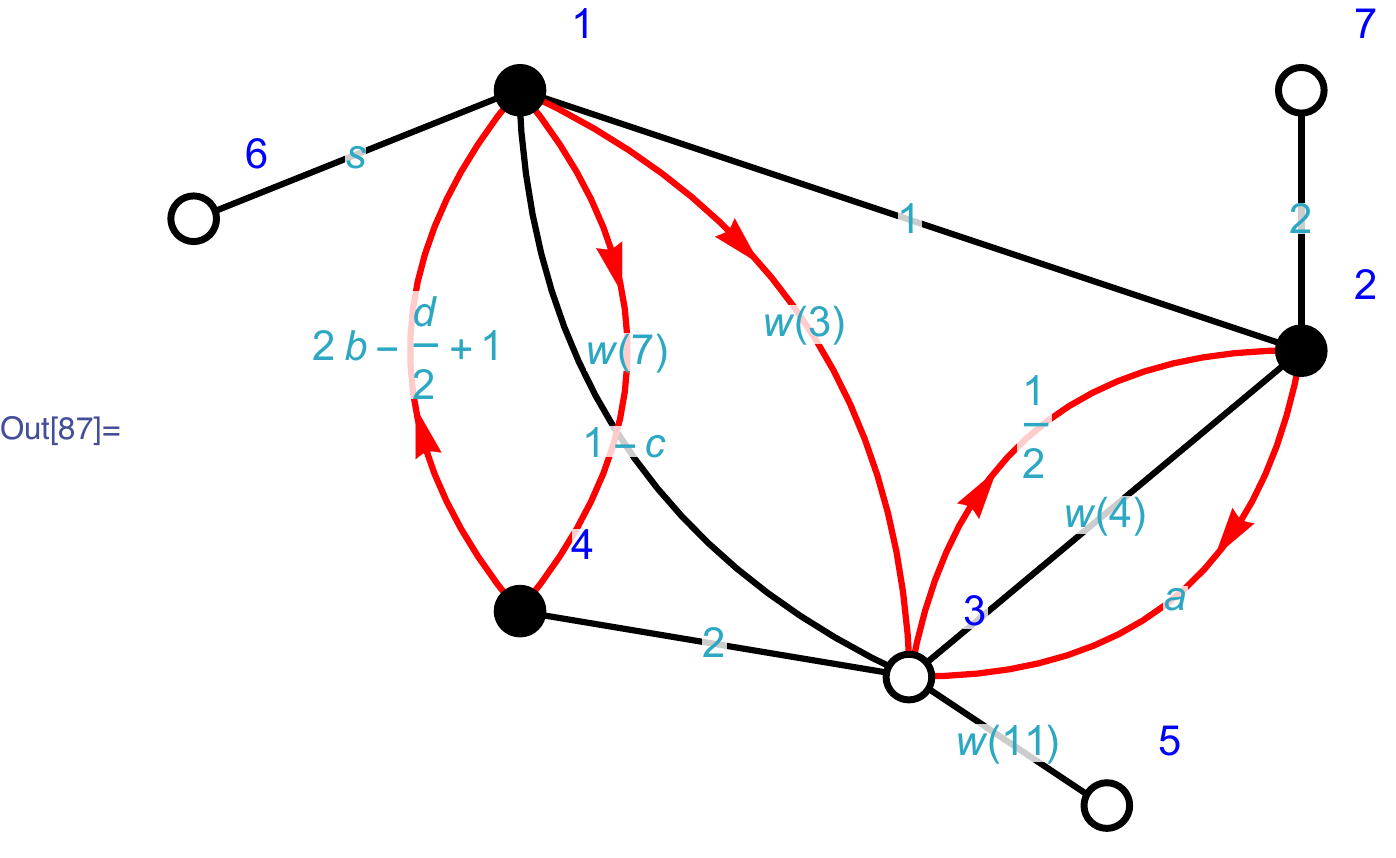}}}
\end{equation*}
where we solved the integral in the position $x_5$ with the rule \eqref{chain2} in dimension $D=d$. Notice that in the resulting diagram some lines are overlapping but obviously they are not crossing.
\item \textbf{Triangle-star}:
Choosing this tool, the user can select a triangle in the diagram turning it into a star adding an integration point (black dots) according to the star-triangle relations \eqref{STRbos} and \eqref{STRferm}. The procedure to select the triangle is exactly the same presented for the tool \texttt{Add triangle relations}. 
If the selected triangle is not unique, a pop-up window will notify it, otherwise it will turn in a star. In the case in which the selected vertices identify more than one triangle, as in the case examined in \texttt{Add triangle relations} section, the user has to choose between all the possible sub-triangles in a pop-up window.
Then the graph will be modified and the functions \texttt{STRprefactor}, \texttt{STRintegral} and \texttt{STRgraph} will be updated accordingly. For example
\begin{equation*}
\vcenter{\hbox{\includegraphics[trim={1.2cm 0 0 0},clip,width=4.5cm]{graph_relationT}}}\qquad\Longrightarrow\qquad
\vcenter{\hbox{\includegraphics[trim={1.2cm 0 0 0},clip,width=4.5cm]{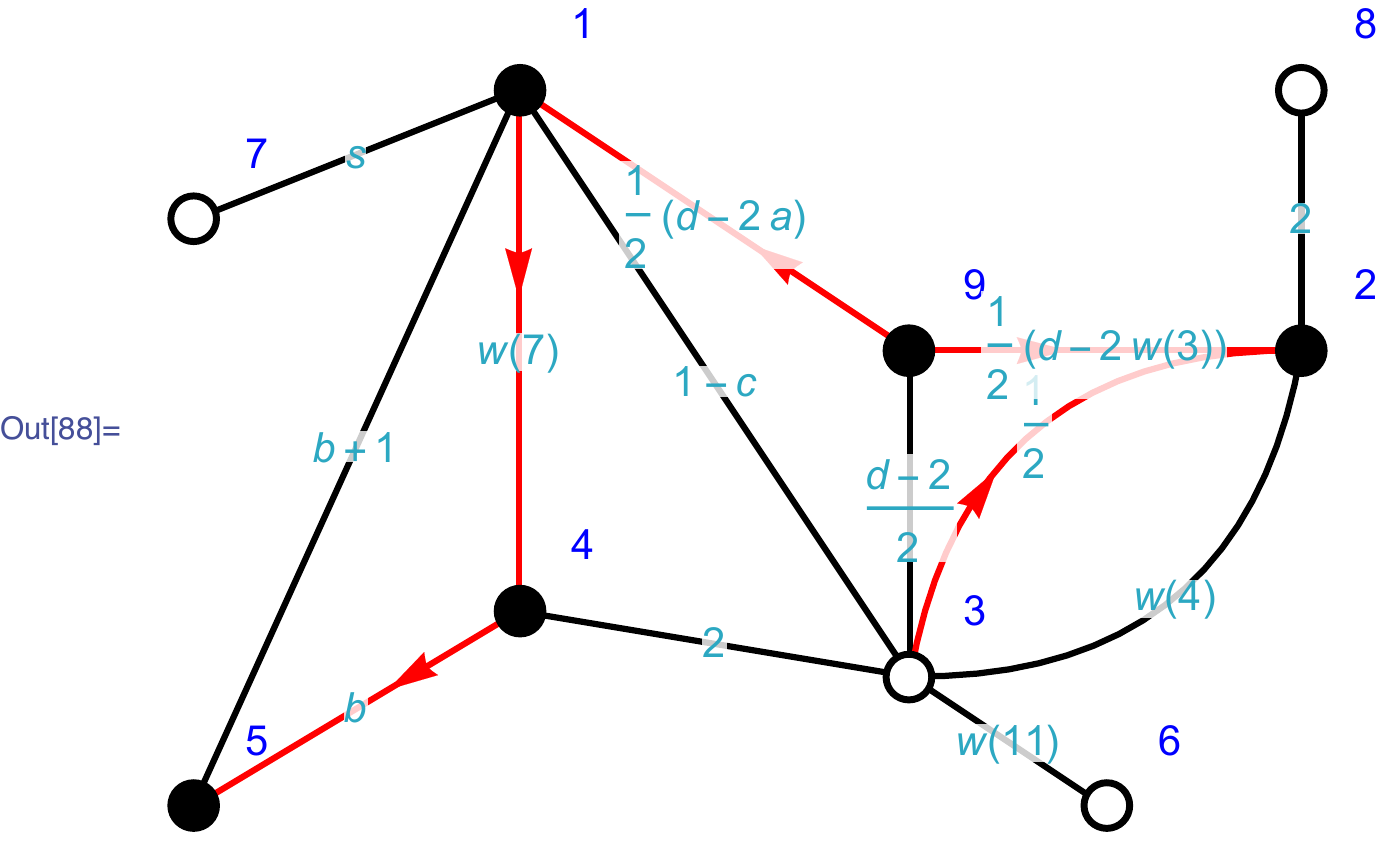}}}
\end{equation*}
where we selected the three vertices at the positions $x_1$, $x_2$ and $x_3$ and then the triangle with weights $1$, $a$ and $w(3)$ that we set to be unique with \texttt{Add triangle relations}. To compute the new diagram, the package is using the (inverse) star-triangle relation \eqref{STRferm} in dimension $D=d$ adding a new integrated vertices in position $x_9$.
\item \textbf{Star-triangle}:

With this tool, the user can select a star in the diagram turning it into a triangle solving the integration in the internal vertex (black dot) according to the star-triangle relations \eqref{STRbos} and \eqref{STRferm}. The procedure to select the star is exactly the same presented for the tool \texttt{Add star relations}. 
If the selected star is not unique, a pop-up window will notify it, otherwise it will turn in a triangle. 
Then the graph will be modified and the functions \texttt{STRprefactor}, \texttt{STRintegral} and \texttt{STRgraph} will be updated accordingly. For example
\begin{equation*}
\vcenter{\hbox{\includegraphics[trim={1.2cm 0 0 0},clip,width=4.5cm]{graph_relationS}}}\qquad\Longrightarrow\qquad
\vcenter{\hbox{\includegraphics[trim={1.2cm 0 0 0},clip,width=4.5cm]{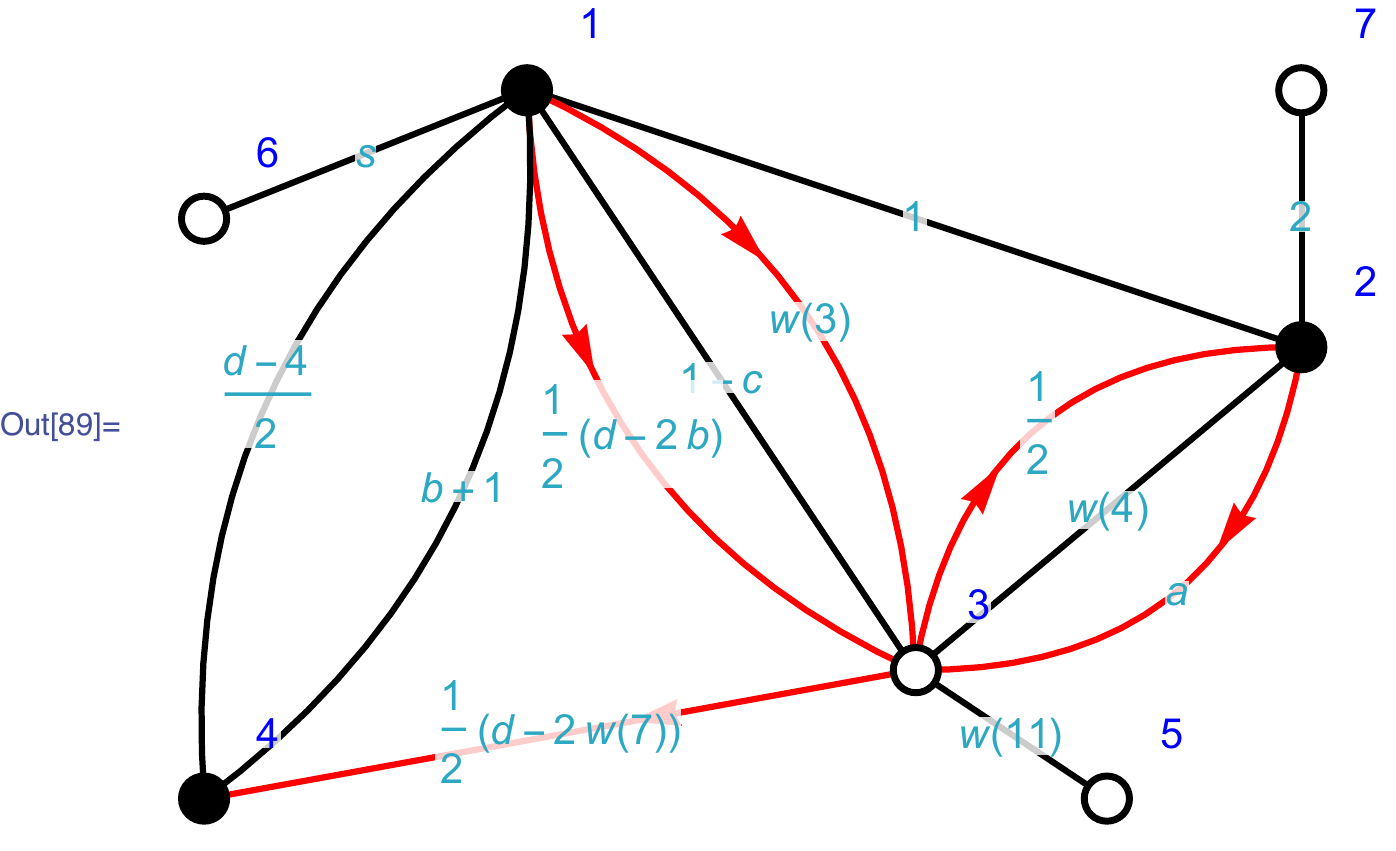}}}
\end{equation*}
where we selected the vertex at the position $x_4$ identifying the star with weights $2$, $b$ and $w(7)$ that we set to be unique with \texttt{Add star relations}. To compute the new diagram, the package is using the star-triangle relation \eqref{STRferm} in dimension $D=d$ solving the integral in position $x_4$.
\item \textbf{Clear prefactor}:
Clicking on this button the user can reset the function \texttt{STRprefactor} at its initial value 1.
\end{itemize}

\item \textbf{Edit buttons}: 
Those buttons are self-explanatory, in particular
\begin{itemize}
\item \textbf{Undo/Redo}: The \texttt{Undo} button erases the most recent action the user did in the interactive panel. The \texttt{Redo} button is needed to re-apply or undo what the user just undid one time. Both of the buttons can be used repeatedly to apply their effects multiple times. 
\item \textbf{Clear All}: This button is needed to reset the working-space to the initial conditions: the diagram in the graphical environment is erased and the \texttt{STRprefactor}, \texttt{STRintegral}, \texttt{STRrelations} and \texttt{STRgraph} function restored to their original value.
\end{itemize}
\item \textbf{Output tools}: 
These buttons allow the user to print or export the data computed using all the previous tools.
\begin{itemize}
\item  \textbf{Print relations}: This button prints below the \texttt{STR} interactive panel the list of uniqueness relations encoded in the function \texttt{STRrelations}.
\item  \textbf{Print prefactor}: This button prints below the \texttt{STR} interactive panel the result of the operations made on the diagram via the \texttt{Computation tools} encoded in the function \texttt{STRprefactor}.
\item  \textbf{Print integral}: This button prints below the \texttt{STR} interactive panel the integral representation of the diagram currently drawn in the graphical environment also encoded in the function \texttt{STRintegral}.
\item  \textbf{Export}: This button will open a menu in which is possible to choose to export data or the diagram. Choosing \texttt{Data}, all the functions \texttt{STRprefactor}, \texttt{STRintegral} and \texttt{STRrelations} will be exported and they will be ready to be modified in the \textit{Mathematica}\textsuperscript{\textregistered} notebook. Choosing \texttt{Graph}, the function \texttt{STRgraph} will be updated containing exactly the diagram drawn in the graphical environment.
\end{itemize}
\end{enumerate}

\paragraph{STRrelations:} 
This function doesn't need arguments. Writing in the notebook \texttt{STRrelations} and pressing Shift-Enter the output will be a list of equations involving the propagator weights and the spacetime dimension registered from the \texttt{STR} interactive window using \texttt{Add triangle relations} and \texttt{Add star relations}. This function is updated in real-time but to use it in the \textit{Mathematica}\textsuperscript{\textregistered} notebook one needs to press \texttt{Export} and then \texttt{Data} in the \texttt{STR} window. 
For example the two relations added in the previous examples are registered in  \texttt{STRrelations} as follows

$
\qquad\quad\vcenter{\hbox{\includegraphics[trim={1.5cm 0 0 0},clip,width=2.5cm]{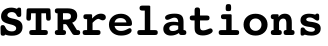}}\hbox{\includegraphics[trim={1.4cm 0 0 0 },clip,width=6.5cm]{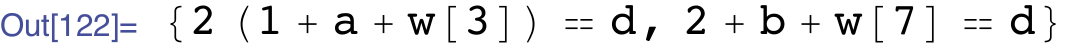}}}
$

\paragraph{STRprefactor and STRintegral:} 
Those functions don't need arguments. They represent together the integral representation of the diagram in the graphical environment. Indeed, for a given graph, its integral representation respect to the initial diagram can be written as \texttt{STRprefactor}$\int$\texttt{STRintegral}. The \texttt{STRprefactor} function contains the numerical prefactor of the integral initially set to 1 and then updated with all the contributions given by the computations on the graph by means of the \texttt{Computation tools}.  The \texttt{STRintegral} function contains the integrand related to the diagram drawn in the interactive panel in terms of product of propagators. The points that are integrated are represented by their measure $d^Dx_i$. If in \texttt{STRintegral} no integration measures are present, means that no integrals are left and all the points are external (white dots). Those functions are updated in real-time but to use them in the \textit{Mathematica}\textsuperscript{\textregistered} notebook one needs to press \texttt{Export} and then \texttt{Data} in the \texttt{STR} window. 
For example, let's start with the same diagram of before
\begin{equation*}
\vcenter{\hbox{\includegraphics[trim={1.2cm 0 0 0},clip,width=4cm]{EXgraph}}}\quad\vcenter{
\hbox{\includegraphics[trim={1.5cm 0 0 0},clip,width=2.5cm]{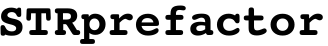}}\hbox{\includegraphics[trim={1.4cm 0 0 0 },clip,width=.3cm]{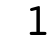}}\hbox{\includegraphics[trim={1.5cm 0 0 0},clip,width=2.5cm]{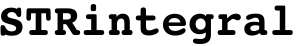}}\;\;\hbox{\includegraphics[trim={1.4cm 0 0 0},clip,width=11cm]{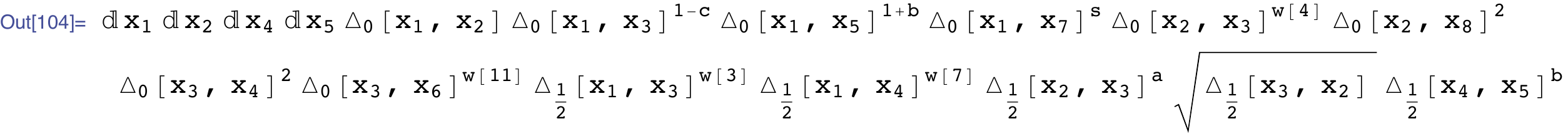}}}
\end{equation*}
where 
\begin{equation}
\vcenter{\hbox{\includegraphics[trim={1.5cm 0 0 0},clip,width=.9cm]{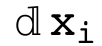}}}\;\Rightarrow\;d^Dx_i\qquad\text{and}\qquad
\vcenter{\hbox{\includegraphics[trim={1.6cm 0 0 0},clip,width=2.5cm]{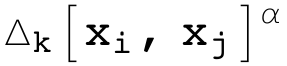}}}\;\Rightarrow\;\frac{{\slashed{x}_{ij}}^{2k}}{(x_{ij}^2)^{\alpha+k}}
\end{equation}
Setting the unique relations between weights as in the previous example in \texttt{STRrelations} with $D=d$, we can use the \texttt{Computation tools} to compute part of the integral as follows
\begin{equation*}
\vcenter{\hbox{\includegraphics[trim={1.2cm 0 0 0},clip,width=3.5cm]{graph_TS}}}\Rightarrow
\vcenter{\hbox{\includegraphics[trim={1.2cm 0 0 0},clip,width=3.5cm]{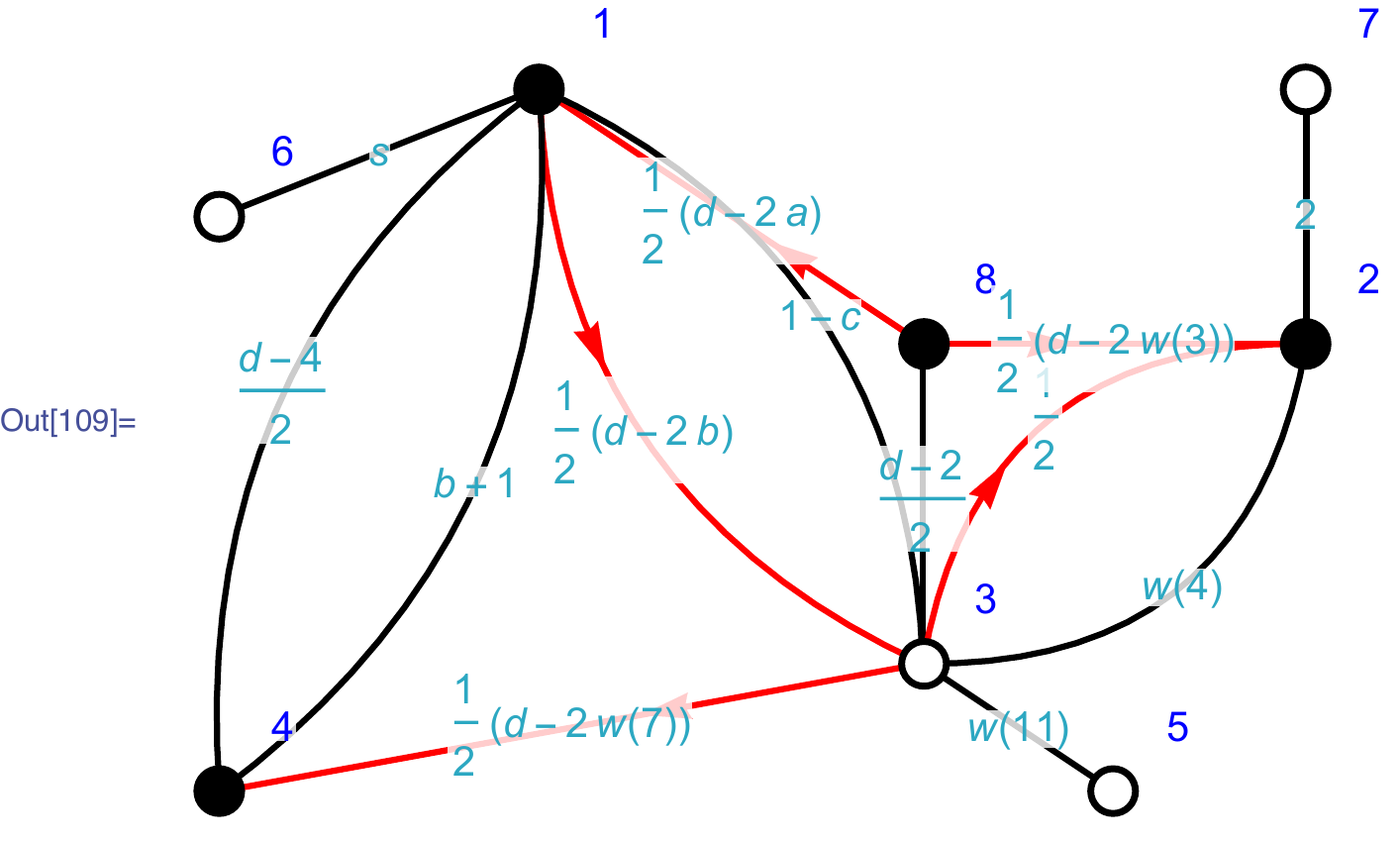}}}\Rightarrow
\vcenter{\hbox{\includegraphics[trim={1.2cm 0 0 0},clip,width=3.5cm]{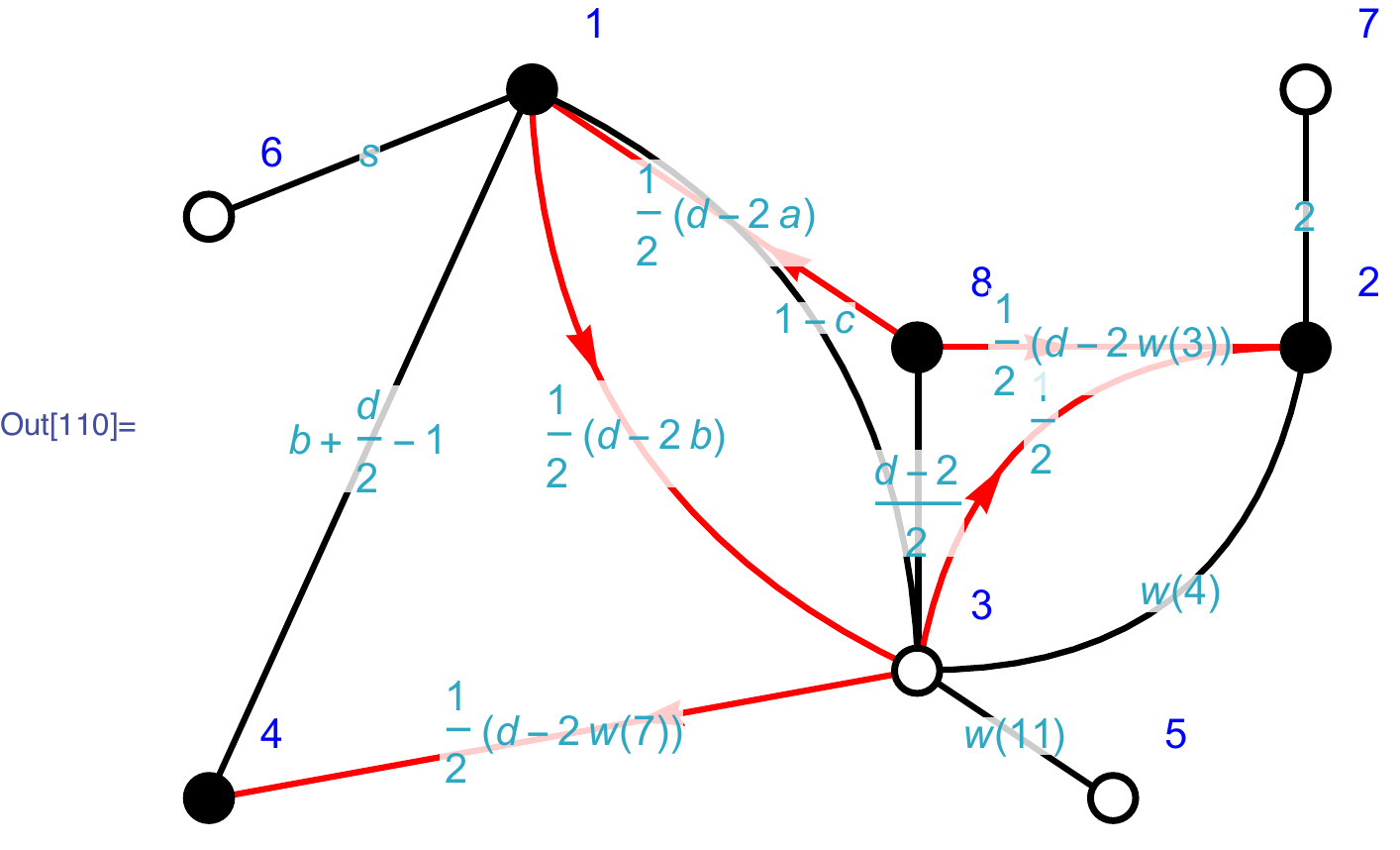}}}\Rightarrow
\vcenter{\hbox{\includegraphics[trim={1.2cm 0 0 0},clip,width=3.5cm]{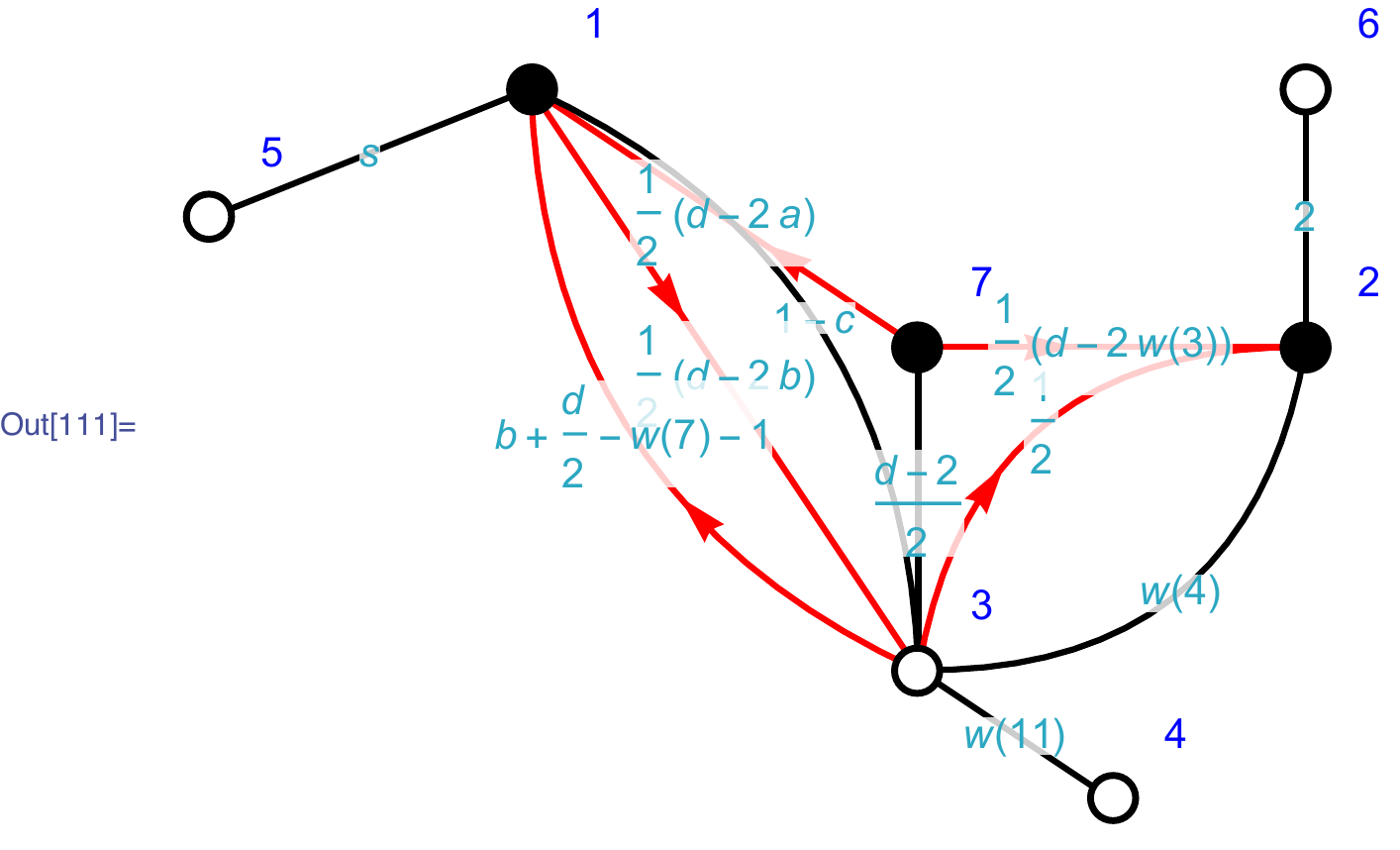}}}
\end{equation*}
and then we can read the result exporting again the updated functions \texttt{STRprefactor} and \texttt{STRintegral} that now are
\begin{equation*}
\vcenter{
\hbox{\includegraphics[trim={1.5cm 0 0 0},clip,width=2.5cm]{STRprefactor}}\hbox{\includegraphics[trim={1.4cm 0 0 0 },clip,width=12cm]{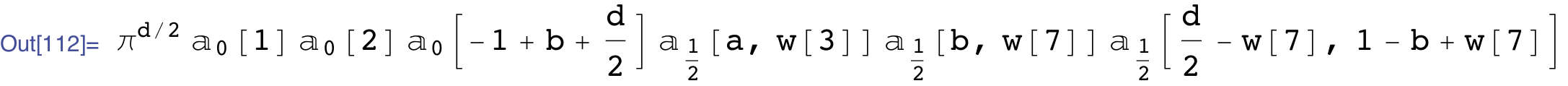}}\hbox{\includegraphics[trim={1.5cm 0 0 0},clip,width=2.5cm]{STRintegral}}\hbox{\includegraphics[trim={1.4cm 0 0 0},clip,width=15cm]{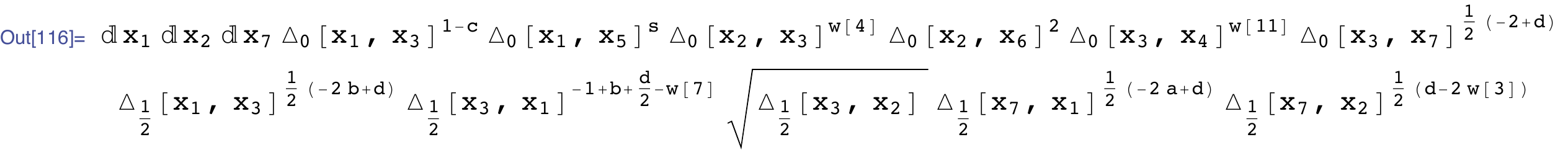}}}
\end{equation*}
where
\begin{equation}
\vcenter{\hbox{\includegraphics[trim={1.5cm 0 0 0},clip,width=4cm]{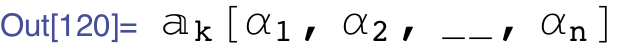}}}\;\Rightarrow\; \mathbb{a}_{k}(\alpha_1,\alpha_2,...,\alpha_n)\,,
\end{equation}
defined in \eqref{defa} and \eqref{amultiple}.

In the output, it's possible that the following symbols appear
\begin{equation}
\vcenter{\hbox{\includegraphics[trim={1.5cm 0 0 0},clip,width=1cm]{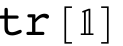}}}=\text{tr}(\mathbb{1})
\qquad\text{and}\qquad
\vcenter{\hbox{\includegraphics[trim={1.5cm 0 0 0},clip,width=1.3cm]{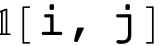}}}=\mathbb{1}\,,
\end{equation}
where the first is the trace of the identity computed following appendix \ref{sec:appendixA} and the second is the identity inserted in the diagram between the points $i$ and $j$. Notice that in the computation of the diagram the name of the positions $i$ and $j$ can change, than the user has to track the original position of the identity to the new one.
\paragraph{STRgraph:} 
This function doesn't need arguments. Writing in the notebook \texttt{STRgraph} and pressing Shift-Enter the output will be a \texttt{Graphics[]} \textit{Mathematica}\textsuperscript{\textregistered} object reproducing the current diagram in the graphical environment of the \texttt{STR} interactive window. This function is updated in real-time but to use it in the \textit{Mathematica}\textsuperscript{\textregistered} notebook one needs to press \texttt{Export} and then \texttt{Graph} in the \texttt{STR} window. 
\paragraph{STRSimplify:} 
This function, given the spacetime dimension, will rewrite the output of the function  \texttt{STRprefactor} in terms of known functions following the definitions \eqref{defa} and \eqref{amultiple}.

\section{Some explicit examples}
\label{sec:example}
In this section we will present two examples of computation of Feynman diagrams with the method of uniqueness by means of \texttt{STR} package.

\subsection{Example 1: The kite self-energy diagram}
Let's start with a very basic example of computation of a Feynman diagram using the \texttt{STR} package.
We want to compute the two loop scalar \textit{kite} self-energy diagram in dimensional regularization $D=4-2\epsilon$ in the case in which all the propagators weights are $\alpha_i=1$ with $i=1,...,5$.
This diagram was computed in \cite{Kazakov:1983ns} by means of the uniqueness method.
It is well known that this diagram can be written as follows
\begin{equation}\label{kite}
\vcenter{\hbox{\includegraphics[width=3cm]{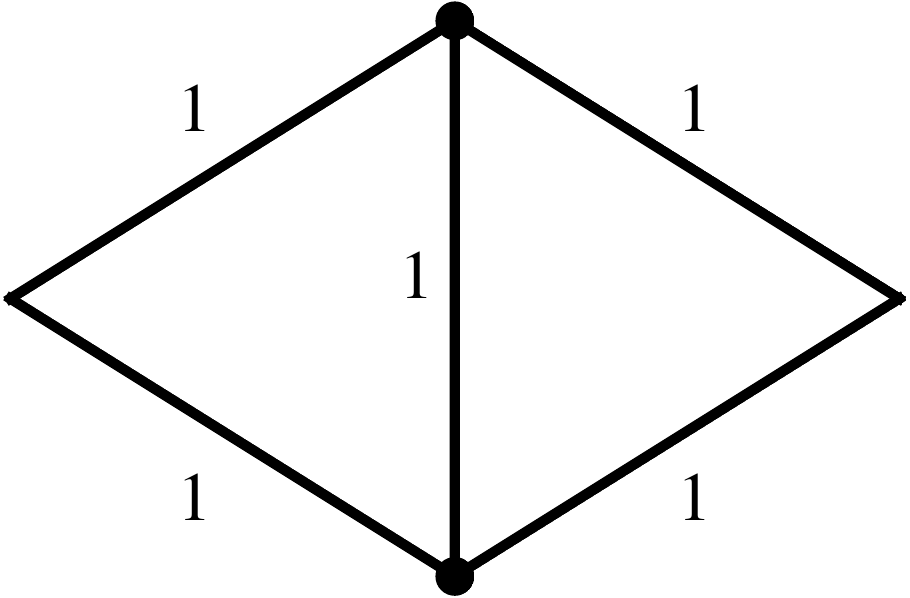}}}=-\frac{1}{\epsilon}\,\biggl[\;
\vcenter{\hbox{\includegraphics[width=3cm]{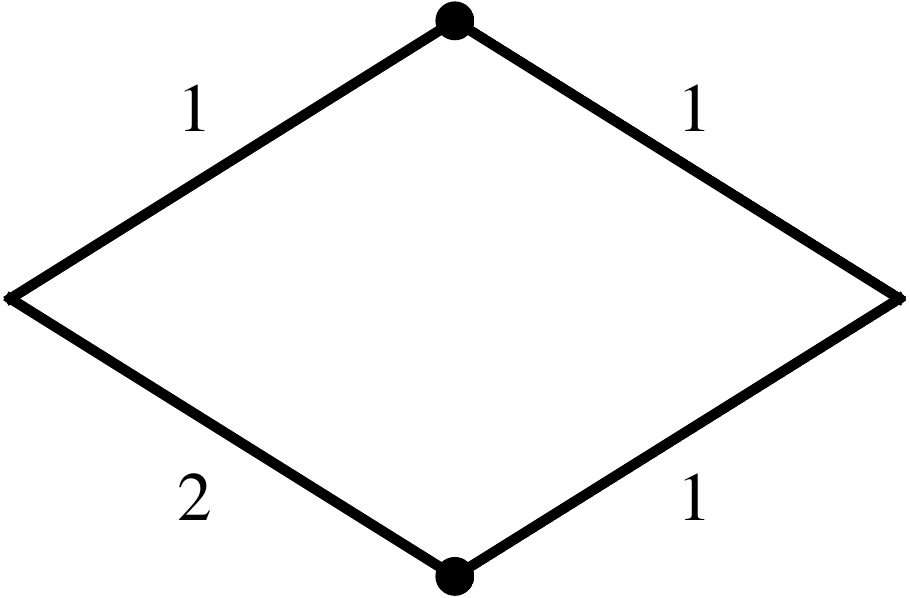}}}\,-\;
\vcenter{\hbox{\includegraphics[width=3cm]{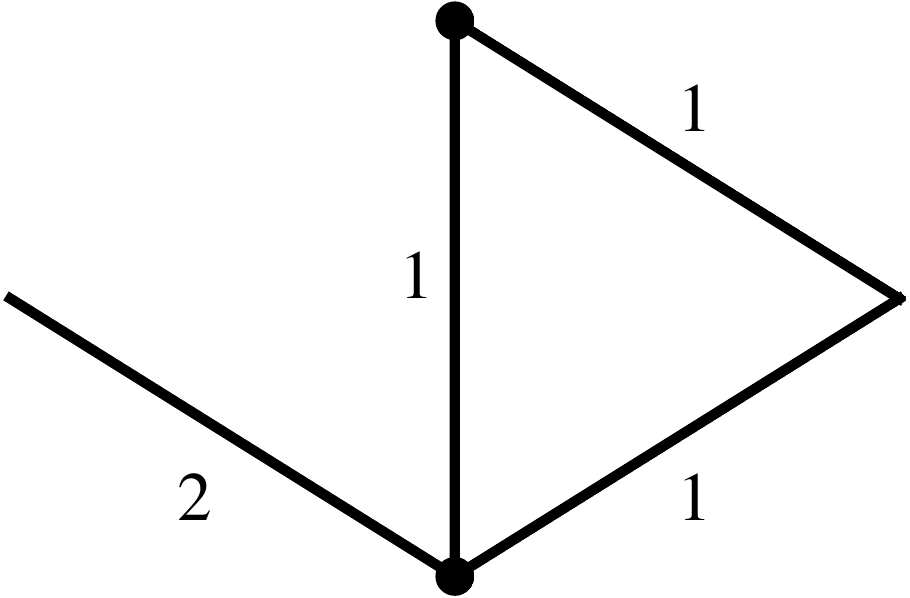}}}\,\biggl]\,,
\end{equation}
using the semi-uniqueness method (see footnote \ref{foot:semiunique}). Both of the diagrams in the right side can be computed with \texttt{STR} running the command \texttt{STR[4-2eps]} where \texttt{eps} is the dimensional regulator. Indeed the first one, using \texttt{Chain rule} and \texttt{Merge} can be transformed as follows
\begin{equation*}
\vcenter{\hbox{\includegraphics[trim={1.2cm 0 0 0},clip,width=3.5cm]{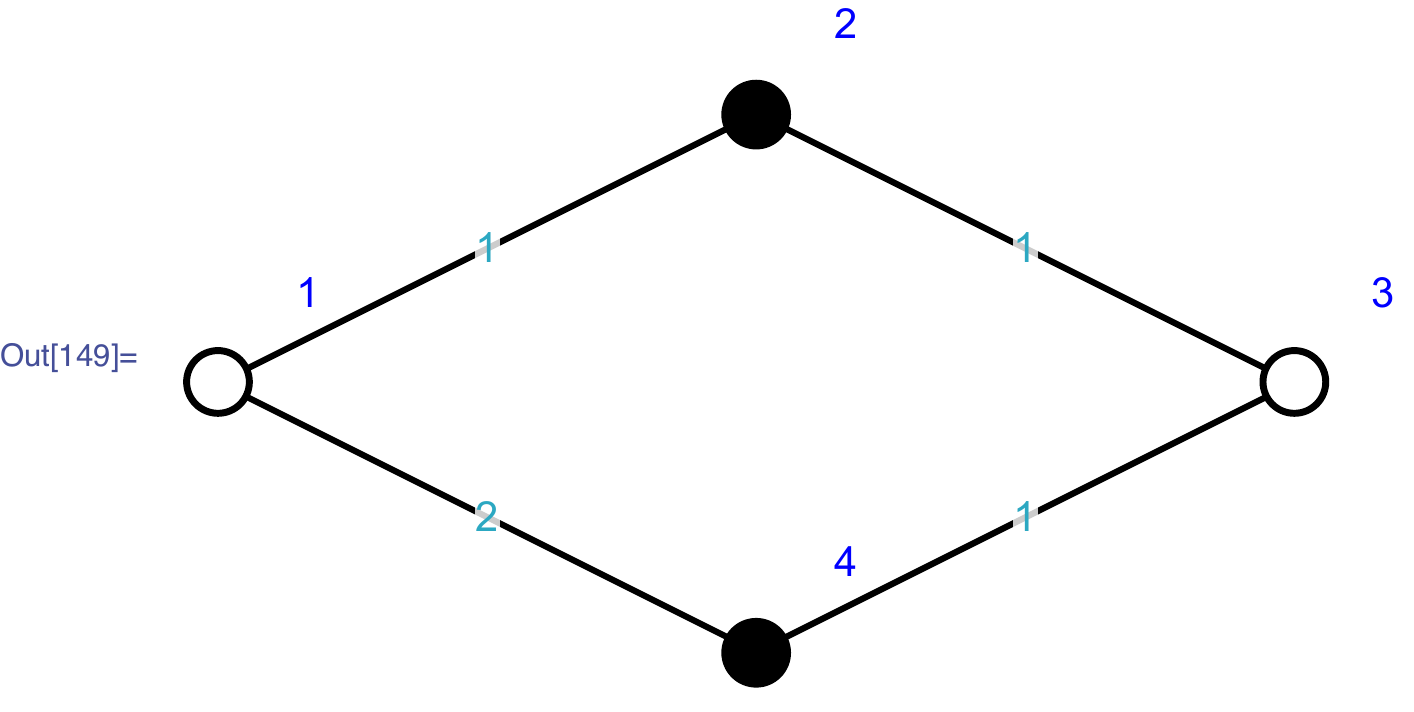}}}\Rightarrow
\vcenter{\hbox{\includegraphics[trim={1.2cm 0 0 0},clip,width=3.5cm]{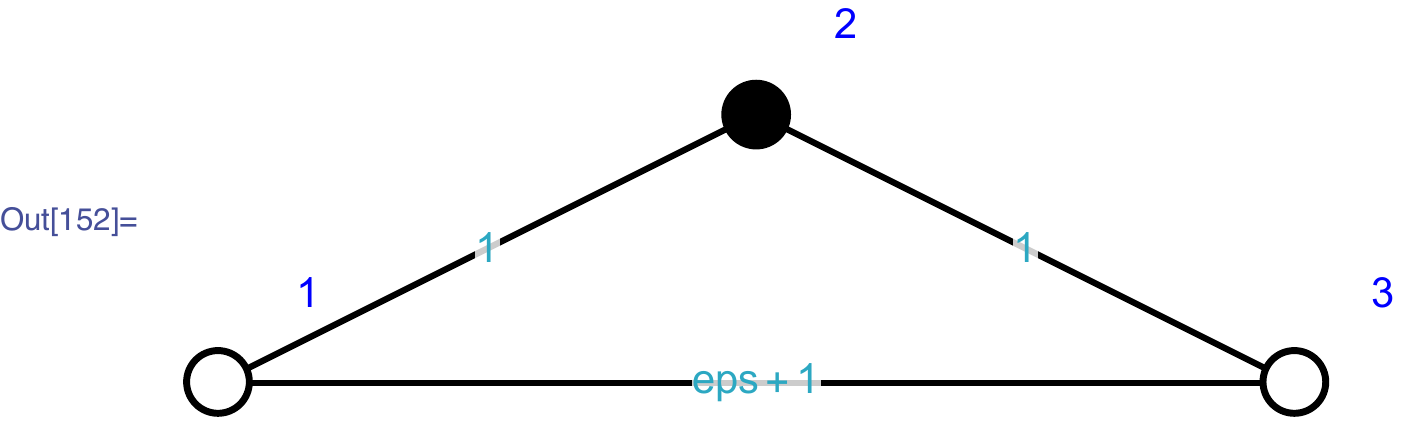}}}\Rightarrow
\vcenter{\hbox{\includegraphics[trim={1.2cm 0 0 0},clip,width=3.5cm]{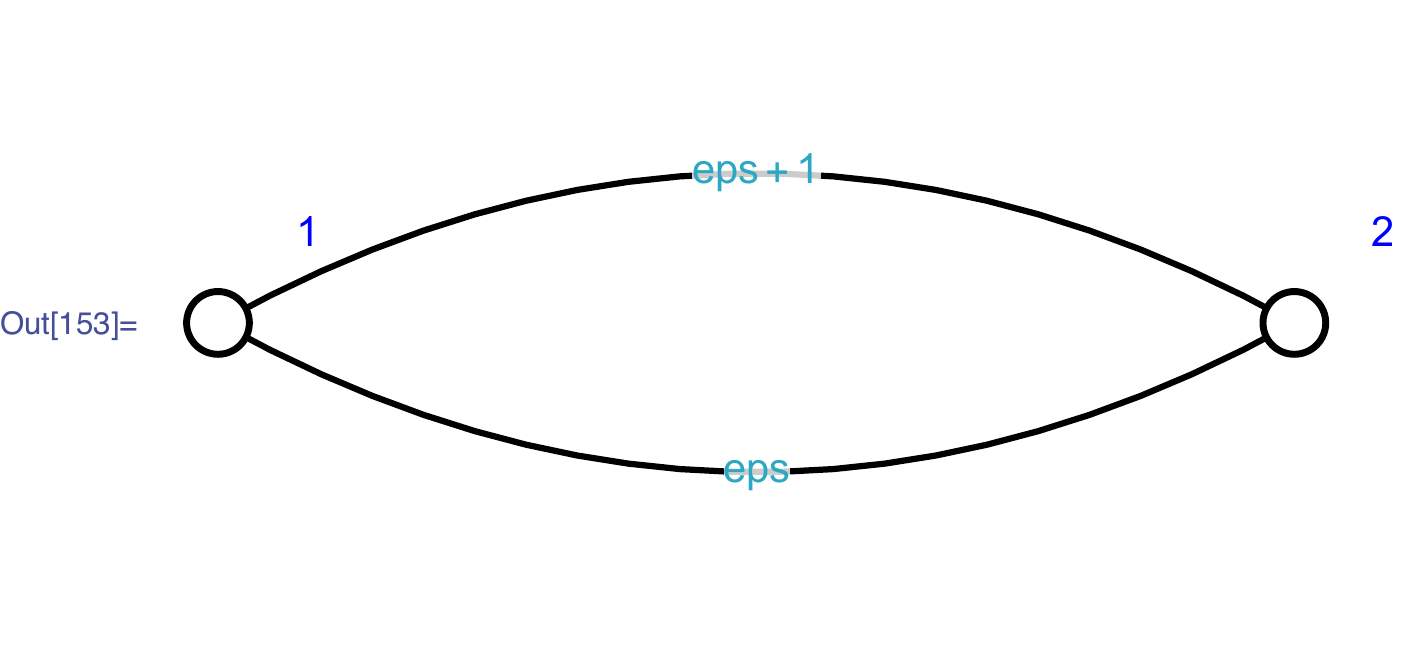}}}\Rightarrow
\vcenter{\hbox{\includegraphics[trim={1.2cm 0 0 0},clip,width=3.5cm]{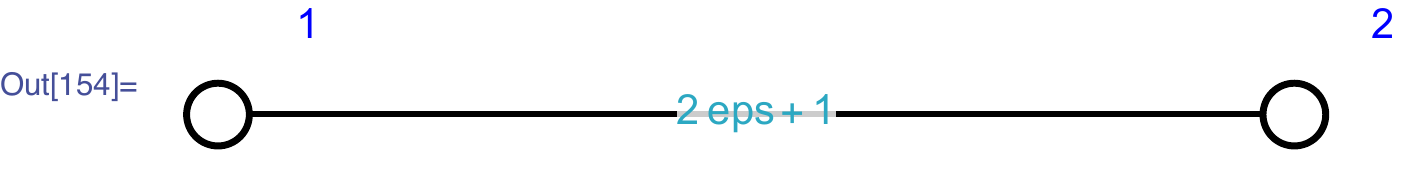}}}\,,
\end{equation*}
where in the last step we have

$
\qquad\quad
\vcenter{
\hbox{\includegraphics[trim={1.5cm 0 0 0},clip,width=4.2cm]{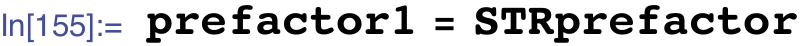}}\hbox{\includegraphics[trim={1.4cm 0 0 0 },clip,width=7cm]{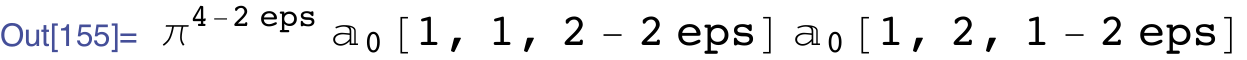}}\hbox{\includegraphics[trim={1.5cm 0 0 0},clip,width=3cm]{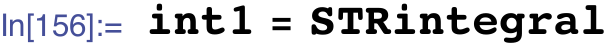}}\hbox{\includegraphics[trim={1.4cm 0 0 0},clip,width=2.5cm]{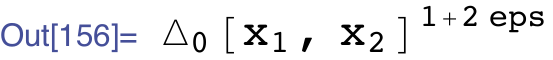}}}
$

Similarly we can compute the second diagram with the following steps
\begin{equation*}
\vcenter{\hbox{\includegraphics[trim={1.2cm 0 0 0},clip,width=3.5cm]{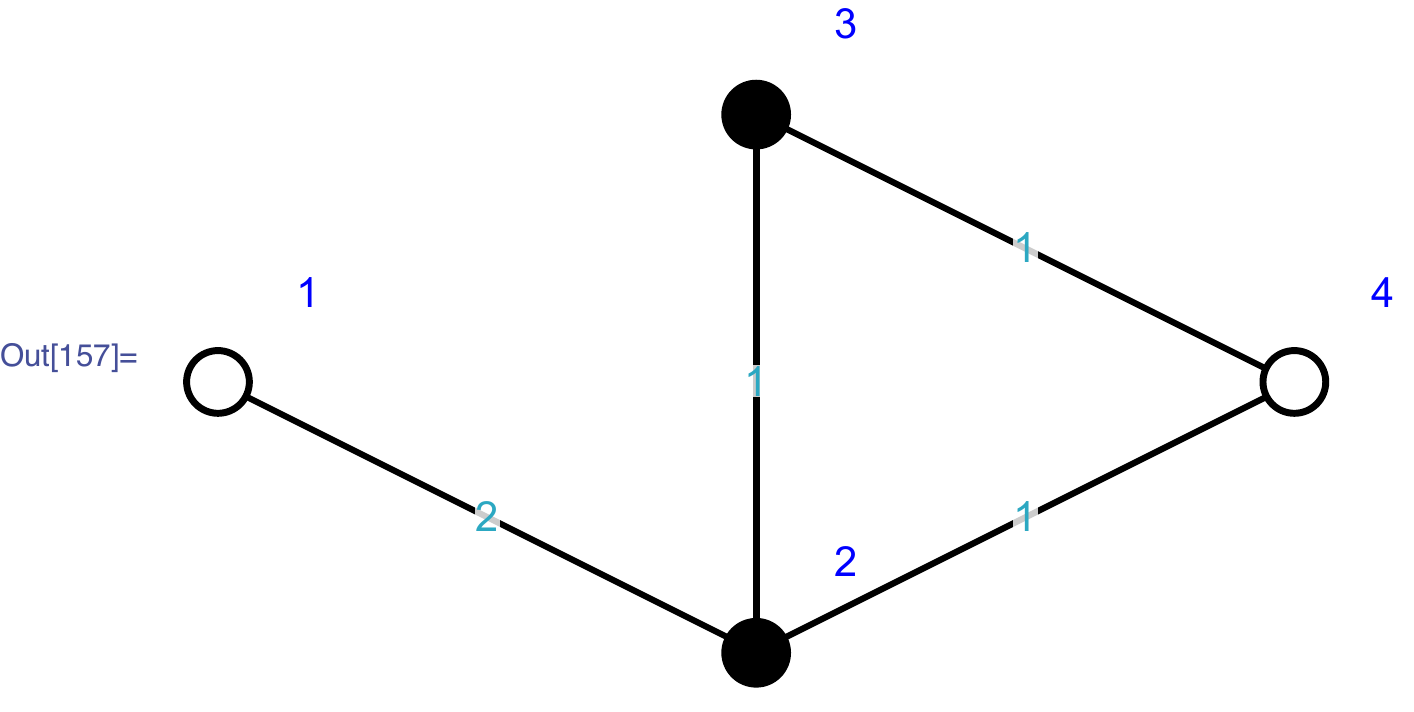}}}\Rightarrow
\vcenter{\hbox{\includegraphics[trim={1.2cm 0 0 0},clip,width=3.5cm]{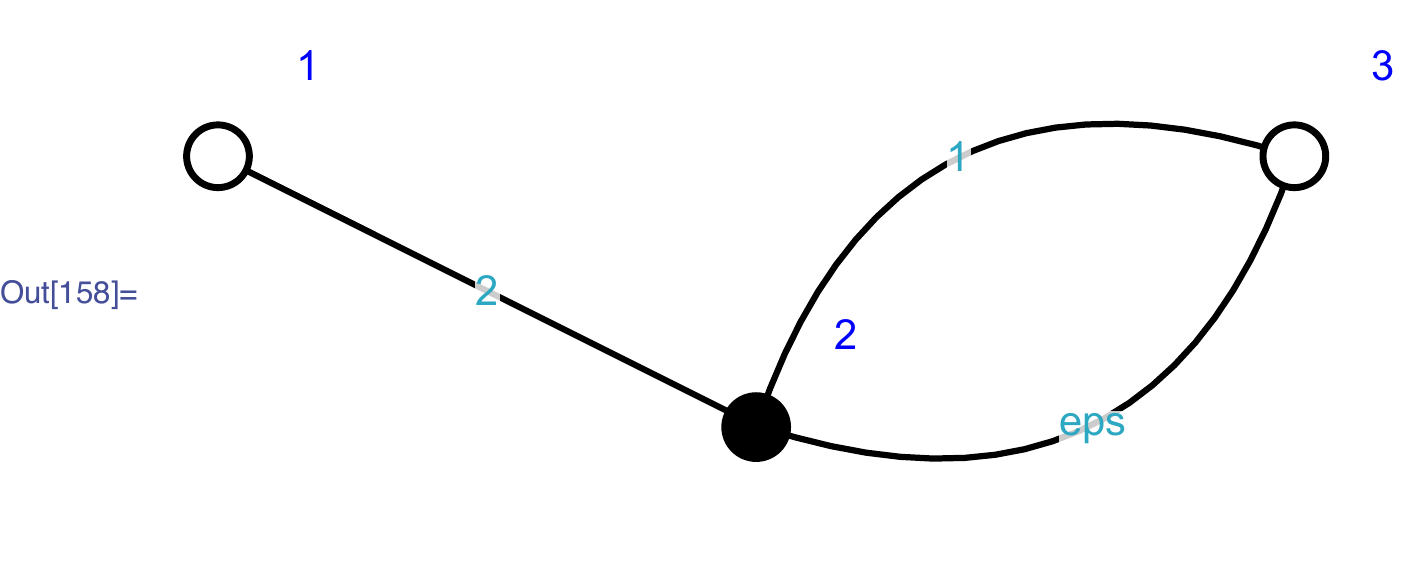}}}\Rightarrow
\vcenter{\hbox{\includegraphics[trim={1.2cm 0 0 0},clip,width=3.5cm]{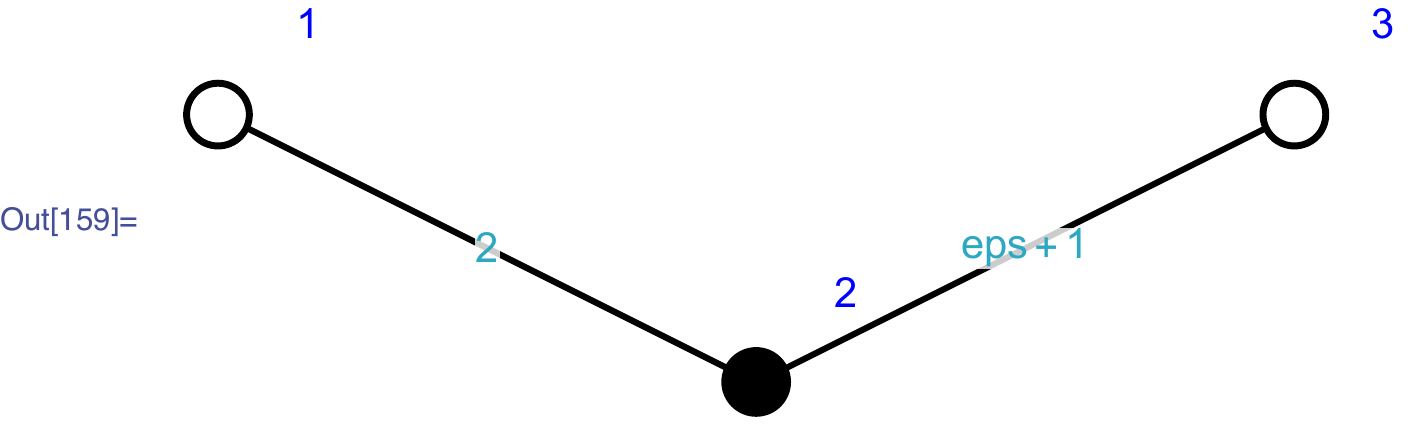}}}\Rightarrow
\vcenter{\hbox{\includegraphics[trim={1.2cm 0 0 0},clip,width=3.5cm]{STRkite14}}}\,,
\end{equation*}
where in the last step we have

$
\qquad\quad
\vcenter{
\hbox{\includegraphics[trim={1.5cm 0 0 0},clip,width=4.2cm]{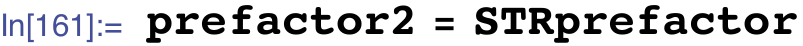}}\hbox{\includegraphics[trim={1.4cm 0 0 0 },clip,width=7.5cm]{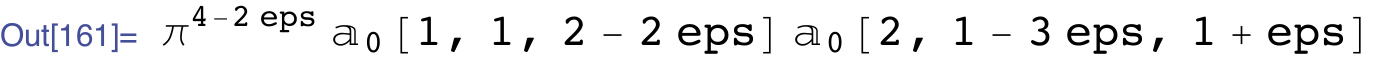}}\hbox{\includegraphics[trim={1.5cm 0 0 0},clip,width=3cm]{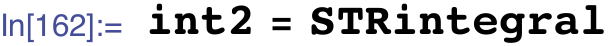}}\hbox{\includegraphics[trim={1.4cm 0 0 0},clip,width=2.5cm]{intkite2}}}
$

Summing up the two contributions according to \eqref{kite}, we obtain

$
\qquad\quad
\vcenter{
\hbox{\includegraphics[trim={1.5cm 0 0 0},clip,width=11cm]{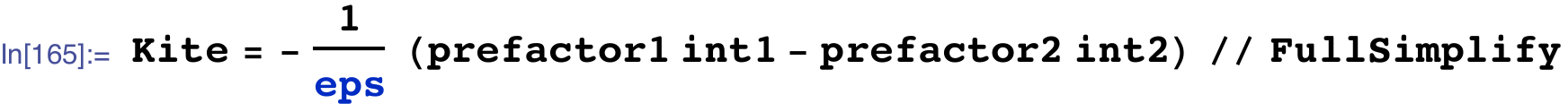}}\hbox{\includegraphics[trim={1.4cm 0 0 0 },clip,width=14cm]{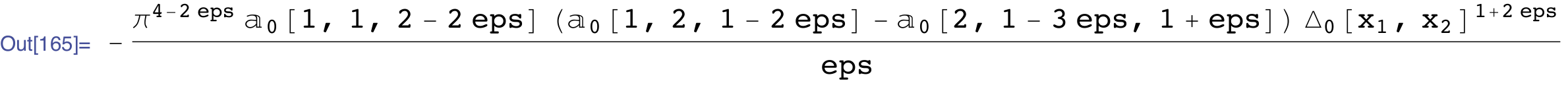}}}
$

matching the result of \cite{Kazakov:1983ns}. Now, using the function \texttt{STRSimplify} and expanding around \texttt{eps=0}, we can compute the value of the Feynman diagram in $D=4$
\begin{equation*}
\vcenter{\hbox{\includegraphics[width=2.5cm]{kite}}}\quad=\quad
\vcenter{
\hbox{\includegraphics[trim={1.5cm 0 0 0},clip,width=11cm]{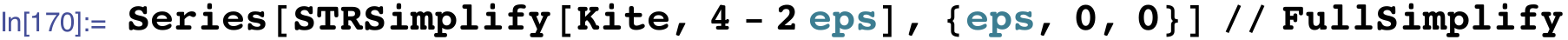}}\hbox{\includegraphics[trim={1.4cm 0 0 0 },clip,width=6cm]{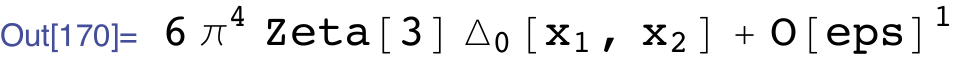}}}\,.
\end{equation*}

\subsection{Example 2: Diagram with a fermionic loop}
Here we present an example of computation of a Feynman diagram involving Yukawa star-triangle relations in $D=4$. The diagram is the following
\begin{equation*}
\vcenter{\hbox{\includegraphics[width=4.5cm]{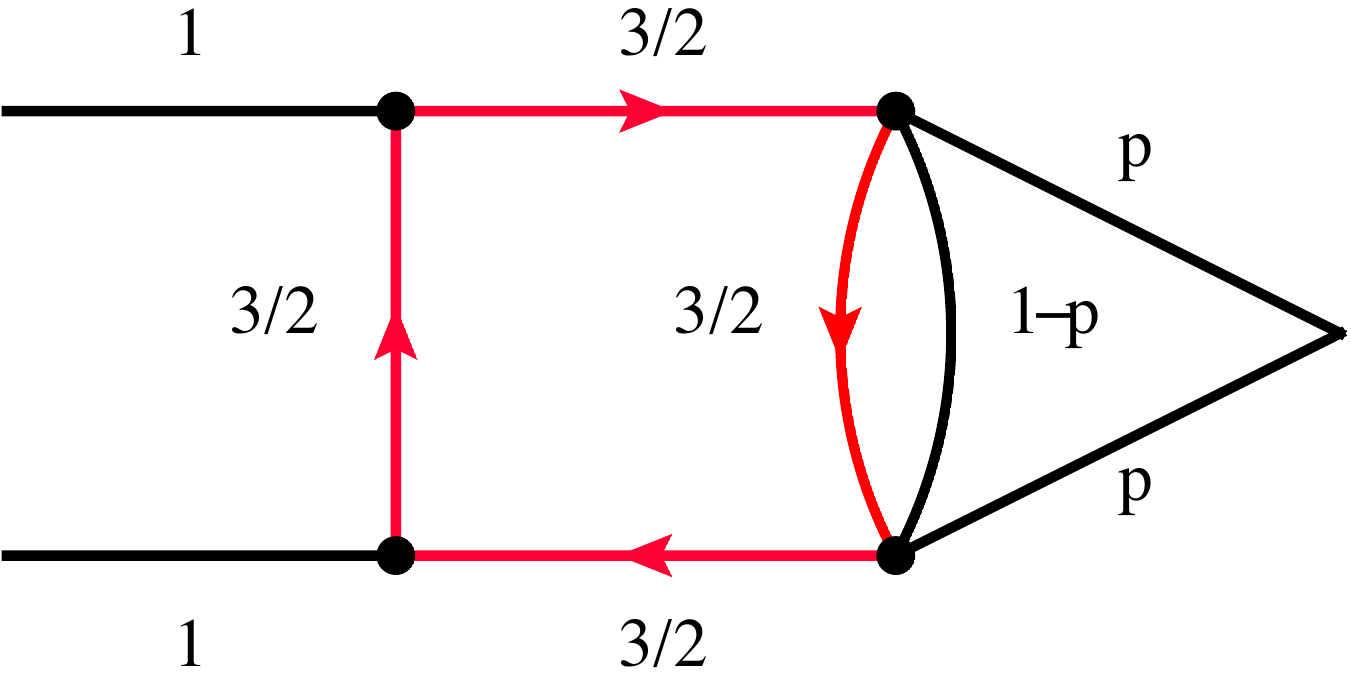}}}
\end{equation*}
where $p$ is real number and the chiral fermionic propagators are contracted in the spin indices forming a trace of $\sigma$-matrices. 

This graph arises in the computation of the spectrum of length-2 operators in the $\gamma$-deformed $\mathcal{N}=4$ SYM theory in the double scaling limit \cite{Kazakov:2018}.
This theory, proposed in \cite{Frolov:2005dj,Beisert:2005if}, is a 3-parameters deformation of the usual $\mathcal{N}=4$ SYM theory which breaks the $R$-symmetry down to $U(1)^3$ subgroup. In this framework, it is possible to consider a double scaling limit\footnote{Similar double scaling limits were also studied in \cite{Correa:2012nk,Bonini:2016fnc,deLeeuw:2016vgp,Aguilera-Damia:2016bqv,Preti:2017fhw} in different contexts.}, combining large imaginary twists and the weak coupling limit \cite{Gurdogan:2015csr}. Then the gauge fields decouple and we obtain a theory of three scalars and fermions interacting with quartic and Yukawa couplings. In order to compute the spectrum of the unprotected operators of length-2, we have to solve the diagram above that it can be computed straightforwardly with the uniqueness method with the help of the package \texttt{STR}. 

Running the command \texttt{STR[4]} and drawing the diagram above, we can solve it with the following steps
\begin{equation*}\begin{split}
&\vcenter{\hbox{\includegraphics[trim={1.75cm 0 0 0},clip,width=3cm]{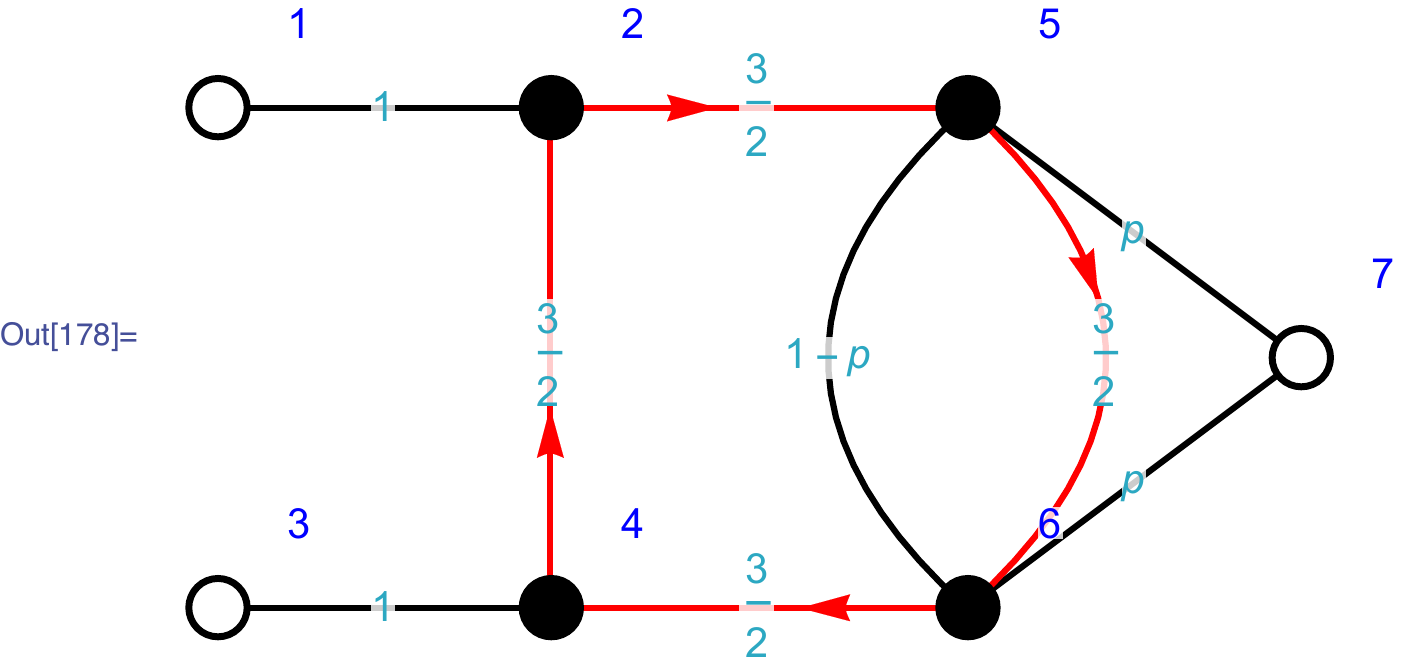}}}\!\!\Rightarrow\!\!
\vcenter{\hbox{\includegraphics[trim={1.75cm 0 0 0},clip,width=3cm]{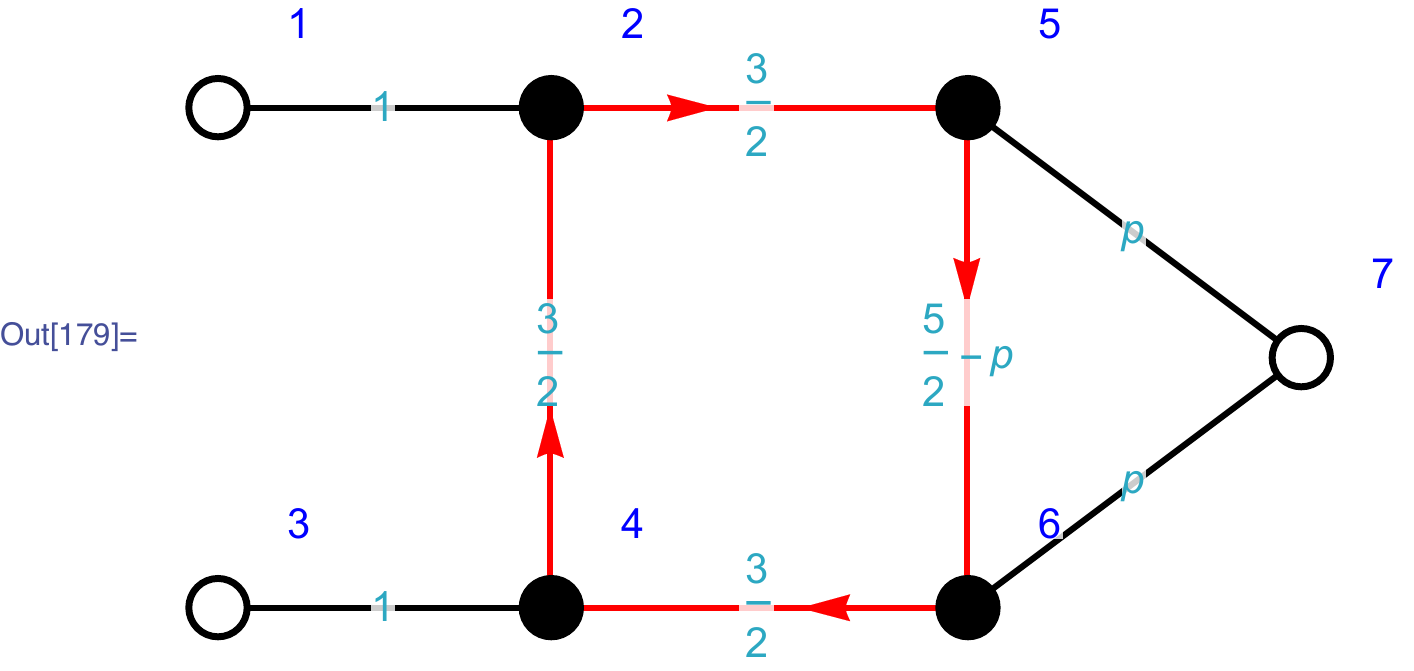}}}\!\!\Rightarrow\!\!
\vcenter{\hbox{\includegraphics[trim={1.75cm 0 0 0},clip,width=3cm]{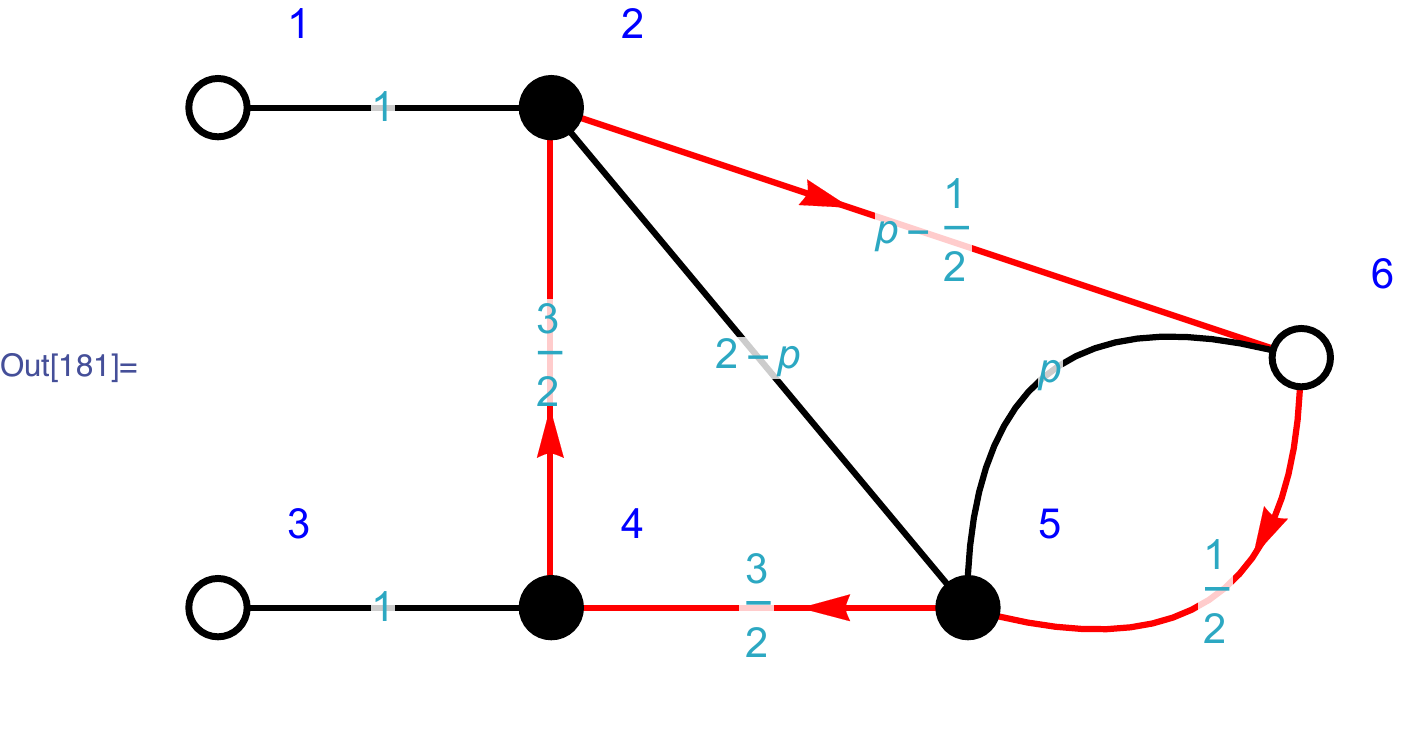}}}\!\!\Rightarrow\!\!
\vcenter{\hbox{\includegraphics[trim={1.75cm 0 0 0},clip,width=3cm]{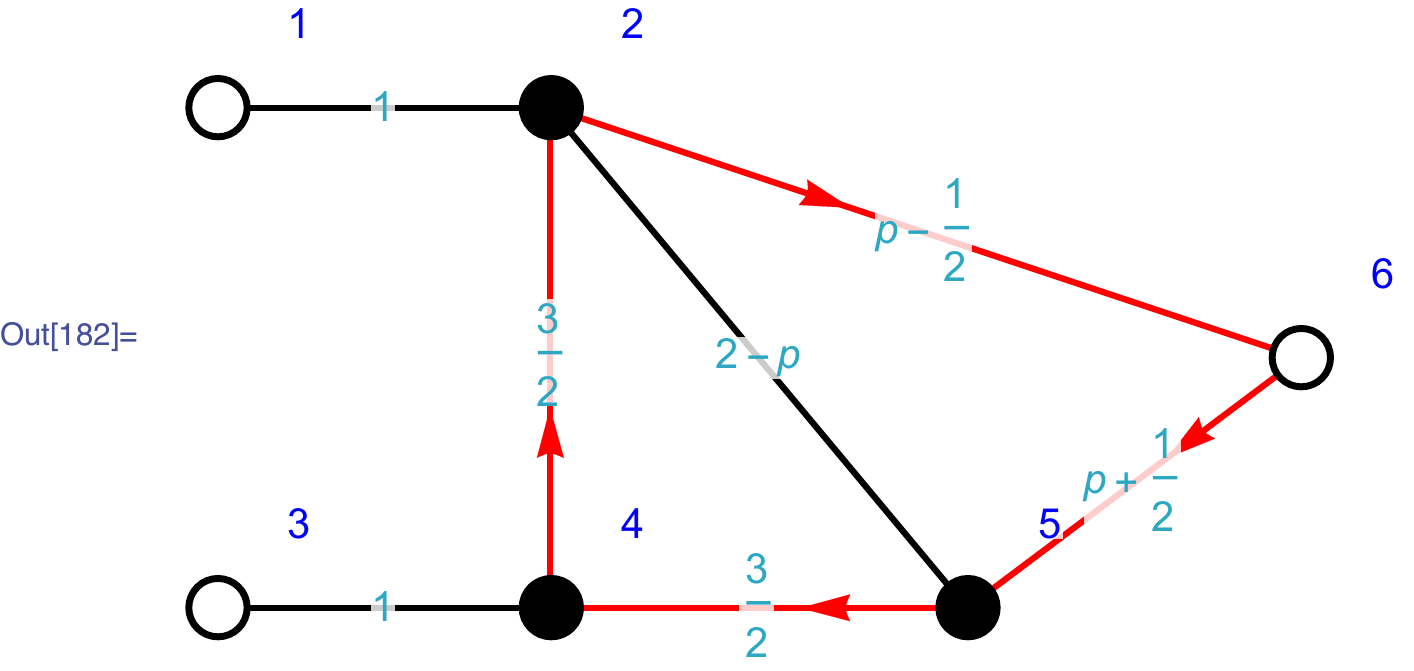}}}\!\!\Rightarrow\!\!\\
\!\!\Rightarrow&
\vcenter{\hbox{\includegraphics[trim={1.75cm 0 0 0},clip,width=3cm]{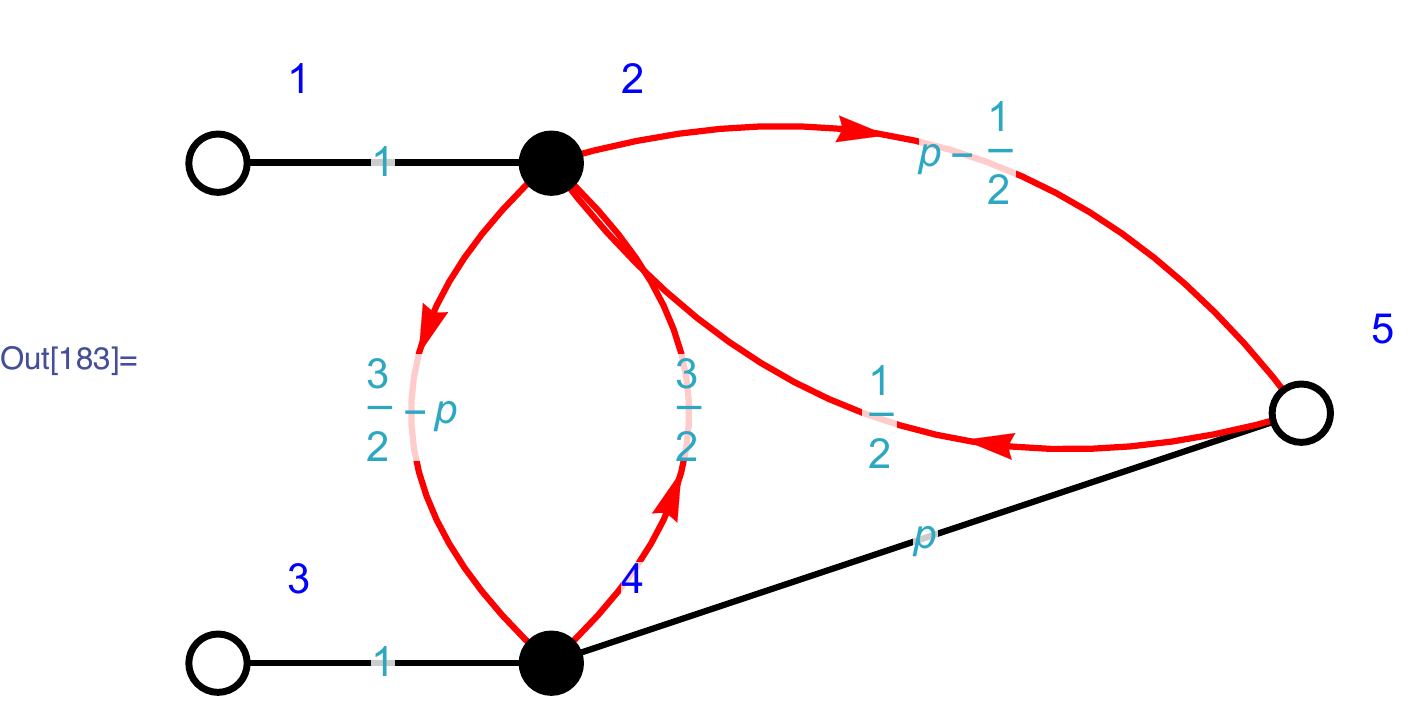}}}\!\!\Rightarrow\!\!
\vcenter{\hbox{\includegraphics[trim={1.75cm 0 0 0},clip,width=2.2cm]{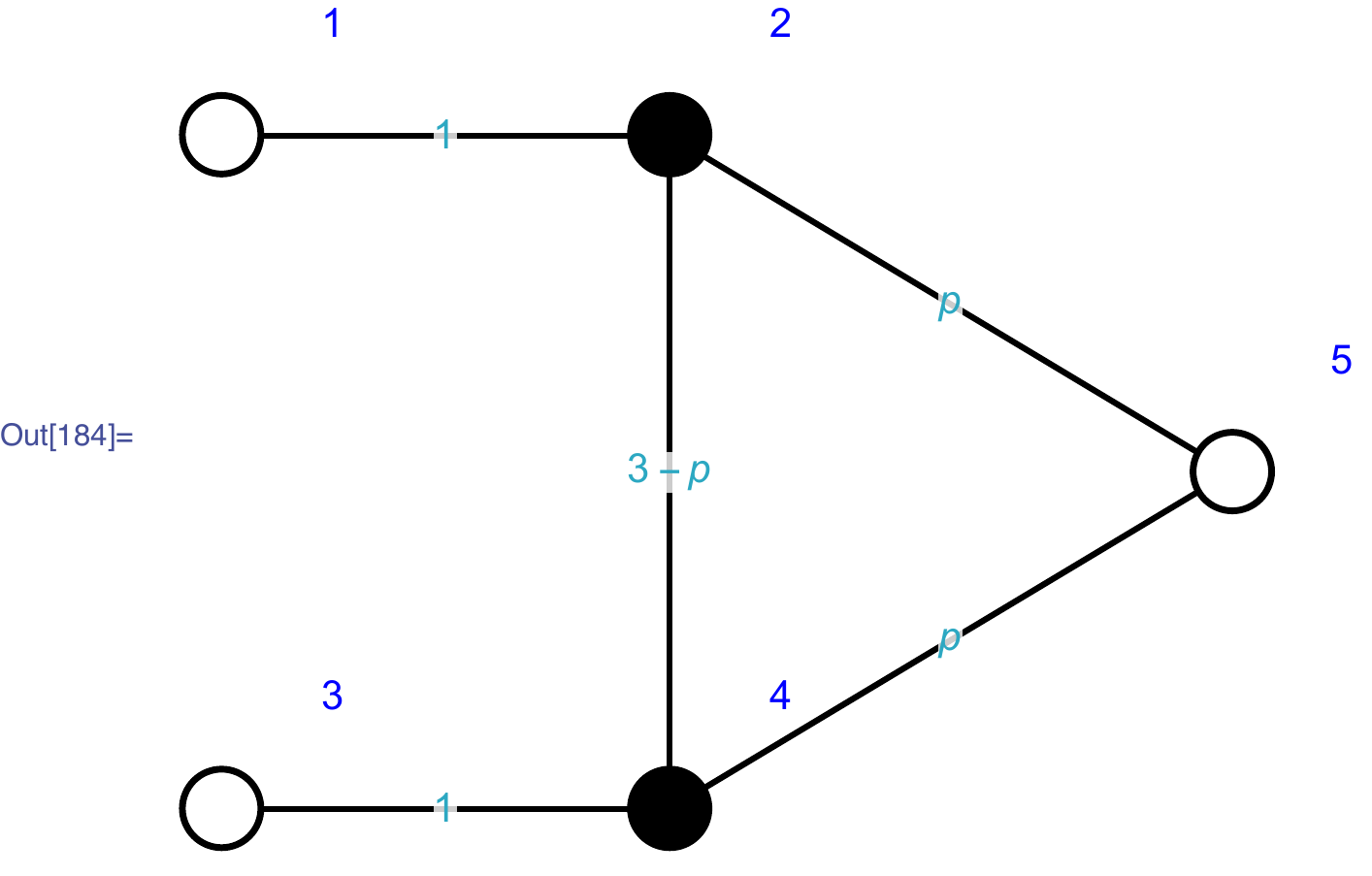}}}\!\!\Rightarrow\!\!
\vcenter{\hbox{\includegraphics[trim={1.75cm 0 0 0},clip,width=2.2cm]{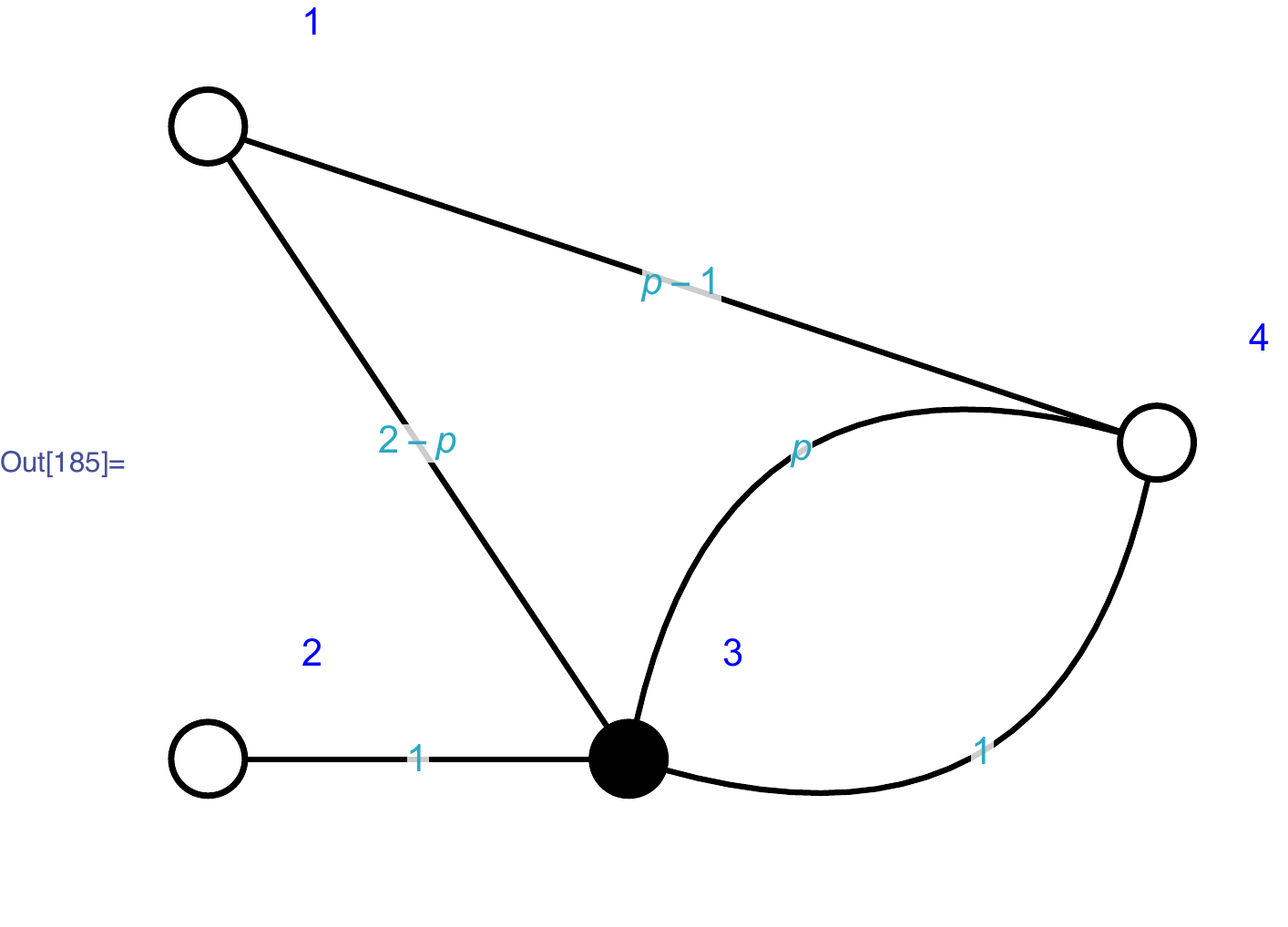}}}\!\!\Rightarrow\!\!
\vcenter{\hbox{\includegraphics[trim={1.75cm 0 0 0},clip,width=2.2cm]{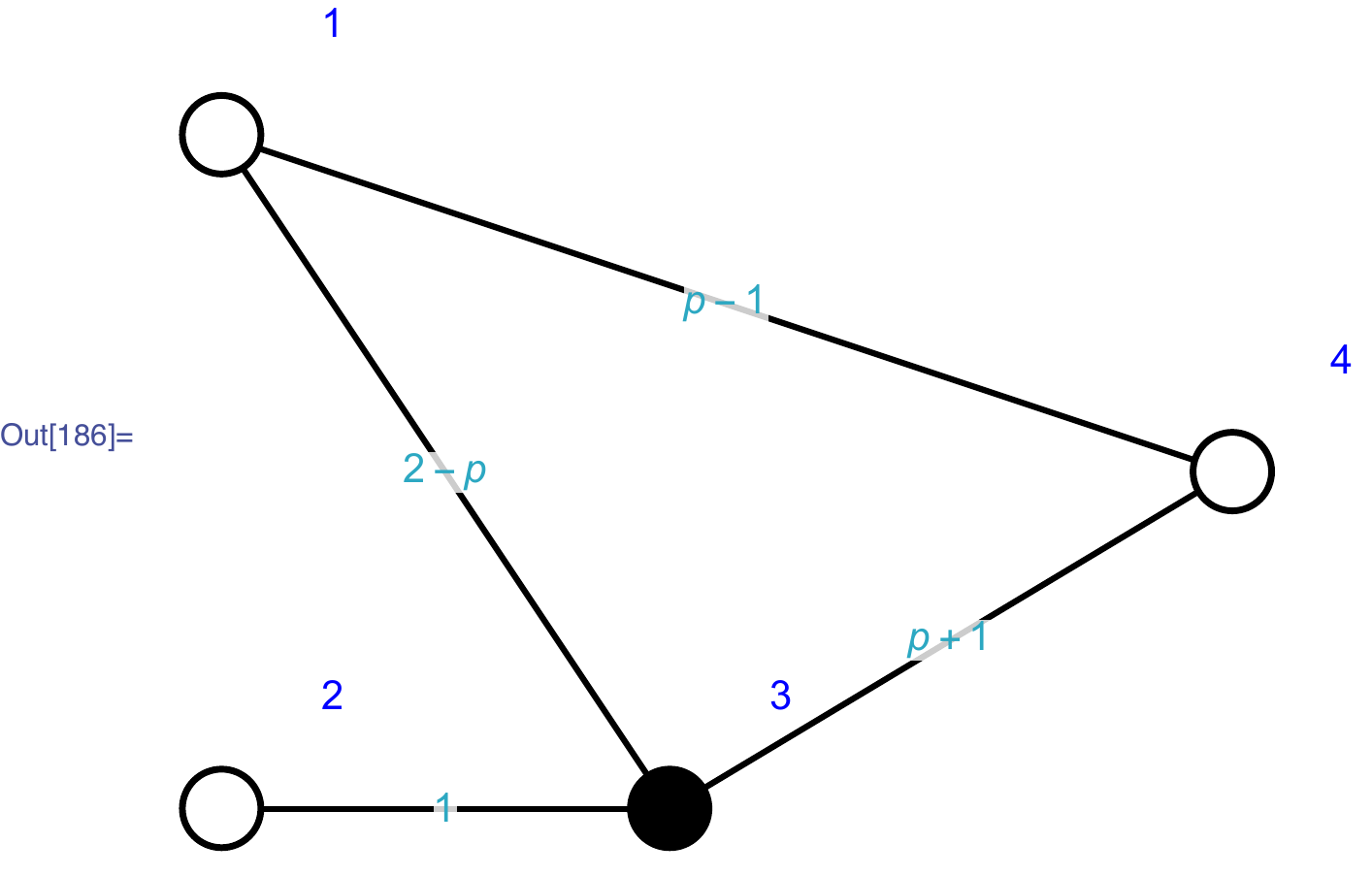}}}\!\!\Rightarrow\!\!
\vcenter{\hbox{\includegraphics[trim={1.75cm 0 0 0},clip,width=2.2cm]{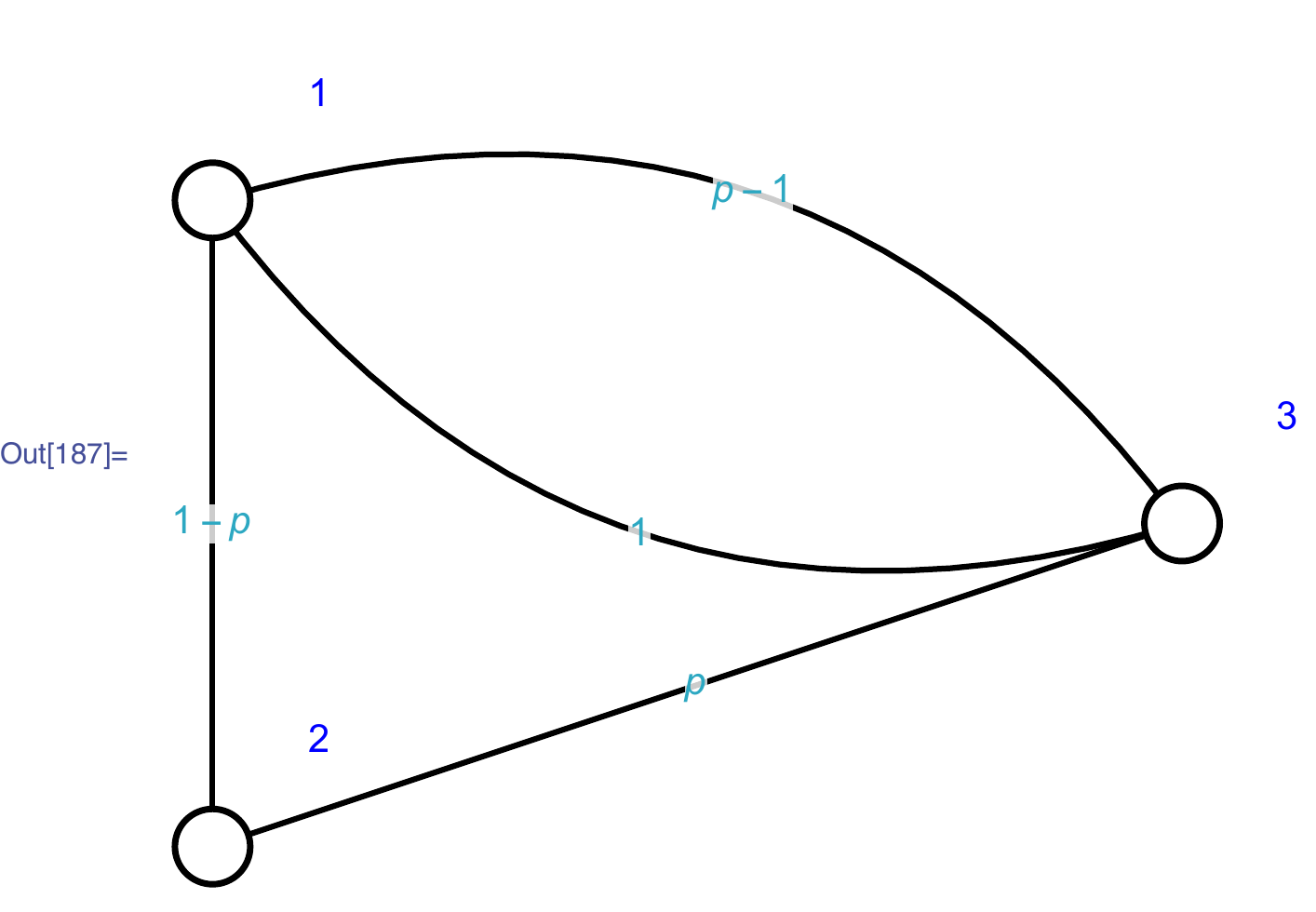}}}\!\!\Rightarrow\!\!
\vcenter{\hbox{\includegraphics[trim={1.75cm 0 0 0},clip,width=1.5cm]{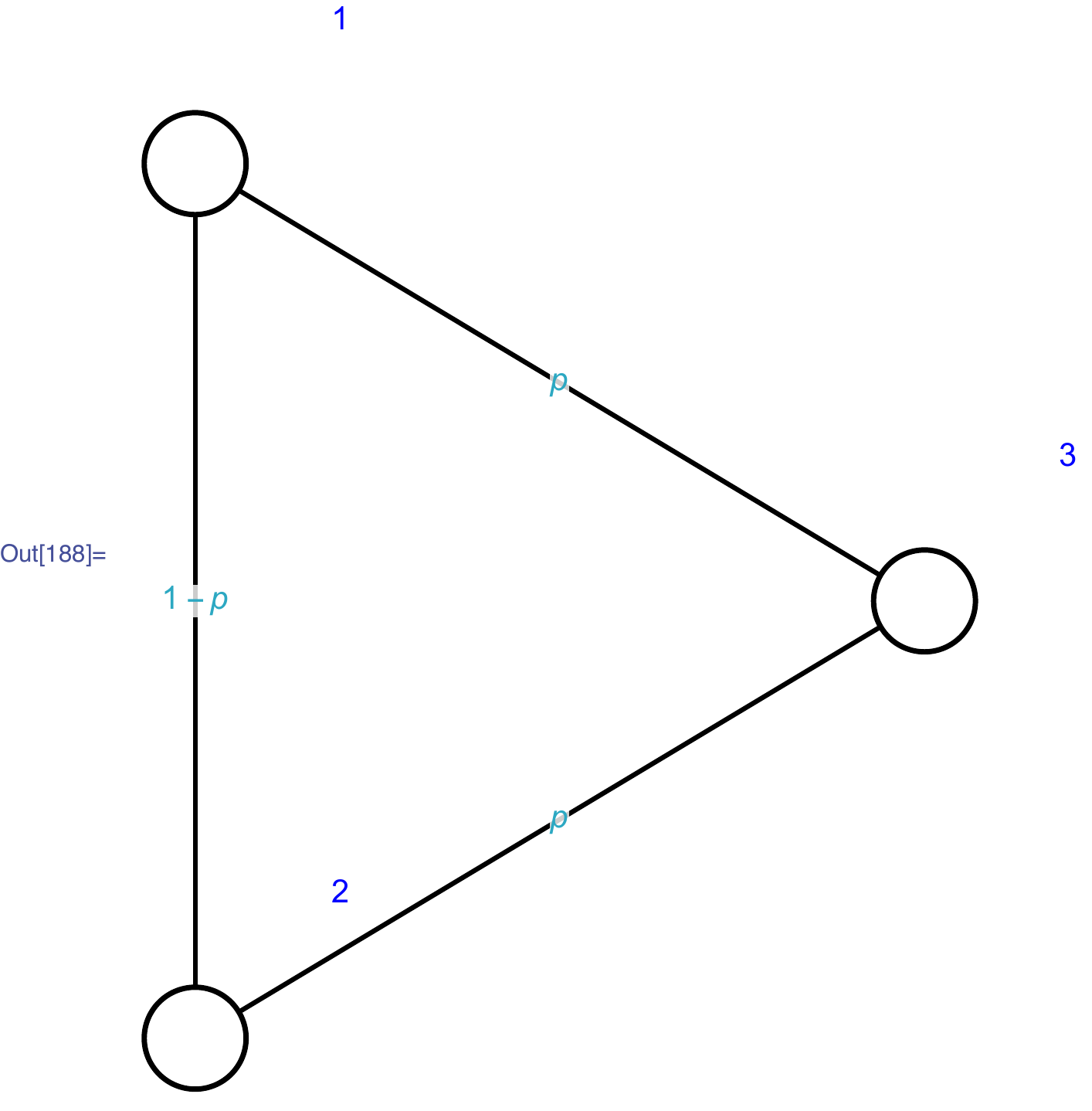}}}
\end{split}\end{equation*}
where we used the tools \texttt{Star-triangle} and \texttt{Merge multiedge} several times. In the last step we don't have any residual integration and we obtain

$
\qquad\quad
\vcenter{
\hbox{\includegraphics[trim={1.5cm 0 0 0},clip,width=2.5cm]{STRprefactor}}\hbox{\includegraphics[trim={1.4cm 0 0 0 },clip,width=12cm]{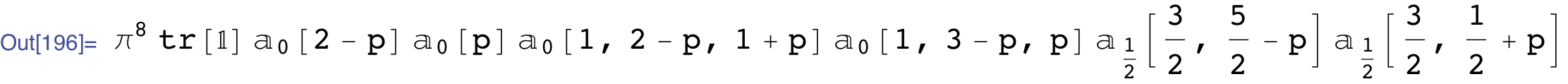}}\hbox{\includegraphics[trim={1.5cm 0 0 0},clip,width=2.5cm]{STRintegral}}\hbox{\includegraphics[trim={1.4cm 0 0 0},clip,width=7cm]{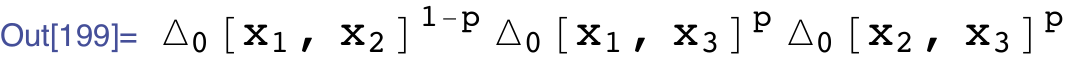}}}
$

Now using the function \texttt{STRSimplify}, we can conclude that
\begin{equation*}
\vcenter{\hbox{\includegraphics[width=4cm]{fish}}}=
\vcenter{
\hbox{\includegraphics[trim={1.5cm 0 0 0},clip,width=12cm]{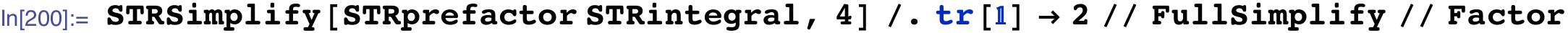}}\hbox{\includegraphics[trim={1.4cm 0 0 0 },clip,width=8cm]{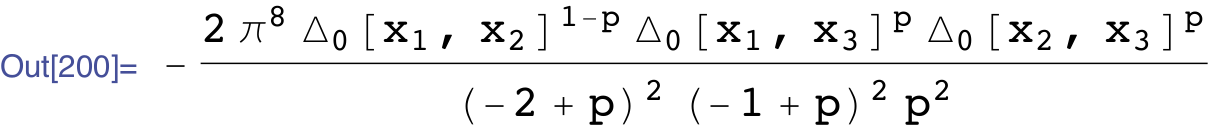}}}
\end{equation*}
where we used that $\text{tr}(\mathbb{1})=2$ in $D=4$ (see appendix \ref{sec:appendixA}).

\section*{Acknowledgments}
We thank Sergei Derkachov, Vladimir Kazakov and Enrico Olivucci for discussions and the critical reading of the draft. The work of MP is supported by the European Research Council (Programme "Ideas" ERC-2012-AdG 320769 AdS-CFT-solvable).

\appendix

\section{Notation and conventions}\label{sec:appendixA}

In this paper the metric tensor of the $D$ dimensional Euclidean space is taken to be
\begin{equation}
g_{\mu\nu}=\delta_{\mu\nu}=\text{diag}(\;\underbrace{1,1,...,1,...,1}_{D \text{ times}}\;)
\end{equation}
where $\mu,\nu=0,1,...,D-1$ are spacetime vector indices.
We can introduce the $D$ dimensional gamma matrices ${(\gamma^\mu)_{\alpha}}^\beta$ satisfying the Clifford algebra
\begin{equation}\label{clifford}
\gamma^\mu\gamma^\nu+\gamma^\nu\gamma^\mu=2\delta^{\mu\nu}\mathbb{1}
\end{equation}
The gamma matrices are hermitian and their size depends on the spacetime dimension: if $D$ is even they are $2^{D/2}\times 2^{D/2}$ matrices and if $D$ is odd they are $2^{(D-1)/2}\times 2^{(D-1)/2}$. When the spin indices of the gamma matrices are not explicit we are considering the matrices contracted as $\gamma^\mu\gamma^\nu={(\gamma^\mu)_{\alpha}}^\beta{(\gamma^\nu)_{\beta}}^\gamma$. The trace over the product of gamma matrices is dimension-dependent if $D$ is odd, for instance the trace over an odd number of matrices is not-zero in general
\begin{equation}\begin{split}
&\text{tr}(\gamma^\mu\gamma^\nu\gamma^\rho)\neq 0\qquad\text{if}\quad D=1,3\\
&\text{tr}(\gamma^\mu\gamma^\nu\gamma^\rho\gamma^\sigma\gamma^\lambda)\neq 0\qquad\text{if}\quad D=1,3,5\qquad\quad\text{etc.}
\end{split}\end{equation}
However when the spacetime dimension $D$ is even, the traces are dimensional independent, for instance
\begin{equation}\begin{split}
&\text{tr}(\text{\textit{odd number of $\gamma$'s}})= 0\\
&\text{tr}(\gamma^\mu\gamma^\nu)=\delta^{\mu\nu}\text{tr}(\mathbb{1})\\
&\text{tr}(\gamma^\mu\gamma^\nu\gamma^\rho\gamma^\sigma)=(\delta^{\mu\sigma}\delta^{\nu\rho}-\delta^{\mu\rho}\delta^{\nu\sigma}+\delta^{\mu\nu}\delta^{\rho\sigma})\text{tr}(\mathbb{1})\qquad\quad\text{etc.}
\end{split}\end{equation}
and the $D$-dependence resides entirely in $\text{tr}(\mathbb{1})=2^{D/2}$.

The gamma matrices in even dimensions can be chosen to have the "off-block
diagonal" form
\begin{equation}
\gamma^\mu=\begin{pmatrix} 
0 & (\sigma^\mu)_{\alpha\dot{\beta}} \\
(\bar{\sigma}^\mu)^{\dot{\alpha}\beta} & 0 
\end{pmatrix}
\end{equation}
by introducing the $2^{{D/2}-1}\times 2^{{D/2}-1}$ Euclidean $\sigma$ matrices
\begin{equation}
\sigma^\mu=(-i\vec{\boldsymbol{\sigma}},\mathbb{1}_{2^{{D/2}-1}})\qquad\text{and}\qquad
\bar{\sigma}^\mu=(i\vec{\boldsymbol{\sigma}},\mathbb{1}_{2^{{D/2}-1}})
\end{equation}
where $\vec{\boldsymbol{\sigma}}$ is a $D-1$ dimensional vector of \textit{Pauli matrices} and $\mathbb{1}_{2^{{D/2}-1}}$ is the $2^{{D/2}-1}\times 2^{{D/2}-1}$ identity matrix.
Writing the identity matrix appearing in \eqref{clifford} as
\begin{equation}
\mathbb{1}=\begin{pmatrix} 
{\delta_\alpha}^\beta\equiv\mathbb{1}_{2^{{D/2}-1}} & 0 \\
 0 & {\delta^{\dot{\alpha}}}_{\dot{\beta}} \equiv\mathbb{1}_{2^{{D/2}-1}}
\end{pmatrix}
\end{equation}
the $\sigma$ matrices satisfy
\begin{equation}\label{symferm}
\bar{\sigma}_\mu\sigma_\nu +\bar{\sigma}_\nu \sigma_\mu=2\delta_{\mu\nu} \mathbb{1}_{2^{{D/2}-1}} \qquad\text{and}\qquad
\sigma_\mu\bar{\sigma}_\nu +\sigma_\nu \bar{\sigma}_\mu=2\delta_{\mu\nu} \mathbb{1}_{2^{{D/2}-1}} 
\end{equation}
The traces identities are
\begin{equation}\begin{split}
&\text{tr}(\text{\textit{odd number of $\sigma$'s}})= 0\\
&\text{tr}(\sigma^\mu\bar{\sigma}^\nu)=\text{tr}(\bar{\sigma}^\mu\sigma^\nu)=\delta^{\mu\nu}\text{tr}( \mathbb{1}_{2^{{D/2}-1}})\\
&\text{tr}(\sigma^\mu\bar{\sigma}^\nu\sigma^\rho\bar{\sigma}^\sigma)+\text{tr}(\bar{\sigma}^\mu\sigma^\nu\bar{\sigma}^\rho\sigma^\sigma)=2(\delta^{\mu\sigma}\delta^{\nu\rho}-\delta^{\mu\rho}\delta^{\nu\sigma}+\delta^{\mu\nu}\delta^{\rho\sigma})\text{tr}( \mathbb{1}_{2^{{D/2}-1}})\qquad\quad\text{etc.}
\end{split}\end{equation}
where $\text{tr}( \mathbb{1}_{2^{{D/2}-1}})=2^{D/2-1}$.

\bibliographystyle{nb}

\bibliography{biblio}

\begin{thebibliography}{10}
\ifx\href\asklfhas\newcommand{\href}[2]{#2}\fi
\ifx\arxivref\asklfhas\newcommand{\arxivref}[2]{\href{http://arxiv.org/abs/#1}{#2}}\fi
\ifx\doiref\asklfhas\newcommand{\doiref}[2]{\href{http://dx.doi.org/#1}{#2}}\fi
\raggedright
\small
\parskip 0pt

\bibitem{Vasiliev:1981yc}
A.~N.~Vasiliev, {\relax Yu}.~M.~Pismak and {\relax Yu}.~R.~Khonkonen,
\textit{``{Simple Method of Calculating the Critical Indices in the 1/$N$
  Expansion}''},
\textsf{\doiref{10.1007/BF01030844}{Theor.~Math.~Phys.~46,~104~(1981)}},
[Teor. Mat. Fiz.46,157(1981)].

\bibitem{Tkachov:1981wb}
F.~V.~Tkachov,
\textit{``{A Theorem on Analytical Calculability of Four Loop Renormalization
  Group Functions}''},
\textsf{\doiref{10.1016/0370-2693(81)90288-4}{Phys.~Lett.~100B,~65~(1981)}}.

\bibitem{Chetyrkin:1981qh}
K.~G.~Chetyrkin and F.~V.~Tkachov,
\textit{``{Integration by Parts: The Algorithm to Calculate beta Functions in 4
  Loops}''},
\textsf{\doiref{10.1016/0550-3213(81)90199-1}{Nucl.~Phys.~B192,~159~(1981)}}.

\bibitem{Smirnov:2014hma}
A.~V.~Smirnov,
\textit{``{FIRE5: a C++ implementation of Feynman Integral REduction}''},
\textsf{\doiref{10.1016/j.cpc.2014.11.024}{Comput.~Phys.~Commun.~189,~182~(2015)}},
\texttt{\arxivref{1408.2372}{arXiv:1408.2372}}.

\bibitem{Laporta:2001dd}
S.~Laporta,
\textit{``{High precision calculation of multiloop Feynman integrals by
  difference equations}''},
\textsf{\doiref{10.1016/S0217-751X(00)00215-7,
  10.1142/S0217751X00002157}{Int.~J.~Mod.~Phys.~A15,~5087~(2000)}},
\texttt{\arxivref{hep-ph/0102033}{hep-ph/0102033}}.

\bibitem{Chetyrkin:1980pr}
K.~G.~Chetyrkin, A.~L.~Kataev and F.~V.~Tkachov,
\textit{``{New Approach to Evaluation of Multiloop Feynman Integrals: The
  Gegenbauer Polynomial x Space Technique}''},
\textsf{\doiref{10.1016/0550-3213(80)90289-8}{Nucl.~Phys.~B174,~345~(1980)}}.

\bibitem{Bergere:1973fq}
M.~C.~Bergere and Y.-M.~P.~Lam,
\textit{``{Asymptotic expansion of Feynman amplitudes. Part 1: The convergent
  case}''},
\textsf{\doiref{10.1007/BF01609168}{Commun.~Math.~Phys.~39,~1~(1974)}}.

\bibitem{Usyukina:1975yg}
N.~I.~Usyukina,
\textit{``{On a Representation for Three Point Function}''},
\textsf{\doiref{10.1007/BF01037795}{Teor.~Mat.~Fiz.~22,~300~(1975)}}.

\bibitem{Grozin:2014hna}
A.~Grozin, J.~M.~Henn, G.~P.~Korchemsky and P.~Marquard,
\textit{``{Three Loop Cusp Anomalous Dimension in QCD}''},
\textsf{\doiref{10.1103/PhysRevLett.114.062006}{Phys.~Rev.~Lett.~114,~062006~(2015)}},
\texttt{\arxivref{1409.0023}{arXiv:1409.0023}}.

\bibitem{Grozin:2015kna}
A.~Grozin, J.~M.~Henn, G.~P.~Korchemsky and P.~Marquard,
\textit{``{The three-loop cusp anomalous dimension in QCD and its
  supersymmetric extensions}''},
\textsf{\doiref{10.1007/JHEP01(2016)140}{JHEP~1601,~140~(2016)}},
\texttt{\arxivref{1510.07803}{arXiv:1510.07803}}.

\bibitem{Bianchi:2017svd}
M.~S.~Bianchi, L.~Griguolo, A.~Mauri, S.~Penati, M.~Preti and D.~Seminara,
\textit{``{Towards the exact Bremsstrahlung function of ABJM theory}''},
\textsf{\doiref{10.1007/JHEP08(2017)022}{JHEP~1708,~022~(2017)}},
\texttt{\arxivref{1705.10780}{arXiv:1705.10780}}.

\bibitem{Remiddi:1997ny}
E.~Remiddi,
\textit{``{Differential equations for Feynman graph amplitudes}''},
\textsf{Nuovo~Cim.~A110,~1435~(1997)},
\texttt{\arxivref{hep-th/9711188}{hep-th/9711188}}.

\bibitem{Smirnov:2012gma}
V.~A.~Smirnov,
\textit{``{Analytic tools for Feynman integrals}''},
\textsf{\doiref{10.1007/978-3-642-34886-0}{Springer~Tracts~Mod.~Phys.~250,~1~(2012)}}.

\bibitem{DEramo:1971hnd}
M.~D'Eramo, G.~Parisi and L.~Peliti,
\textit{``{Theoretical predictions for critical exponents at the $\lambda$
  point of Bose liquids}''},
\textsf{\doiref{10.1007/BF02774121}{Lett.~Nuovo~Cim.~2,~878~(1971)}}.

\bibitem{Fradkin:1978pp}
E.~S.~Fradkin and M.~{\relax Ya}.~Palchik,
\textit{``{Recent Developments in Conformal Invariant Quantum Field Theory}''},
\textsf{\doiref{10.1016/0370-1573(78)90172-2}{Phys.~Rept.~44,~249~(1978)}}.

\bibitem{Usyukina:1983gj}
N.~I.~Usyukina,
\textit{``{Calculation of many loop diagrams of perturbation theory}''},
\textsf{\doiref{10.1007/BF01017127}{Theor.~Math.~Phys.~54,~78~(1983)}},
[Teor. Mat. Fiz.54,124(1983)].

\bibitem{Kazakov:1984km}
D.~I.~Kazakov,
\textit{``{The method of uniqueness, a new powerful technique for multiloop
  calculations}''},
\textsf{\doiref{10.1016/0370-2693(83)90816-X}{Phys.~Lett.~133B,~406~(1983)}}.

\bibitem{Kazakov:1983ns}
D.~I.~Kazakov,
\textit{``{Calculation of Feynman integrals by the method of
  ‘uniqueness'}''},
\textsf{\doiref{10.1007/BF01018044}{Theor.~Math.~Phys.~58,~223~(1984)}},
[Teor. Mat. Fiz.58,343(1984)].

\bibitem{Kazakov:1983pk}
D.~I.~Kazakov,
\textit{``{Multiloop Calculations: Method of Uniqueness and Functional
  Equations}''},
\textsf{\doiref{10.1007/BF01034829}{Theor.~Math.~Phys.~62,~84~(1985)}},
[Teor. Mat. Fiz.62,127(1984)].

\bibitem{Zamolodchikov:1980mb}
A.~B.~Zamolodchikov,
\textit{``{'Fishnet' diagrams as a completely integrable system}''},
\textsf{\doiref{10.1016/0370-2693(80)90547-X}{Phys.~Lett.~97B,~63~(1980)}}.

\bibitem{Derkachov:2001yn}
S.~E.~Derkachov, G.~P.~Korchemsky and A.~N.~Manashov,
\textit{``{Noncompact Heisenberg spin magnets from high-energy QCD: 1. Baxter Q
  operator and separation of variables}''},
\textsf{\doiref{10.1016/S0550-3213(01)00457-6}{Nucl.~Phys.~B617,~375~(2001)}},
\texttt{\arxivref{hep-th/0107193}{hep-th/0107193}}.

\bibitem{Lipatov:1993qn}
L.~N.~Lipatov,
\textit{``{High-energy asymptotics of multicolor QCD and two-dimensional
  conformal field theories}''},
\textsf{\doiref{10.1016/0370-2693(93)90951-D}{Phys.~Lett.~B309,~394~(1993)}},
in: \textit{``{QCD and high-energy hadronic interactions. Proceedings, Hadronic
  Session of the 28th Rencontres de Moriond, Moriond Particle Physics Meeting,
  Les Arcs, France, March 20-27, 1993}''},
394-396p,
[,147(1993)].

\bibitem{Lipatov:1993yb}
L.~N.~Lipatov,
\textit{``{Asymptotic behavior of multicolor QCD at high energies in connection
  with exactly solvable spin models}''},
\textsf{JETP~Lett.~59,~596~(1994)},
\texttt{\arxivref{hep-th/9311037}{hep-th/9311037}},
[Pisma Zh. Eksp. Teor. Fiz.59,571(1994)].

\bibitem{Symanzik:1972wj}
K.~Symanzik,
\textit{``{On Calculations in conformal invariant field theories}''},
\textsf{\doiref{10.1007/BF02824349}{Lett.~Nuovo~Cim.~3,~734~(1972)}}.

\bibitem{Isaev:2003tk}
A.~P.~Isaev,
\textit{``{Multiloop Feynman integrals and conformal quantum mechanics}''},
\textsf{\doiref{10.1016/S0550-3213(03)00393-6}{Nucl.~Phys.~B662,~461~(2003)}},
\texttt{\arxivref{hep-th/0303056}{hep-th/0303056}}.

\bibitem{Chicherin:2012yn}
D.~Chicherin, S.~Derkachov and A.~P.~Isaev,
\textit{``{Conformal algebra: R-matrix and star-triangle relation}''},
\textsf{\doiref{10.1007/JHEP04(2013)020}{JHEP~1304,~020~(2013)}},
\texttt{\arxivref{1206.4150}{arXiv:1206.4150}}.

\bibitem{Belokurov:1983km}
V.~V.~Belokurov and N.~I.~Usyukina,
\textit{``{Calculation of ladder diagrams in arbitrary order}''},
\textsf{\doiref{10.1088/0305-4470/16/12/026}{J.~Phys.~A16,~2811~(1983)}}.

\bibitem{Preti:2017fjb}
M.~Preti,
\textit{``{WiLE: a Mathematica package for weak coupling expansion of Wilson
  loops in ABJ(M) theory}''},
\textsf{\doiref{10.1016/j.cpc.2017.12.011}{Comput.~Phys.~Commun.~227,~126~(2018)}},
\texttt{\arxivref{1707.08108}{arXiv:1707.08108}}.

\bibitem{Kazakov:2018}
V.~Kazakov, E.~Olivucci and M.~Preti,
\textit{``{Four-point correlation functions of double-scaled N=4 SYM}''},
\textsf{(to~appear)~,~~(2018)}.

\bibitem{Frolov:2005dj}
S.~Frolov,
\textit{``{Lax pair for strings in Lunin-Maldacena background}''},
\textsf{\doiref{10.1088/1126-6708/2005/05/069}{JHEP~0505,~069~(2005)}},
\texttt{\arxivref{hep-th/0503201}{hep-th/0503201}}.

\bibitem{Beisert:2005if}
N.~Beisert and R.~Roiban,
\textit{``{Beauty and the twist: The Bethe ansatz for twisted N=4 SYM}''},
\textsf{\doiref{10.1088/1126-6708/2005/08/039}{JHEP~0508,~039~(2005)}},
\texttt{\arxivref{hep-th/0505187}{hep-th/0505187}}.

\bibitem{Correa:2012nk}
D.~Correa, J.~Henn, J.~Maldacena and A.~Sever,
\textit{``{The cusp anomalous dimension at three loops and beyond}''},
\textsf{\doiref{10.1007/JHEP05(2012)098}{JHEP~1205,~098~(2012)}},
\texttt{\arxivref{1203.1019}{arXiv:1203.1019}}.

\bibitem{Bonini:2016fnc}
M.~Bonini, L.~Griguolo, M.~Preti and D.~Seminara,
\textit{``{Surprises from the resummation of ladders in the ABJ(M) cusp
  anomalous dimension}''},
\textsf{\doiref{10.1007/JHEP05(2016)180}{JHEP~1605,~180~(2016)}},
\texttt{\arxivref{1603.00541}{arXiv:1603.00541}}.

\bibitem{deLeeuw:2016vgp}
M.~de~Leeuw, A.~C.~Ipsen, C.~Kristjansen and M.~Wilhelm,
\textit{``{One-loop Wilson loops and the particle-interface potential in
  AdS/dCFT}''},
\textsf{\doiref{10.1016/j.physletb.2017.02.047}{Phys.~Lett.~B768,~192~(2017)}},
\texttt{\arxivref{1608.04754}{arXiv:1608.04754}}.

\bibitem{Aguilera-Damia:2016bqv}
J.~Aguilera-Damia, D.~H.~Correa and V.~I.~Giraldo-Rivera,
\textit{``{Circular Wilson loops in defect Conformal Field Theory}''},
\textsf{\doiref{10.1007/JHEP03(2017)023}{JHEP~1703,~023~(2017)}},
\texttt{\arxivref{1612.07991}{arXiv:1612.07991}}.

\bibitem{Preti:2017fhw}
M.~Preti, D.~Trancanelli and E.~Vescovi,
\textit{``{Quark-antiquark potential in defect conformal field theory}''},
\textsf{\doiref{10.1007/JHEP10(2017)079}{JHEP~1710,~079~(2017)}},
\texttt{\arxivref{1708.04884}{arXiv:1708.04884}}.

\bibitem{Gurdogan:2015csr}
O.~Gurdogan and V.~Kazakov,
\textit{``{New Integrable 4D Quantum Field Theories from Strongly Deformed
  Planar $\mathcal N = $ 4 Supersymmetric Yang-Mills Theory}''},
\textsf{\doiref{10.1103/PhysRevLett.117.201602,
  10.1103/PhysRevLett.117.259903}{Phys.~Rev.~Lett.~117,~201602~(2016)}},
\texttt{\arxivref{1512.06704}{arXiv:1512.06704}},
[Addendum: Phys. Rev. Lett.117,no.25,259903(2016)].

\end{thebibliography}

\end{document}